\documentclass[a4paper]{article}
\usepackage[utf8x]{inputenc}
\usepackage{color}
\usepackage{amsmath}
\usepackage{amssymb}
\allowdisplaybreaks 
\usepackage{tikz}
\usepackage{dcolumn}
\usepackage{array}
\usepackage{caption}
\usepackage{pbox}
\usepackage{geometry}
\numberwithin{equation}{section}

\begin{document}

\thispagestyle{empty}

\rightline{\small}

\vskip 3cm
\noindent
\noindent
{\LARGE \bf Duality and Modularity in Elliptic Integrable Systems  
}
\vskip .2cm
\noindent
{\LARGE \bf   and Vacua of  ${\cal N}=1^\ast$ Gauge Theories }
\vskip .8cm
\begin{center}
\linethickness{.06cm}
\line(1,0){447}
\end{center}

\vskip .8cm

\noindent
{\large \bf Antoine Bourget and Jan Troost}
\vskip 0.2cm
{\em \hskip -.05cm Laboratoire de Physique Th\'eorique\footnote{Unit\'e Mixte du
CNRS et
    de l'Ecole Normale Sup\'erieure associ\'ee \`a l'universit\'e Pierre et
    Marie Curie 6, UMR
    8549.}}
    \vskip -.05cm
{\em \hskip -.05cm Ecole Normale Sup\'erieure}
 \vskip -.05cm
{\em \hskip -.05cm 24 rue Lhomond, 75005 Paris, France}

\vskip 1cm

\vskip 0.6 cm

\noindent {\sc Abstract :}

\vskip  0.15cm

\noindent

We study complexified elliptic Calogero-Moser integrable systems.  We
determine the value of the potential at isolated extrema,  as a function of the modular
parameter of the torus on which the integrable system lives. We  calculate the extrema
 for low rank $B,C,D$ root systems using
a mix of analytical and numerical tools.  For $so(5)$ we find convincing evidence that the
extrema constitute a vector valued modular form for the $\Gamma_0(4)$ congruence subgroup of
the modular group. For $so(7)$ and $so(8)$, the extrema split into two sets. One set contains extrema
that make up   vector valued 
modular forms for congruence subgroups (namely $\Gamma_0 (4)$, $\Gamma(2)$ and $\Gamma(3)$), and a second set contains extrema that exhibit  monodromies around 
points in the interior of the fundamental domain. The former set can be described analytically, while for the latter, we provide an analytic value for the point of monodromy for $so(8)$, as well as
extensive numerical predictions for the Fourier coefficients of the extrema. Our results on the extrema provide a rationale
for integrality properties observed
in integrable models, and embed these into the theory of vector valued modular forms. Moreover, using the data we gather on the modularity of complexified integrable system extrema,  we analyse the
massive vacua of mass deformed ${\cal N}=4$
supersymmetric Yang-Mills theories with low rank gauge group of type $B,C$ and $D$.
We map out their transformation properties under the infrared 
electric-magnetic duality group as well as under triality for ${\cal N}=1^\ast$ with gauge algebra $so(8)$. 
 We compare the exact massive vacua on $\mathbb{R}^3 \times S^1$
to those found in a semi-classical analysis.
We identify several intriguing features of the quantum gauge theories.

\newpage

{
\tableofcontents }


\section{Introduction}

Four-dimensional gauge theories accurately describe  forces of
nature. Since solving them is hard, we may revert
to studying supersymmetric four-dimensional gauge theories, in which
the power of holomorphy lends a helping hand. Twenty years
ago, we realised how to solve for the low-energy effective action on
the Coulomb branch of ${\cal N}=2$ gauge theories in four dimensions
\cite{Seiberg:1994rs,Seiberg:1994aj}.  The solution techniques
were soon recognised to lie close to those studied in integrable systems
 \cite{Gorsky:1995zq,Martinec:1995by}. It is the bridge between integrable models 
 and supersymmetric gauge theories that we will further explore in this paper.
 We also attempt to reinforce both sides separately, and present results in a manner such that the
 contributions to these two domains may be read independently.

 The link between supersymmetric gauge theories and integrable systems was useful
 in writing down
 the low-energy effective action for 
 ${\cal N}=2^\ast$ gauge theory, namely ${\cal N}=4$ super
Yang-Mills theory with gauge group $G$, broken to ${\cal N}=2$  supersymmetry by
adding a mass term for one hypermultiplet.  For the gauge group
$G=SU(N)$ this program was completed in terms of a Hitchin integrable system with
$SL(N,\mathbb{C})$ bundle over a torus with puncture
\cite{Donagi:1995cf}. The associated elliptic Calogero-Moser system
permits generalisations to any root system, and allows for twists,
which were  used to provide Seiberg-Witten curves and differentials for
${\cal N}=2^\ast$ theory with
general gauge group $G$ \cite{D'Hoker:1998yi}. The generalisation 
was non-trivial since the elegant technique of lifting to
M-theory \cite{Witten:1997sc} is difficult to implement in the presence of orientifold planes (see
e.g. \cite{Uranga:1998uj,Ennes:1999fb}), while the relevant
generalised Hitchin integrable system has a gauge group which is 
 related to the  gauge group of the Yang-Mills theory in an intricate manner
\cite{Hurtubise:1999fw}. For a review of part of the history, see  the lectures \cite{D'Hoker:1999ft}.

We will be interested in breaking supersymmetry further, from ${\cal
  N}=2$ to ${\cal N}=1$ by adding another mass term for the
remaining chiral multiplet (providing us with three massive chiral
multiplets of arbitrary mass). We will study this ${\cal N}=1^\ast$ gauge theory with generic gauge group
$G$. With ${\cal N}=1$ supersymmetry, we hope to calculate the
effective superpotential ${W}$ at low energies exactly. For an
adjoint mass deformation from ${\cal N}=2$ to ${\cal N}=1$ this was
done in the original work
\cite{Seiberg:1994rs} in certain cases. For ${\cal N}=1^\ast$ and gauge
group $G=SU(N)$, the exact superpotential was proposed in
\cite{Dorey:1999sj} following the techniques of
\cite{Seiberg:1994rs,Seiberg:1996nz}. The superpotential is the
 potential of the complexified elliptic Calogero-Moser integrable
system associated to the root lattice of type $A_{N-1}$. In \cite{Kumar:2001iu} the
exact superpotential for ${\cal N}=1^\ast$ with more general gauge algebra was argued to be
the potential of the twisted elliptic Calogero-Moser system with root
lattice associated to the Lie algebra of the gauge group
$G$.  See \cite{Kim:2004xx} for further generalizations to ${\cal N}=1^\ast$ theories with twisted boundary
conditions on $\mathbb{R}^3 \times S^1$.

In this paper, we wish to analyse the proposed
exact superpotential in more detail. This involves a study of the properties
of the isolated extrema of the complexified and twisted elliptic Calogero-Moser integrable system. The results are of independent interest, and we have
therefore dedicated a first part of this paper to the study of the integrable systems per se.

The paper is structured as follows. In section \ref{ellint}, we 
review the relevant elliptic Calogero-Moser models. We pause to demonstrate a
Langlands duality between the $B$ and $C$ type integrable systems.
We then analyse the isolated extrema of the
complexified potential of low rank integrable systems of $B,C$ and $D$ type, and their  modular properties.
We observe the strong connection to vector valued modular forms. The
latter in turn provide a natural backdrop for integrality properties of
integrable systems (see e.g. \cite{CP,Corrigan:2002th,P}).  Section \ref{ellint} is the technical heart of the paper, and we will lay bare
many properties of the vector valued modular forms, using a combination of analytical work and extensive numerics. We will  analytically describe 
the potential in certain classes of extrema. 
We  also find sets of extrema that exhibit a monodromy in the interior of the fundamental domain.
In these cases we are able to calculate the monodromy, as well as to provide extensive numerical data for the integer valued coefficients describing the value of the potential at the extrema.

Finally, in section \ref{gaugetheory}, we reinterpret the
results we obtained  in terms of the physics of massive vacua of
${\cal N}=1^\ast$ theories. We compare our results for the quantum theory on $\mathbb{R}^3 \times S^1$ to semi-classical results for massive vacua
and discuss electric-magnetic
duality properties in the infrared under the modular group as well as the Hecke group. For $so(8)$, we also detail  the action of the global triality symmetry on the massive
vacua. We will encounter
 several interesting phenomena. We conclude in section
\ref{conclusions} and argue that we have only scratched the surface of  a broad field of open problems.

\section{Elliptic Integrable Systems and Modularity}
\label{ellint}
It is interesting to identify and study dynamical systems
that are integrable. Often they form solvable
subsectors of more complicated theories of even more physical interest.  
There exist one-dimensional
models of particles with  interactions that are integrable,
and the Calogero-Moser models of our interest are one such class
\cite{Calogero:1970nt,Sutherland:1971ks,Moser:1975qp}. These
models are associated to root systems of Lie algebras (amongst others). See e.g.
\cite{Olshanetsky:1981dk,Olshanetsky:1983wh} for a review.
Integrable systems are also known to have certain integrality properties. Namely, their
minimal energy, frequencies of small oscillations as well as eigenvalues of
Lax matrices are often expressible in terms of a series of integers \cite{CP,Corrigan:2002th,P}.

In this section, we study properties of (twisted) elliptic Calogero-Moser systems. We analyse the complexified model, defined on a torus with modular parameter $\tau$. In particular, we examine the extrema of the complexified potential, and exhibit their curious characteristics.

\subsection{The Elliptic Calogero-Moser models}
The member of the pyramid of Calogero-Moser integrable systems we concentrate on is
the elliptic Calogero-Moser model. We concentrate
on the models associated with a root system $\Delta$, as well as their
twisted counterparts. These models have a Hamiltonian with rank $r$ variables, with canonical
kinetic term, and a potential of the form:
\begin{eqnarray}
V_\Delta &=& g \sum_{\alpha \in \Delta} \wp ( \alpha (X) ; \omega_1,\omega_2)
\, ,
\end{eqnarray}
where $\wp$ is the Weierstrass elliptic function on a  torus 
with periods $2 \omega_1$ and $2 \omega_2$ and $g$ is a coupling constant. We choose the half-periods such that the
imaginary part of the modular parameter $\tau = \omega_2/\omega_1$ is positive.\footnote{See appendix \ref{ellipticfunctions} for more on 
our conventions for elliptic functions.} The vector $X$ lives in the space dual to the
root lattice of rank $r$ and the sum in the potential is over all the roots $\alpha$ of the root system $\Delta$.\footnote{
See appendix \ref{Lie} for our conventions and a compendium of properties of Lie algebras and Lie groups.} The model is integrable
for all Lie algebra root systems. 
The twisted elliptic Calogero-Moser model is defined in terms of twisted
Weierstrass functions:
\begin{eqnarray}
\wp_n (x ; \omega_1,\omega_2)
&=& \sum_{k \in \mathbb{Z}_n} \wp ( x + \frac{k}{n} 2 \omega_1 ; \omega_1,\omega_2) \, ,
\end{eqnarray}
which are summed over shifts by fractions of periods (thus in effect modifying 
that period).
We have a twisted elliptic Calogero-Moser model for all non-simply laced root systems 
and the value of $n$ is then given by the ratio of the length squared of the long versus the short roots.
We will be interested in the twisted elliptic Calogero-Moser model with potential:
\begin{eqnarray}
V_{\Delta,tw} &=& g_l  \sum_{\alpha_l \in \Delta_l} \wp ( \alpha_l (X) ; \omega_1,\omega_2)
+ g_s  \sum_{\alpha_s \in \Delta_s} \wp_n ( \alpha_s (X) ; \omega_1,\omega_2) \, ,
\end{eqnarray}
where $\alpha_l$ denote the long and $\alpha_s$ the short roots in the root system $\Delta=\Delta_l \cup \Delta_s$, and $g_{l}$ and $g_s$ are two coupling constants.
We will concentrate on the root systems $A_r$, $B_r$, $C_r$ and $D_r$ corresponding
to the classical algebras $su(r+1)$, $so(2r+1)$, $sp(2r)$ and $so(2r)$.
We allow complex values for the components of the vector $X$ (i.e. $X \in \mathbb{C}^r$).

\subsubsection*{The symmetries of the potential}
Let us discuss in detail the symmetries of the twisted elliptic Calogero model that act on the set of variables $X$. 
We first observe that
the Weyl group action leaves invariant the scalar product $\alpha(X) = (\alpha, X)$ and that 
the root system is Weyl invariant.\footnote{We mostly follow \cite{OV} for our conventions on 
Lie algebras. See also appendix \ref{Lie} for the definitions of the different lattices discussed hereafter. } This implies that the Weyl group action on $X$ leaves the potential
invariant. Secondly, we note that the outer automorphisms of the Lie algebra, which correspond to symmetries
of the Dynkin diagram, also leave the set of roots and the scalar product invariant. Therefore,
outer automorphisms as well form a symmetry of the model. 

Moreover, the periodicities of the model in the two directions
of the torus are as follows. By the definition of the 
dual weight, or co-weight lattice, we have that $\alpha(\lambda^\vee) \in \mathbb{Z}$ for all roots
$\alpha$. This implies that shifts of $X$ by $2 \omega_2  \, P^\vee$, namely shifts by periods times co-weights,
leave the potential invariant. 

To discuss the periodicity in the $\omega_1$ direction, we concentrate for simplicity on the algebras
$A,B,C$ and $D$, and normalize their long roots to have length squared two. We then have that for a long root $\alpha_l$ and a weight $\lambda$,
the equation $( \alpha_l , \lambda ) \in
\mathbb{Z}$ holds while for a short root $\alpha_s$ of the $B$ or $C$ algebras we have $( \alpha_s , \lambda ) \in
\frac{1}{2}\mathbb{Z}$, for all weights $\lambda$. 
As a consequence, the periodicity in the (twisted) $\omega_1$ direction is the lattice $ 2 \omega_1 \, P$ where 
$P$ is the weight lattice. 
The group of all symmetries is a semi-direct product of the lattice shifts, the Weyl group as well as the outer automorphism
group.

\subsection{Langlands Duality}
\label{dualityint}
Beyond the many features of these integrable systems already discussed in the literature,
the first supplementary property  that will be pertinent to our study of isolated extrema, 
is their behaviour under an inversion of the modular parameter $\tau$. We therefore briefly
digress in this subsection to discuss a few of the details of the duality.
Models associated to simply laced Lie algebras map to themselves under the
modular S-transformation 
$S : \tau \rightarrow -1/\tau$. This is easily confirmed using the transformation rule
(\ref{wpmod})
of the Weierstrass $\wp$ function under modular transformations.
We do have a non-trivial Langlands or short-long root duality between the
twisted elliptic Calogero-Moser model of B-type and the twisted model of C-type.
In order to exhibit the duality,  we make
the potential for the (twisted) $B_r=so(2r+1)$ theory more explicit:\footnote{For the non-simply laced cases, we will
always work with the twisted model, and we will drop the corresponding subscript on the potential from
now on.}
\begin{eqnarray}
V_{B} &=& b_l \left[
\sum_{i<j} \wp (x_i-x_j;\omega_1,\omega_2) + \wp(x_i+x_j;\omega_1,\omega_2) \right]+ b_s
\left[ \sum_{i=1}^r \wp(x_i;\omega_1,\omega_2) + \wp(x_i+\omega_1;\omega_1,\omega_2) \right] \, ,
\nonumber
\end{eqnarray}
and for the  $C_r=sp(2r)$ theory as well:
\begin{eqnarray}
V_{C} &=&  {c}_s \left[
\sum_{i<j} \wp (y_i-y_j;\omega'_1,\omega'_2) + \wp(y_i+y_j;\omega'_1,\omega'_2)
+  \wp (y_i-y_j + \omega'_1 ;\omega'_1,\omega'_2) + \wp(y_i+y_j +
\omega'_1;\omega'_1,\omega'_2) \right]
\nonumber \\
& & 
 + {c}_l   \sum_{i=1}^r \wp(2 y_i;\omega'_1,\omega'_2)  \, .
 \label{Cpotential}
\end{eqnarray}
We have chosen a standard parameterisation of the vector $X$ as well as the root systems, and we have assigned half-periods
$\omega_i$ to the B-system and $\omega_i'$ to the C-system. We have also made explicit the twisted Weierstrass functions 
$\wp_2$ with twisting index $2$, which is the ratio of lengths squared of the long and short roots.
To demonstrate the duality between these models, we use the 
elliptic function identities (\ref{halfperiod}) 
to manipulate the $so(2r+1)$ potential such that it becomes of the form of
the $sp(2r)$ potential:
\begin{eqnarray}
V_{B} &=& b_l \left[
\sum_{i<j} \wp (x_i-x_j; 2 \omega_2 , - \omega_1) +\wp (x_i-x_j + 2  \omega_2; 2
\omega_2 , - \omega_1)+ \wp(x_i+x_j;2 \omega_2 , - \omega_1) + \right. \nonumber \\ 
 & & \left. \wp(x_i+x_j + 2 \omega_2 ;2 \omega_2 , - \omega_1) \vphantom{\sum_{i<j}} \right]
+  b_s 
\sum_{i=1}^r  \wp(x_i; \omega_2 , - \omega_1/2)
\nonumber \\
& & - \frac{ \pi^2 r(r-1)}{24 \omega_2^2}  b_l \left[ 2 E_2
\left(-\frac{\omega_1}{\omega_2}\right)-E_2 \left(-\frac{\omega_1}{2 \omega_2} \right) \right]
+ \frac{\pi^2 r}{6 \omega_1^2}  b_s  \left[ 2 E_2
\left(2 \frac{\omega_2}{\omega_1} \right)-E_2 \left(\frac{\omega_2}{ \omega_1} \right) \right]
\nonumber \\
& = &  b_l \left[
\sum_{i<j} \wp (x_i-x_j; 2 \omega_2 , - \omega_1) +\wp (x_i-x_j + 2  \omega_2; 2
\omega_2 , - \omega_1)+ \wp(x_i+x_j;2 \omega_2 , - \omega_1) + \right. \nonumber \\ 
 & & \left. \wp(x_i+x_j + 2 \omega_2 ;2 \omega_2 , - \omega_1) \vphantom{\sum_{i<j}} \right]
+  b_s 
\sum_{i=1}^r  \wp(x_i; \omega_2 , - \omega_1/2)
\nonumber \\
& & 
+ \frac{\pi^2}{12 \omega_1^2} (2 r b_s +  r (r-1) b_l)  \left[2 E_2
\left(2 \frac{\omega_2}{\omega_1} \right)-E_2 \left(\frac{\omega_2}{ \omega_1} \right) \right]
\, .
\label{transformedB}
\end{eqnarray}
In the last equality, we used the modular transformation rule (\ref{S2E2}) for a combination of
second Eisenstein series.
We observe that the end result (\ref{transformedB}) can be identified with the $C_r$ 
potential (\ref{Cpotential}), provided we match parameters as follows:
\begin{eqnarray}
\omega'_1 =2 \omega_2
\qquad
\omega'_2 = - \omega_1
\qquad 
y_i = x_i & & 
\nonumber \\
c_s = b_l
\qquad
c_l = 4 b_s  \, ,
\qquad  & &  
\label{dualitymap}
\end{eqnarray}
and we allow for a $\tau$-dependent shift of the potential that invokes the second
Eisenstein series $E_2$. These identifications imply a duality (which we will denote $S_2$) between the modular
parameters of the $B$ and $C$-type integrable systems:
\begin{eqnarray}
\tau^B & \equiv & - \frac{1}{2 \tau^C} \, . 
\label{bccoupling}
\end{eqnarray}
In the following, we will be interested in $B$ and $C$ models in which the ratio of the long to short 
root coupling
constants is equal to two, i.e. we put $b_l=b=2 b_s$ and $c_l=c=2c_s$.\footnote{
Various particular choices of parameters and observables that we make
in section \ref{ellint} are
  motivated by the gauge theory applications that we will discuss in section \ref{gaugetheory}. It is also of interest
  to study the integrable systems more generally.}
It is important that this relation is  compatible
with the duality map (\ref{dualitymap}).
We rewrite the identity of the potentials for this specific ratio
of parameters:
\begin{eqnarray*}
\sum_{i<j} \wp (x_i-x_j;\omega_1,\omega_2) + \wp(x_i+x_j;\omega_1,\omega_2) +
\frac{1}{2}
\left( \sum_i \wp(x_i;\omega_1,\omega_2) + \wp(x_i+\omega_1;\omega_1,\omega_2) \right) = 
\nonumber \\
 \sum_{i<j} \wp (x_i-x_j; 2 \omega_2 , - \omega_1) +\wp (x_i-x_j + 2  \omega_2;
2 \omega_2 , - \omega_1)+ \wp(x_i+x_j;2 \omega_2 , - \omega_1) +
\nonumber \\
 \\ 
\wp(x_i+x_j + 2 \omega_2 ;2 \omega_2 , - \omega_1) + 
2 \sum_i  \wp(2 x_i;2 \omega_2 , - \omega_1)
+ \frac{ \pi^2 r^2}{12 \omega_1^2} \left[ 2 E_2
\left( 2\frac{\omega_2}{\omega_1} \right) -E_2 \left(\frac{\omega_2}{\omega_1} \right) \right]  \, , \nonumber
\end{eqnarray*}
and the integrable system duality can be summarised as: 
\begin{equation}
V_B (x_i , \tau) = \frac{1}{2 \tau ^2} V_C \left( \frac{x_i}{2 \tau} , - \frac{1}{2 \tau} \right) + \frac{ \pi^2 r^2}{3} \left[ 2 E_2
(2 \tau)-E_2 (\tau) \right] \, ,
\label{BCduality}
\end{equation}
when we use the rescaling (\ref{periodstotau}).
The duality may be viewed as a standard Langlands duality. We went through its detailed derivation
since the $\tau$-dependent shift in the duality transformation (\ref{BCduality}) is important for later purposes.

\subsubsection{Langlands duality at rank two}
There is a further special case of low rank which is of 
particular interest to us in the following.  The $B$ and $C$ type Lie algebras
of rank two are identical: $so(5) \equiv sp(4)$. If we apply the duality of $B$ and $C$ type 
potentials to this special case, we derive that the
following transformations leave the potential invariant:
\begin{eqnarray}
\omega_1' = 2 \omega_2
\qquad  \qquad
\omega'_2 = - \omega_1
& & \quad c'  = 2 b
\nonumber \\
 x_2'-x_1' = 2 x_1
 \qquad  
x_1'+x_2' = 2 x_2 \, . & & 
\end{eqnarray}
If we parameterise the potential in terms of the modular parameter $\tau=\omega_2/\omega_1$, the duality
transformation for $so(5)$ reads:
\begin{equation}
 V_{so(5)} (x_1,x_2,\tau) = \frac{1}{2 \tau^2}  V_{so(5)}
\left(\frac{x_1+x_2}{2 \tau},\frac{x_1-x_2}{2 \tau},-\frac{1}{2 \tau}\right) +
\frac{4 \pi^2}{3} [ 2 E_2
(2 \tau)-E_2 (\tau) ] \, .
\label{duality2}
\end{equation}
In summary,
we derived a Langlands duality between $B$ and $C$ type (twisted) elliptic Calogero-Moser
models. The resulting identities captured in equations  (\ref{BCduality}) and (\ref{duality2}) and the shifts appearing in these duality transformations will be 
useful. We return to the more general discussion of the integrable systems, and in particular their extrema.

\subsection{Integrable Models at Extrema}
There have been many studies of classical integrable models at
equilibrium. These have uncovered remarkable properties, like the
integrality of the minimum of the potential and of the frequencies of
small oscillations around the minimum, amongst others (see e.g. \cite{CP,Corrigan:2002th,P}). We will analyse
the potential of certain elliptic integrable systems evaluated at
generalised equilibrium positions. We show that they give rise to interesting
vector valued modular forms as well as more general non-analytic modular
vectors.  Modularity provides a more conceptual
way of understanding the integrality properties of the integrable
system. This rationale then continues to hold for the
integrable systems that can be obtained from the elliptic
Caloger-Moser systems by limiting procedures (e.g. the trigonometric models). Thus, studying {\em elliptic} integrable
systems, depending on a modular parameter, is found to have an additional pay-off.

It is known that $A$-type integrable systems often have simpler properties
than do the integrable systems associated with other root systems.
As a relevant example, let us quote the fact that the (real) Calogero-Moser (Sutherland)
system with trigonometric potential of $A$-type has equally spaced equilibrium positions along the real axis,
while the $B,C,D$-type  potentials have minima associated to zeroes of Jacobi polynomials \cite{Corrigan:2002th},
which satisfy known relations \cite{Szego}, but are not known explicitly in general.  The elliptic Calogero-Moser
systems that we examine show a similar dichotomy. Extrema of the (complex) elliptic $A$-model are equally
spaced. This fact leads to relatively easily constructable values for the potential at extrema, for any rank
\cite{Donagi:1995cf,Dorey:1999sj,Aharony:2000nt}.
For the $B,C,D$-type models that we study in this paper, much less is known, and we need to combine numerical
searches with analytic approaches to determine the extremal values of the potential, for low rank cases.

To be more precise, we will be interested in extrema of the complexified potential,
satisfying:\footnote{This will correspond, in section \ref{gaugetheory}, to a supersymmetric vacuum in the ${\cal N}=1^\ast$ gauge theory, where the effective superpotential $W$
is identified with the potential $V$ of the integrable system. }
\begin{eqnarray}
\partial_{X_i} V (X_j) = 0 \qquad \forall i \, , \label{extr}
\end{eqnarray}
and we moreover demand that at the extremum (\ref{extr}) the function
\begin{eqnarray}
\sum_{i=1}^r | \partial_{X^i} V(X^j) |^2
\label{nonflat}
\end{eqnarray}
not posses any flat directions.\footnote{This condition implies that the vacuum is massive in the 
supersymmetric gauge theory. We briefly comment on massless vacua later on.}

Recall that the group of symmetries acting on the variables $X$ were a lattice group of translations,
the Weyl group as well as the outer automorphisms of the Lie algebra. Using these symmetries, we will introduce
a notion of equivalence on the variables $X$. We will consider the vector $X$ to be identified by the periodicities
of the model.
The periodicity in the $\omega_1$ direction is given by the weight lattice $P$, while in the
$\omega_2$ direction it is the co-weight lattice $P^\vee$. Furthermore, we will consider extrema that are related by the action of the Weyl group of the Lie algebra to be equivalent. By contrast, 
outer automorphisms are taken to be global
symmetries of the problem. When the global symmetry group is broken by a given extremum,
 the global symmetries will generate a set of degenerate extrema.

\subsection{The Case $A_r = su(r+1)$}
The extrema of the elliptic Calogero-Moser model of type $A_r$ have been studied in
great detail, mostly in the context of supersymmetric gauge theory
dynamics (see
e.g. \cite{Donagi:1995cf,Dorey:1999sj,Aharony:2000nt}).
Firstly, we remark that in this case, the equivalence relations that follow from
the periodicity of the potential as well as the Weyl symmetry group of the Lie
algebra are straightforwardly implemented. We use the
parameterisation of simple roots in terms of orthogonal vectors
$\alpha_i = e_i -e_{i+1}$, and the fundamental weights then read
$\pi_i = \sum_{j=1}^i e_j$, with weight lattice spanned by the vectors
$e_i$.  We can parameterise the coordinates of our integrable system
by a vector $X_j e^j$ living in the dual to the root space (and $e^j(e_i)={\delta^j}_i$).  The Weyl
group $S_n$ acts by permuting the components 
$X_j$. We can shift one of the components $X_j$ to
zero by convention. The equivalence under shifts by fundamental
weights is identical to the toroidal periodicity relations for the individual
coordinates $X_j$.  The inequivalent extrema of the $su(n)$ potential
(satisfying the additional condition (\ref{nonflat}) of non-flatness) are then argued
to correspond one-to-one to sublattices of order $n$ of the torus with
modular parameter $\tau$ \cite{Donagi:1995cf,Dorey:1999sj}. These extrema are classified by two integers
$p$ and $k$ satisfying that $p$ is a divisor of $n$ and $k \in \{ 0,1,\dots,\frac{n}{p}-1 \}$.
The number of extrema
is equal to the
sum of the divisors of $n$.
The $\mathbb{Z}_2$ outer automorphism of
$A_{r>1}$ acts trivially on the minima, since it acts by permutation, combined with 
a sign flip for all $X_j$, which leaves a sublattice ankered at the origin
invariant.

The value of the potential at
one of these extrema is (with a given choice of coupling constant):
\begin{eqnarray}
V_{A_{n-1}}(\tau) &=& \frac{n^3}{24} \left( E_2 (\tau) - \frac{p}{q} E_2 \left( \frac{p}{q} \tau + \frac{k}{q}\right)\right) \, .
\label{Aextrema}
\end{eqnarray}
Under the $SL(2,\mathbb{Z})$ action on the torus modular parameter $\tau$,
the sublattices of order $n$ of the torus are permuted into each other (in a way that 
depends intricately on the integer $n$). The permutation of the sublattices also entails the 
permutation of the values (\ref{Aextrema}) at these extrema
under $SL(2,\mathbb{Z})$. The list of extremal
values of the elliptic Calogero-Moser model therefore form a vector valued modular form
(see e.g. \cite{KM,Bantay:2007zz,Gannon:2013jua})
of weight two under the group $SL(2,\mathbb{Z})$.
The associated representation of the modular group is a representation in terms of permutations
specified by the $SL(2,\mathbb{Z})$ action on sublattices of order $n$. One can identify a subgroup of the modular group under which a given
component of the vector-valued modular form is invariant, and then use minimal data
to fix it \cite{Ritz:2006ji}. 

In summary, the extrema of the Calogero-Moser model of type $A_r=su(r+1)$
are under analytic control. The positioning of the extrema can be
expressed linearly in terms of the periods of the model, and the
vector valued modular form of extremal values for the potential
has an automorphy factor that can be characterised by sublattice permutation 
properties. The extremal values are generalised Eisenstein series of weight two under congruence
subgroups of the modular group.

\subsection{The $B,C,D$ Models}
\label{BCDmodels}

For other algebras, we are at the moment only able to study low rank cases. {From} the analysis,
it is clear that crucial simplifying properties of the $A_r$ case are
absent. Nevertheless, generic features of the $A_r$ case persist in a subclass of extrema, in that we find
vector-valued  modular forms as extremal values for the potential. We also
find a class of extremal values that exhibit new features. 

To describe in detail which extrema are considered to be equivalent,
we must discuss the equivalence relations that we mod out by for the $B,C$ and $D$ root
systems individually.

\subsubsection*{$D_r=so(2r)$}
For the $D_r$ case, we can parameterise the roots as $\alpha_i = e_i - e_{i+1}$ (for $i
\in \{ 1,2,\dots, r-1 \}$) and $\alpha_r = e_{r-1}+e_r$. We put $X=X_j
e^j$ and imply that the relation $e_i(e^j)={\delta_i}^j$ holds. The equivalence of the vector $X$ under shifts proportional to the weight lattice
implies that each variable $X_j$ lives on a torus with modular parameter
$\tau$. It moreover identifies the vector $X$ with the vector $X$ shifted
by a half-period in each variable simultaneously.
The Weyl
group is $W(so(2r)) = S_r \ltimes \mathbb{Z}_2^{r-1}$, and acts by
permutation of the components $X_j$, as well as the sign change of an even number
of them.  The outer automorphism group (for $r \neq 4$) is
equal to $\mathbb{Z}_2$ and acts as $X_r \rightarrow - X_r$.  For $r=4$, the global symmetry group
is $S_3$ triality.

\subsubsection*{$B_r=so(2r+1)$}
For $B_r$, the roots are $\alpha_i = e_i - e_{i+1}$ (for $i
\in \{ 1,2,\dots, r-1 \}$) and $\alpha_r = e_r$. 
We recall that
the periodicity is the weight lattice in the $\omega_1$ direction (due to the twist), and the co-weight
lattice in the $\omega_2$ direction.
Thus, we can shift
components of the vector $X=X_j e^j$ by periods, or all components simultaneously by a half
period in the $\omega_1$ direction.  In the $\omega_2$ direction, we allow 
shifts of the individual components by periods. The Weyl group acts by
combinations of permutations and any sign flip of the coordinates.

\subsubsection*{$C_r=sp(2r)$}
The roots are $\alpha_i = (e_i - e_{i+1})/\sqrt{2}$ (for $i
\in \{ 1,2,\dots, r-1 \}$) and $\alpha_r = \sqrt{2} e_r$.\footnote{By our conventions, 
we normalise the long roots such that they
have length squared two.} We can shift
components $X_j$ of $X=\sqrt{2} X_j e^j$ by half-periods in the $\omega_1$ direction, while in the $\omega_2$ direction,
we can allow shifts by any period, as well as a half-period shift of all
$X_j$ simultaneously. The Weyl group allows
any permutation and sign flip of the coordinates.
\noindent
The equivalence relations and symmetries in the $B,C$ and $D$ cases, beyond permutation symmetries and toroidal
periodicity, are summarized in the table:
\begin{center}
\renewcommand{\arraystretch}{1.3}
\setlength{\extrarowheight}{2pt}
 \begin{tabular}{|c|l|}
\hline
$B_r$ & \parbox[t]{8cm}{Individual $X_i \rightarrow -X_i$ }
\\
&\parbox[t]{8cm}{Collective $X_i
\rightarrow X_i + \omega_1$} 
\\
\hline
$C_r$ &  \parbox[t]{8cm}{Individual $X_i \rightarrow -X_i$ 
and $X_i \rightarrow X_i + \omega_1$ } \\
& \parbox[t]{8cm}{Collective $X_i \rightarrow X_i + \omega_2$} 
\\
\hline
$D_r$ & \parbox[t]{8cm}{Even number of sign flips $X_i \rightarrow 
-X_i$ } \\  & \parbox[t]{8cm}{ Collective $X_i
\rightarrow X_i + \omega_1$ and $X_i
\rightarrow X_i + \omega_2$ } \\
 & \parbox[t]{8cm}{Global symmetries : $\mathbb{Z}_2$ generically and $S_3$ for $D_4$.}
\\
\hline
\end{tabular}
\end{center}

\vskip.2cm

Armed with this detailed knowledge about the equivalence of configurations,
we programmed a numerical search for isolated extrema. In the following
subsections, we list
the results we found by root system. For simply
laced root systems we studied the elliptic Caloger-Moser model, while results
for non-simply laced root systems correspond to the twisted elliptic Calogero-Moser
model with a coefficient for the short root term which is equal to one half the coefficient
in front of the long root terms (as described below equation (\ref{bccoupling})).

\subsection{The Case $C_2=sp(4)=so(5)$ and Vector Valued Modular Forms}
Since the root system $C_2$ is the first example of our series, we provide a detailed discussion.
We discuss the positions of the isolated extrema, the series expansions relevant
to the potential at these extrema, the action of the duality group, as well as the 
identification of the relevant vector valued modular forms.
\subsubsection{The positions of the extrema}

\begin{figure}
\caption{Extrema at $\tau=i$ for the Lie algebra $so(5)$}
\label{extremaso5}
\begin{minipage}{\linewidth}



\begin{tabular}{p{4.8cm}p{4.8cm}p{4.8cm}}

{%
\setlength{\fboxsep}{8pt}%
\setlength{\fboxrule}{0pt}%
\fbox{\includegraphics[width=3.5cm]{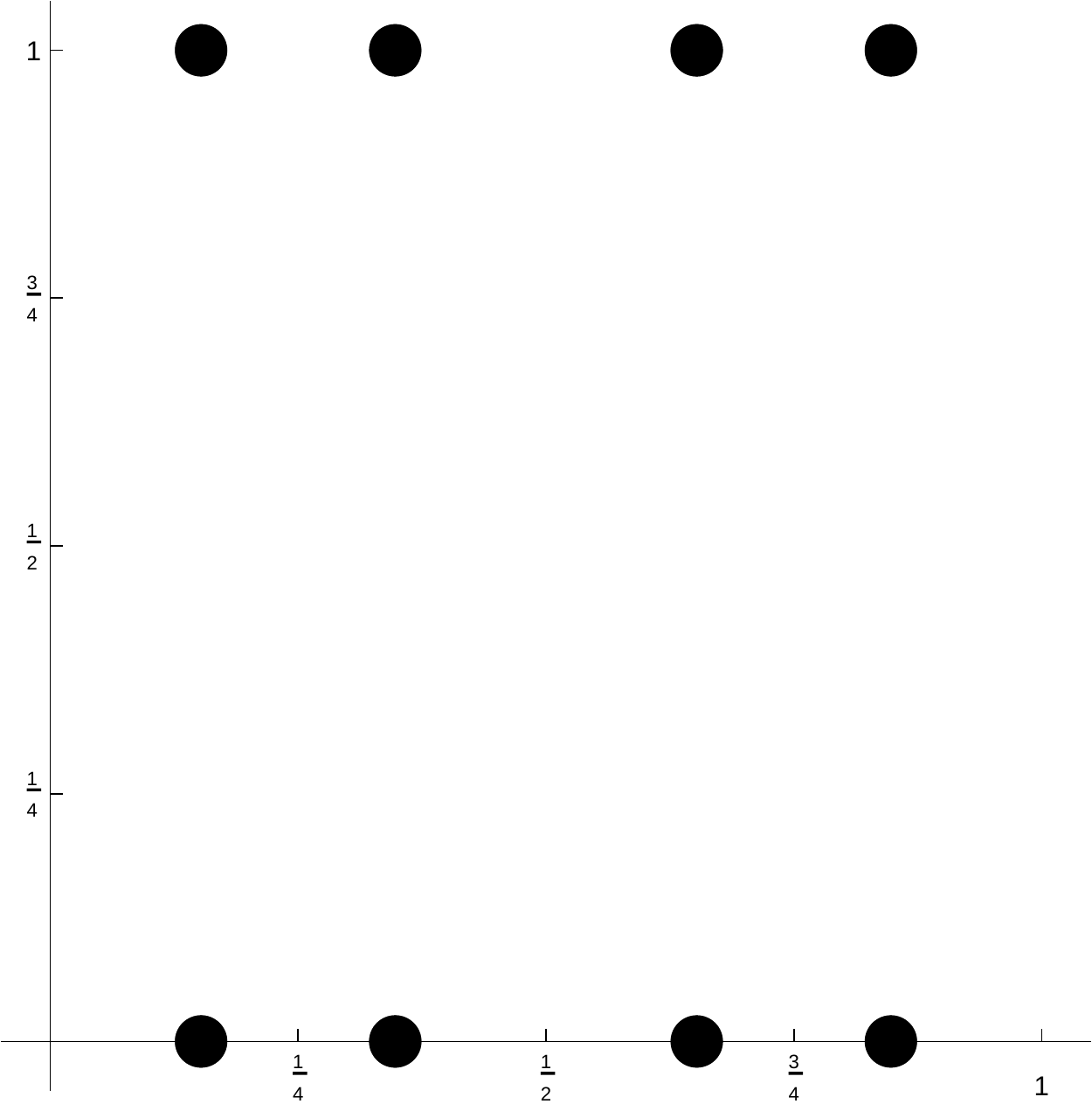}}%
}%

\begin{center}

{Extremum 1}

\end{center}

&
{%
\setlength{\fboxsep}{8pt}%
\setlength{\fboxrule}{0pt}%
\fbox{\includegraphics[width=3.5cm]{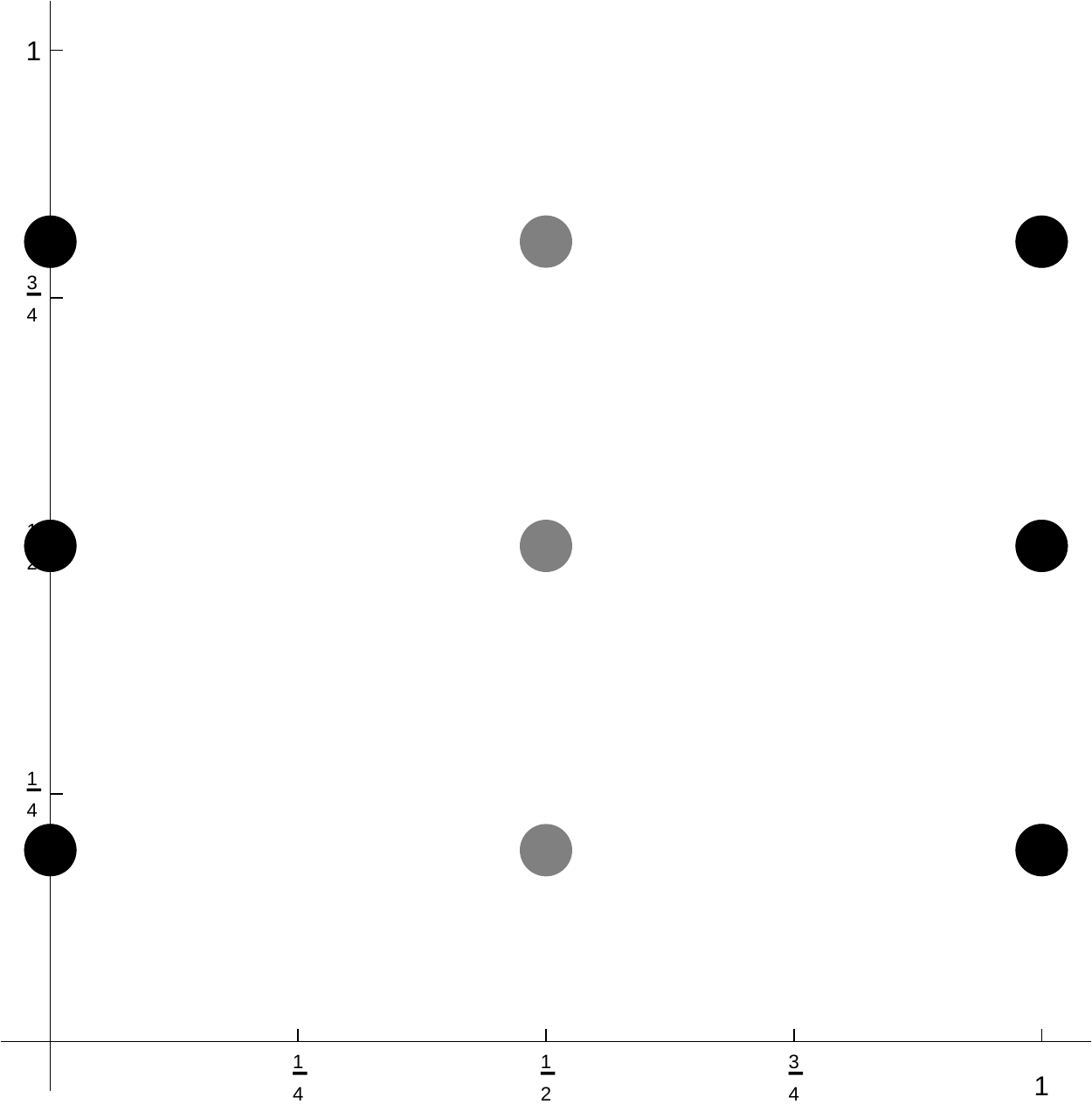}}%
}%

\begin{center}
Extremum 2
\end{center}


%
 &
 {%
\setlength{\fboxsep}{12pt}%
\setlength{\fboxrule}{0pt}%
\fbox{\includegraphics[width=3.5cm]{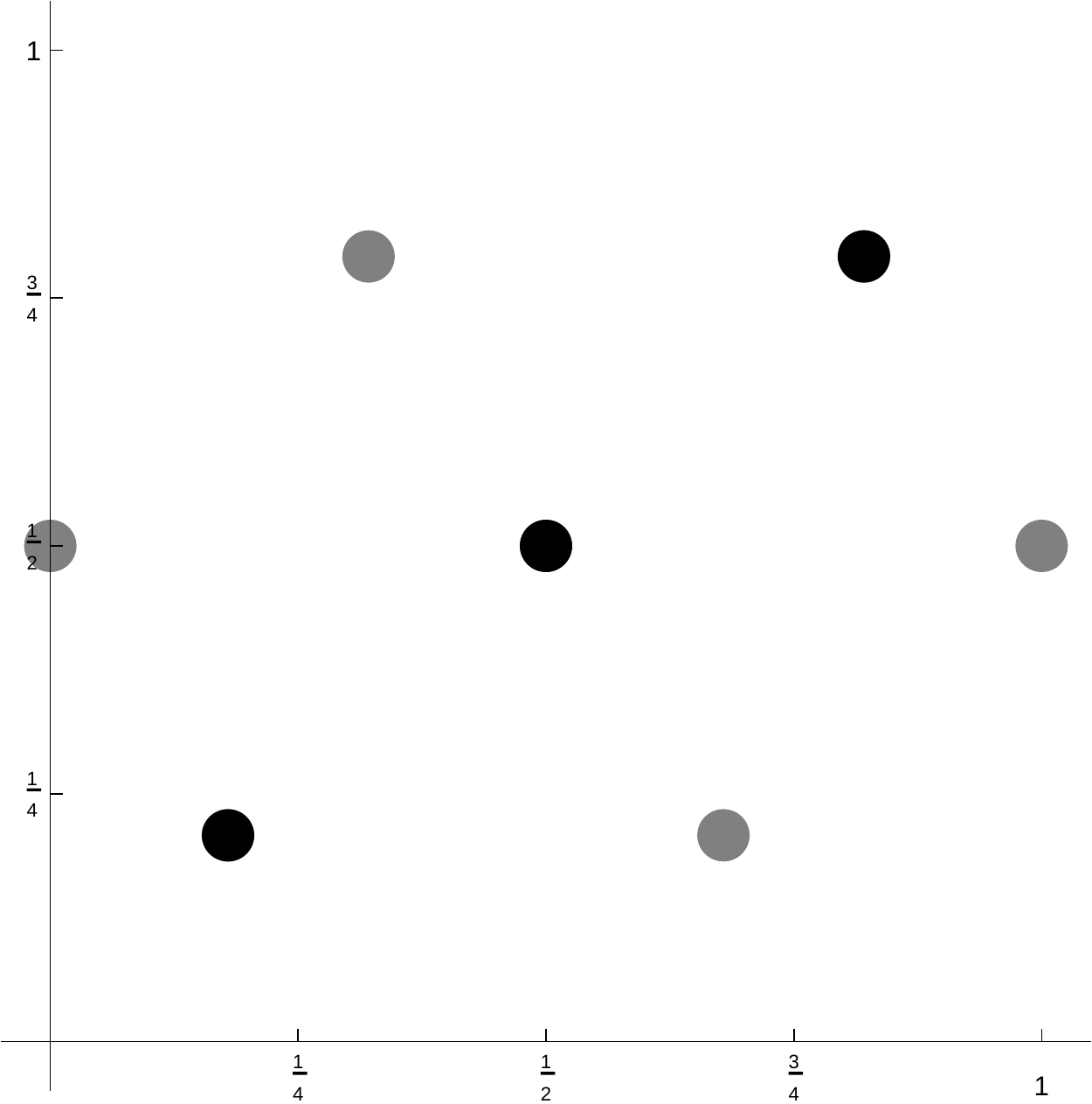}}%
}%

\begin{center}
Extremum 3
\end{center}

\\
{%
\setlength{\fboxsep}{12pt}%
\setlength{\fboxrule}{0pt}%
\fbox{\includegraphics[width=3.5cm]{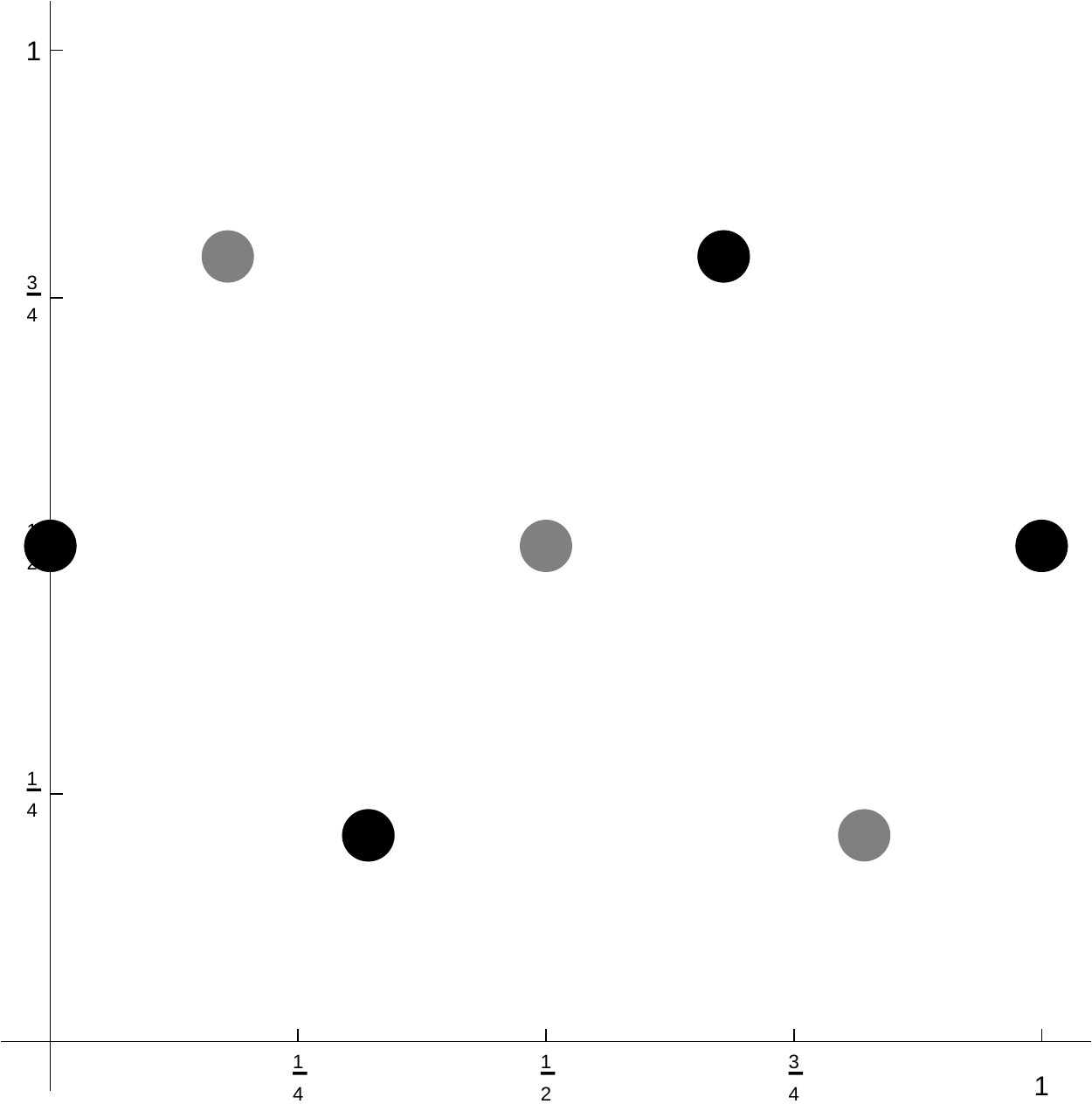}}%
}%

\begin{center}
Extremum 4
\end{center}

 & 
 {%
\setlength{\fboxsep}{12pt}%
\setlength{\fboxrule}{0pt}%
\fbox{\includegraphics[width=4cm]{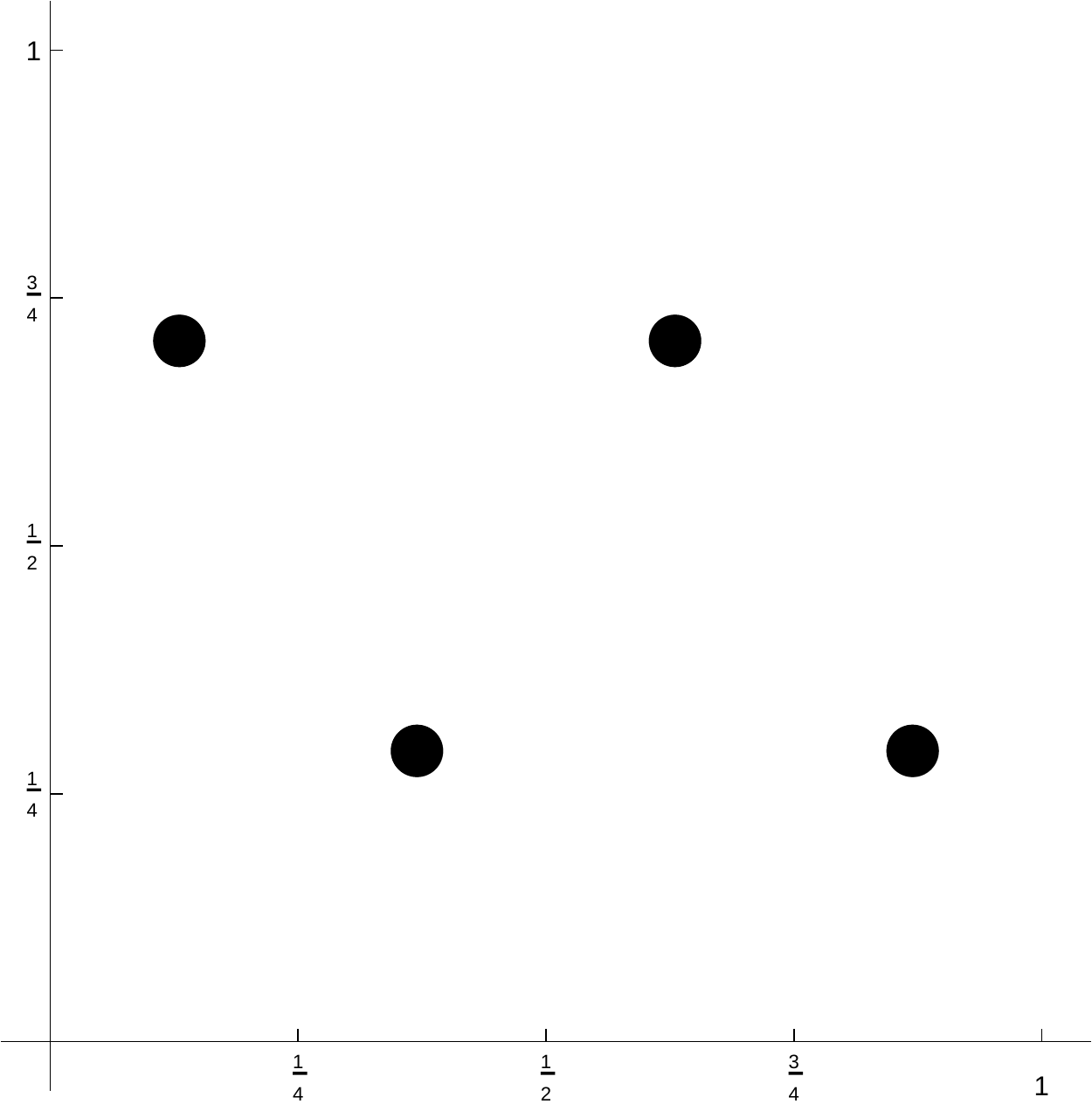}}%
}%

\begin{center}
Extremum 5
\end{center}
 & 
 {%
\setlength{\fboxsep}{12pt}%
\setlength{\fboxrule}{0pt}%
\fbox{\includegraphics[width=4cm]{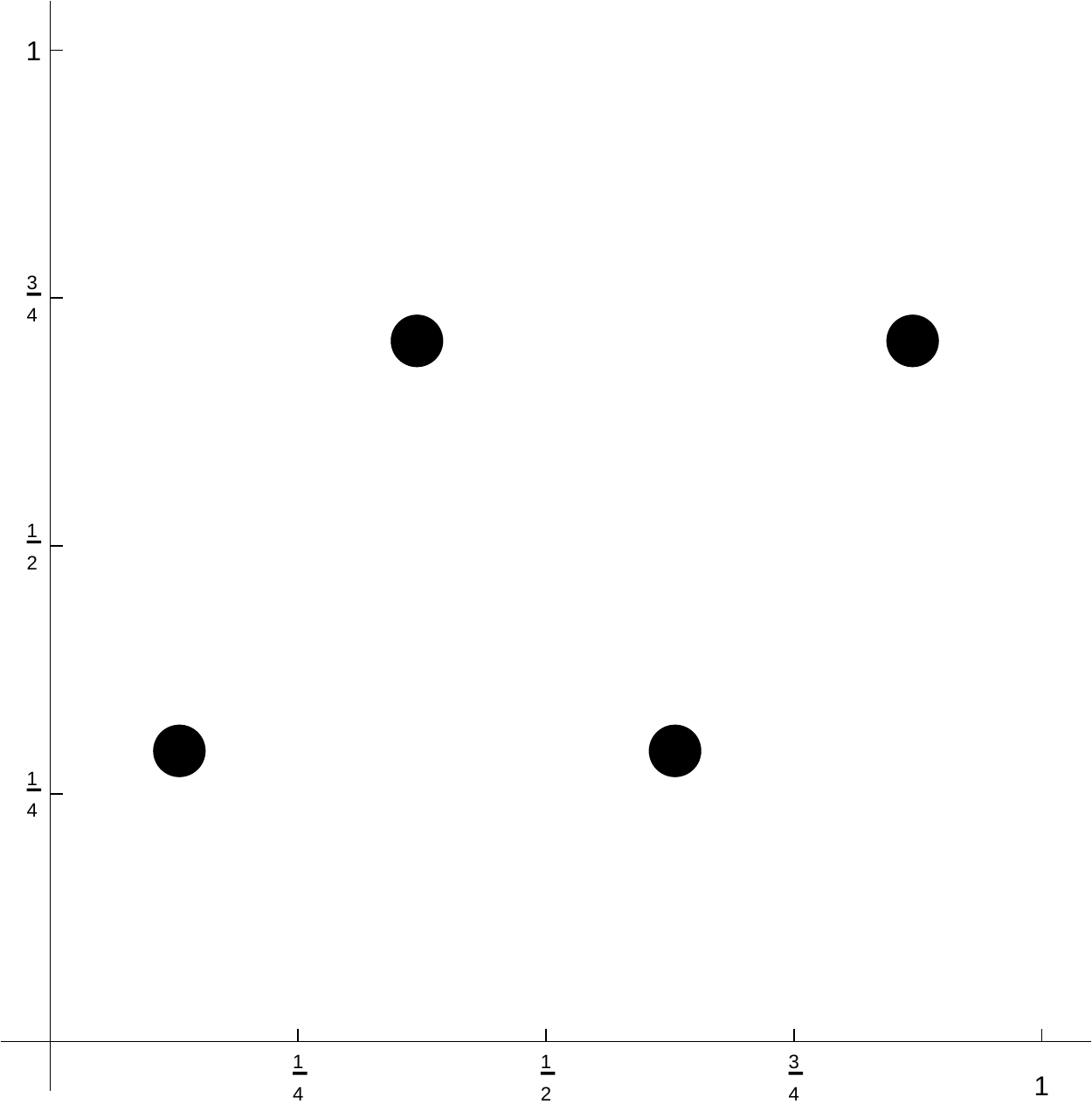}}%
}%

\begin{center}
Extremum 6
\end{center}

\\
{%
\setlength{\fboxsep}{12pt}%
\setlength{\fboxrule}{0pt}%
\fbox{\includegraphics[width=4cm]{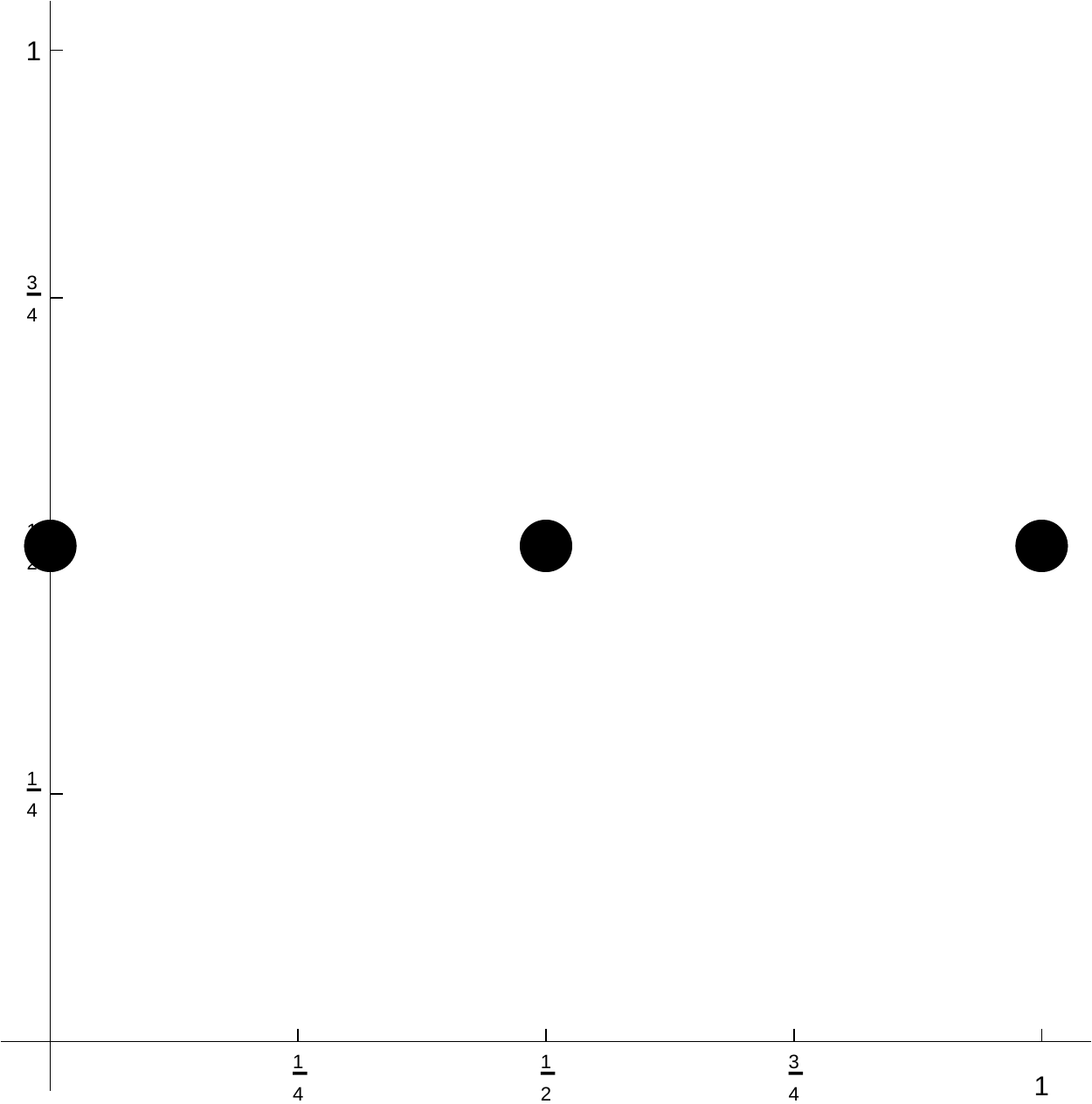}}%
}%

\begin{center}
Extremum 7
\end{center}

& & 
\end{tabular}

\end{minipage}
\end{figure}

For the Lie algebra $so(5)=sp(4)$ we found
7 isolated extrema of the potential. We provide their positioning at $\tau=i$ 
in figure \ref{extremaso5}. 
We have drawn in bold the positions of the extrema as well as their opposites, in a fundamental
cell of the torus.\footnote{We have indicated reflections over other half-periods in grey, to illustrate that
the minima are close to forming sublattice structures.}


These numerical
results were found using a Mathematica program, which was written around 
 the built-in function \texttt{FindMinimum}. Careful programming augments the precision 
of the algorithm to at least two hundred digits. The most costly part of the algorithm is the random search for extrema.
Indeed, the intricate landscape drawn by the potential can hide extrema. 
We gave a drawing of the position of the numbered extrema on the torus with modular
parameter $\tau=i$. 
The positions of the extrema
for other values of the modular 
parameter can be reached by interpolation. We have analytic control over a few extra
properties of the extrema. E.g. if we follow extremum 1 to $\tau = i \infty$,
we find that the equilibrium positions are given by $\frac{1}{2 \pi} \arccos (\pm 1/\sqrt{3})$ where $ \pm 1/\sqrt{3}$ are the zeroes of the Jacobi polynomial
$P^{(0,0)}_2$. 
The first extremum, which we label 1, lies on the real axis and is the equilibrium position of the real integrable
system. The extremum 2 lies on the imaginary axis, while extrema 3 and 4 are 
then approximately obtained by
applying the transformation $\tau \rightarrow \tau +1$.
The extrema 5 and 6 are $S_2$ Langlands
duals of extrema 3 and 4.
It is easy to deduce from the potential that
 the positions of the extrema generically behave non-linearly as a function of  
$\tau$. 

\subsubsection{Series expansions of the extrema}

By numerically evaluating the extrema of the potential for a range of values of 
the modular parameter $\tau$, we are able to write the extrema as an expansion
in terms of a power of the modular parameter $q=e^{2 \pi i \tau}$. The extremal values
can be written in terms of the series:
\begin{eqnarray}
A_0(q) &=& 
 \frac{1}{24} + q + q^2 + 4 q^3 + q^4 + 6 q^5 + 4 q^6 + 8 q^7 + q^8 + 13 q^9 + 
 6 q^{10} + 12 q^{11} \nonumber \\ 
  & & + 4 q^{12} + 14 q^{13} + 8 q^{14}  + \dots
  \label{A0numerics}
  \\
A_1(q) &=& 
  1 + 48 q + 828 q^2 + 8064 q^3 + 109890 q^4 + 1451520 q^5 + 
 11198088 q^6 + 141212160 q^7  \nonumber \\
  & & + 1666682811 q^8 + 9413050176 q^9 + 
 145022264892 q^{10} + 1838450006784 q^{11} \nonumber \\
  & & + 11103941590326 q^{12} + 
 138638111404032 q^{13} + \dots \nonumber \\
A_2(q) &=& 
  2 + 48 q + 576 q^2 + 9792 q^3 + 99576 q^4 + 743904 q^5 + 
 13146624 q^6 + 115737984 q^7 \nonumber \\
  & & + 1015727364 q^8 + 14338442448 q^9 + 
 102050482176 q^{10} + 935515738944 q^{11} \nonumber \\
  & & + 12532363069968 q^{12} + 
 122390111091744 q^{13} + \dots \nonumber \\
A_3(q) &=& \frac{13}{216} + 7 q + 541 q^2 + 24508 q^3 + 939669 q^4 + 19944842
q^5 + 
 764752180 q^6  \nonumber \\
  & & + 21016537080 q^7 + 905672825157 q^8 + 
 38827071780859 q^9 + 827503353279726 q^{10} + \dots \nonumber \\
 A_4(q)&=& 1 + 148 q + 7446 q^2 + 154344 q^3 + 5100349 q^4 + 352720380 q^5 + 
 10627587582 q^6 \nonumber \\
  & & + 166124184888 q^7 + 5419843397586 q^8 + 
 294399334337124 q^9 + \dots \nonumber \\
  A_5(q)&=& - \frac{1}{216} + 29 q + 431 q^2 + 80468 q^3 - 231081 q^4 + 94846414
q^5 + 1301490428 q^6 \nonumber \\
 & & + 90560563752 q^7 - 529100109849 q^8 + 
 93349951292249 q^9 + \dots \, .
\end{eqnarray}
The integer coefficients have been determined up to an accuracy of at least $10^{-6}$. For the
first order terms, the accuracy can be up to  $10^{-200}$. In terms of these series, the potential in extremum number 1, on the real axis is (with a given choice of normalisation):
\begin{equation}
V_1 = 144 \pi^2 A_3\left( \frac{q}{27} \right) \, .
\end{equation}
%
The potential in the other extrema are:
\begin{eqnarray}
 V_2 &=& -12 \pi^2 \left( \frac{8}{3} A_0 (q) + (2q)^{1/3} A_1 (q/9) + (2q)^{2/3} A_2
(q/9) \right)
\nonumber \\
V_3 &=& 
 -12 \pi^2 \left( \frac{8}{3} A_0 (q) + (2q)^{1/3} e^{2 \pi i /3} A_1 (q/9) +
(2q)^{2/3} e^{4 \pi i /3} A_2
(q/9) \right)
\nonumber \\
V_4 &=&
 -12 \pi^2 \left( \frac{8}{3} A_0 (q) + (2q)^{1/3} e^{4 \pi i /3} A_1 (q/9) +
(2q)^{2/3} e^{2 \pi i /3} A_2
(q/9) \right) \, ,
\end{eqnarray}
and
\begin{eqnarray}
V_{5,6} &=&  72 \pi^2 \left(A_5 \left( \frac{q}{27} \right) \pm i \sqrt{\frac{q}{27}} A_4 \left(-
\frac{q}{27} \right) \right) \nonumber \\
V_7 &=& \frac{48}{3} \pi^2 A_0(q) \, .
\end{eqnarray}
The growth properties of these series, as well as the fact that we are dealing
with a physical system living on a torus suggests turning these
numerical data into an analytic understanding, based on the theory of modular forms.
In the following, we show that this is possible for the rank 2 root system $B_2$.

\subsubsection{The extrema as  modular forms of the Hecke group and the $\Gamma_0(4)$ subgroup}
We need to introduce a few
groups related to the modular group. We already noted the duality
transform for the $B,C$-type twisted Calogero-Moser system under the
map $S_2 : \tau \rightarrow -1/(2 \tau)$ (see equation (\ref{bccoupling})). 
For the $so(5)$ Lie algebra, which is
identical to the $sp(4)$ Lie algebra, this transformation maps the
integrable system to itself (up to a $\tau$ dependent shift of the
potential and an overall factor -- see equation (\ref{duality2})). 
The map $T: \tau \rightarrow \tau+1$ also maps the
integrable system to itself.  Together, these transformations generate
the action of a Hecke group dubbed $\Gamma^\ast(2)$ on the modular parameter $\tau$. 
This group contains a subgroup $\Gamma_0(4)$
which is a congruence subgroup of the modular group $SL(2,\mathbb{Z})$.
Generators of the group $\Gamma_0(4)$ can be chosen to be the $2 \times 2$
matrices:
\begin{eqnarray}
T & : & \left( \begin{array}{cc} 1 & 1 \\
                                 0 & 1 \end{array} \right)
\nonumber \\
U & : & \left( \begin{array}{cc} 1 & 0 \\
                                 4 & 1 \end{array} \right) \, .
\end{eqnarray}
The action of these matrices on $\tau$ coincides with the action
of the elements $T$ and $U=S_2 T^{-2} S_2$ of the Hecke group. For more
information on Hecke groups and
 associated modular forms see e.g. the lectures \cite{BK}.

The extremal values of the potential may therefore
form a vector valued modular form with respect to the Hecke group
$\Gamma^\ast(2)$, and as a consequence also with respect to the congruence
subgroup $\Gamma_0(4)$ of the modular group $SL(2,\mathbb{Z})$, since we expect
extrema to be at most permuted and/or rescaled under the group. Here, we assume
analyticity in the interior of the fundamental domain. We 
will mostly exploit the group $\Gamma_0(4)$ in the following,
since the literature on the subject of modular forms with respect to
congruence subgroups is abundant. For starters,
we determine the action of the operations $T$ and $S_2$ on the vector
$V_i$ of extremal values of the twisted Calogero-Moser potential:
\begin{eqnarray}
T & : &  \left( \begin{array}{ccccccc} 1 & 0 & 0 & 0  & 0 & 0 & 0 \\
                                       0 & 0 & 0 & 1  & 0 & 0 & 0 \\
                                       0 & 1 & 0 & 0  & 0 & 0 & 0 \\
                                       0 & 0 & 1 & 0  & 0 & 0 & 0 \\
                                       0 & 0 & 0 & 0  & 0 & 1 & 0 \\
                                       0 & 0 & 0 & 0  & 1 & 0 & 0 \\
                                       0 & 0 & 0 & 0  & 0 & 0 & 1 
\end{array} \right) \, ,
\nonumber \\
S_2 & : & \left( \begin{array}{ccccccc} 0 & 1 & 0 & 0 & 0 & 0  & -2 \\
                                        1 & 0 & 0 & 0 & 0 & 0  & -2 \\
                                        0 & 0 & 0 & 0 & 1 & 0  & -2 \\
                                        0 & 0 & 0 & 0 & 0 & 1  & -2 \\
                                        0 & 0 & 1 & 0 & 0 & 0  & -2 \\
                                        0 & 0 & 0 & 1 & 0 & 0  & -2 \\
                                        0 & 0 & 0 & 0 & 0 & 0  & -1
 \end{array} \right) \, .
\end{eqnarray}

\begin{figure}
\centering
\includegraphics[width=0.5\textwidth]{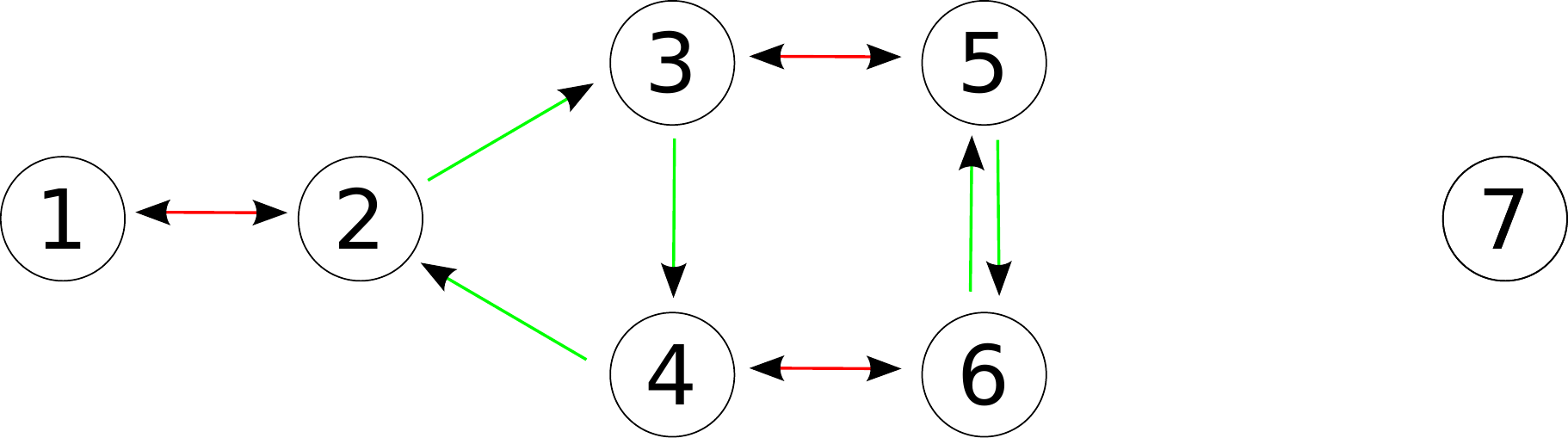}
\caption{The diagram of the action of dualities on the extrema for $B_2=so(5)$. In red, we draw the action of Langlands $S_2$-duality, and in green, $T$-duality (when the action is non-trivial). }
\label{dualities_so5}
\end{figure}

See figure \ref{dualities_so5} for a summary of the action of the duality group.
To this information, we add the last column in the matrix $S_2$, which 
originates in the shift of the potential under Langlands duality.
{From} these data, we easily calculate the action of the generator $U=S_2 T^{-2} S_2$
on the vector valued modular form:
\begin{eqnarray}
U & : &  \left( \begin{array}{ccccccc} 
                                       0 & 0 & 0 & 0  & 1 & 0 & 0 \\
                                       0 & 1 & 0 & 0  & 0 & 0 & 0 \\
                                       0 & 0 & 1 & 0  & 0 & 0 & 0 \\
                                       0 & 0 & 0 & 1  & 0 & 0 & 0 \\
                                       0 & 0 & 0 & 0  & 0 & 1 & 0 \\
                                       1 & 0 & 0 & 0  & 0 & 0 & 0 \\
                                       0 & 0 & 0 & 0  & 0 & 0 & 1 

 \end{array} \right)
 \, .
\end{eqnarray}
 We thus find the action
of $\Gamma_0(4)$ on the vector valued modular form, and we observe the following pattern:
there is one entry (the seventh) which is an ordinary modular form of weight $2$
under $\Gamma_0(4)$, 
and there are two sets of three components (namely $\{ 2,3,4 \}$ and $\{ 1,5,6 \}$) that mix under
$\Gamma_0(4)$. Thus, our vector valued modular form of dimension seven splits into a singlet and
a sextuplet.
Concentrating on the ordinary modular form of weight $2$, we have that
it is a linear combination of Eisenstein series $E_{2,N}$ defined by:
\begin{eqnarray}
E_{2,N} (\tau) &=& E_2(\tau) - N E_2(N \tau) \, .
\end{eqnarray}
Indeed, the dimension of the space ${\cal M}_2 (\Gamma_0(4)) $ of modular forms
of $\Gamma_0(4)$ is two, and it is spanned by $E_{2,2}$ and $E_{2,4}$. We thus
only need two Fourier coefficients to fix the entire modular form, and we find that:
\begin{eqnarray}
A_0 (q) &=& - \frac{1}{24} E_{2,2} (\tau) =  \frac{1}{48} (\theta_3^4 +
\theta_4^4)(\tau)\,  \\
V_7 &=& \frac{\pi^2}{3}  (\theta_3^4 +
\theta_4^4)(\tau) \, . \label{seventhexact}
\end{eqnarray}
We then have a slew of consistency checks on all the other integers
that we determined numerically (see (\ref{A0numerics})). These thirteen checks work out. We do therefore claim that the
result (\ref{seventhexact}) is exact. This is a simple example illustrating our methodology.

Next, we consider the triplet consisting of the components $\{ 2,3,4 \}$. 
We find three eigenvectors of $T$, with eigenvalues corresponding to the cubic
roots of unity. The eigenvector with eigenvalue $1$ is also mapped to itself under
the $U$ transformation, and forms again a modular form of weight $2$ under
$\Gamma_0(4)$. It is indeed proportional to $E_{2,2}$:
\begin{equation}
 V_2 + V_3 + V_4 = -2 \pi^2 (\theta_3^4 + \theta_4^4)(\tau)
 \, .
\end{equation}

The other two eigenvectors, we raise to the power three, such that they become
invariant under the $T$-transformation. These forms belong to the space ${\cal
M}_6 (\Gamma_0(4))$
of weight six modular forms. The dimension of this vector space is $4$ (see theorem 3.5.1 in \cite{Diamond} with $g=\varepsilon_2=\varepsilon_3=0$ and $\varepsilon_{\infty}=3$), and it consists
of three Eisenstein series, and one cusp form. A basis for these vector spaces
is given by:
\begin{eqnarray}
 E_6^1 &=& -\frac{1}{252}  E_6(\tau)  \\
 E_6^2 &=& -\frac{1}{252}  E_6(2 \tau)  \\
 E_6^4 &=& -\frac{1}{252}  E_6(4 \tau)  \\
 S_6 &=& \eta(q^2)^{12} \,  ,
\end{eqnarray}
where $E_6$ is the Eisenstein series of weight six, and $\eta$ is the
$\eta$-function, also recorded in appendix \ref{modularforms}.
We need four coefficients to fix the eigenvectors in terms of this basis and we
find (using the notation $\omega_3 = \exp (2\pi i /3)$) : 
\begin{eqnarray}
 (V_2 + \omega_3 V_3 + \omega_3^2 V_4)^3 &=& -23328 \pi^6 (E_6^1 - E_6^2 - 2 S_6) \\
 (V_2 + \omega_3^2 V_3 + \omega_3 V_4)^3 &=& -23328 \pi^6 (E_6^1 - E_6^2 + 2 S_6)  \, .
\end{eqnarray}
The consistency checks using the numerics work out.

For the second triplet, we diagonalise $U$ first, and proceed very analogously as above,
except that we have to take a higher power for the second combination to find a modular form of weight 12 with respect to
$\Gamma_0(4)$. We find the relations:
\begin{eqnarray}
(V_1 + \omega_3 V_5 + \omega_3^2 V_6)^3
+
(V_1 + \omega_3^2 V_5 + \omega_3 V_6)^3
&=& 5832 \pi^6 (E_6^1(q) - 64 E_6^2)
\nonumber \\
((V_1 + \omega_3 V_5 + \omega_3^2 V_6)^3
-
(V_1 + \omega_3^2 V_5 + \omega_3 V_6)^3)^2
&=& 136048896 \pi^{12} \eta(q)^{24} \, .
\end{eqnarray}
Note that the sum of all potentials is necessarily a modular form with weight 2 of $\Gamma_0(4)$. Indeed, this sum is equal to  $112 \pi ^2 A_0 (q)$ (as follows from the identity $A_5(q/27) + A_3(q/27)=\frac{4}{3} A_0(q)$). 

\subsubsection{A remark on a manifold of extrema}

There are also branches of extrema, namely, non-isolated extrema. These too, we expect to behave well
under a modular subgroup. Although this was not the focus of our investigation, we did find  numerical 
evidence for a  manifold of extrema at which the potential takes the $\Gamma_0(4)$ covariant value $-
\frac{2 \pi^2}{3}  E_{2,2}$.

\subsubsection*{Summary}

In summary, we have full analytic control over the value of the potential for all isolated extrema of
the $so(5)$ twisted Calogero-Moser integrable system. We have found a vector valued
modular form of weight two of $\Gamma_0(4)$, and we were able to explicitly
express its seven components in terms of ordinary modular forms of $\Gamma_0(4)$. 
The vector valued septuplet splits into a singlet modular form and a sextuplet vector
valued modular form.
The plot will thicken at higher rank.

\subsection{The Case $D_4=so(8)$ and the Point of Monodromy}
\label{Sectionso8}
At this stage, we choose to present our results on the rank four $D_4=so(8)$ model first,
since they are simpler than those on the non-trivial rank three cases to be presented in
subsection \ref{thecaseC3}. The $so(8)$ model is simply laced and we therefore expect the ordinary
modular group $SL(2,\mathbb{Z})$ to play the leading role. The integrable system
exhibits a global symmetry group $S_3$
that permutes the three satellite simple roots of the Dynkin diagram of $so(8)$. We will refer to 
the $S_3$ permutation group
as triality. We turn to the enumeration and classification of the extrema of the potential.
We found 34 extrema.
These are listed and labelled in appendix \ref{listofextremaso8}.
If we mod out by the global symmetry group, we are
left with 20 extrema. The latter fall into multiplets of the duality group of size
$1,3,4$ and $12$. We discuss these multiplets in the following paragraphs.

\subsubsection{The singlet}
There is a singlet under $S$ and $T$ duality as well as triality. It has zero potential:
$V_1=0$.

\subsubsection{The triplet}
There is also a triplet under the duality group, labelled $\{ 2,3,4 \}$, and the dualities act as:
\begin{equation*}
T = 
\begin{pmatrix}
1 & 0 & 0 \\
0 & 0 & 1 \\
0 & 1 & 0
\end{pmatrix}
\qquad
\qquad
S = 
\begin{pmatrix}
0 & 1 & 0 \\
1 & 0 & 0 \\
0 & 0 & 1
\end{pmatrix} \, .
\end{equation*}
The relations $S^2=1$ and $(ST)^3=1$
are satisfied.
We note that in these extrema, the positions belong to the lattice generated by $\omega_1 /2$ and $\omega_2/2$. For this multiplet,
T-duality acts geometrically. 

We would like to deduce again from the $S$ and $T$ matrices and from the 
known first coefficients of the series expansions (see appendix \ref{listofextremaso8}) the exact expressions of the 
potentials in these extrema.
 The functions are expected to transform well under some congruence 
subgroup of the modular group. Note that the sum of the three functions must be a full-fledged modular form -- indeed, the
sum  $V_{2}(q) + V_{3}(q) +V_{4}(q)$ vanishes. 
A brute force strategy leading to the identification of the appropriate congruence subgroup is the following. We decompose the generators of congruence subgroups \footnote{There exist algorithms to find the generators. These are for instance implemented in Sage.}
in terms of a product of $S$ and $T$ operations.
We evaluate the product using the representation at hand (here $3 
\times 3$ matrices) and check whether it is trivial for every generator.

It turns out that the subgroup
$\Gamma(2)$ 
acts trivially on the extremal potentials. Hence all the potentials $V_{2}$, $V_{3}$ and $V_{4}$ belong to $\mathcal{M}_2 (\Gamma 
(2)) $. This space has dimension 2, and it is the set of linear 
combinations of the three Eisenstein functions associated to the three vectors 
of order 2 in $(\mathbb{Z}_2)^2$ which have the property that the sum of the three coefficients 
vanishes. (See appendix \ref{modularforms} for details and conventions). Matching a few coefficients,
we find that 
\begin{eqnarray*}
V_{2} &=& 12 \left( 2 G_{2,2} \left[ \begin{matrix} 0 \\ 1 
\end{matrix} \right] - G_{2,2} \left[ \begin{matrix} 1 \\ 1 
\end{matrix} \right] - G_{2,2} \left[ \begin{matrix} 1 \\ 0 
\end{matrix} \right] \right) \\
V_{3} &=& 12 \left( -G_{2,2} \left[ \begin{matrix} 0 \\ 1 
\end{matrix} \right] - G_{2,2} \left[ \begin{matrix} 1 \\ 1 
\end{matrix} \right] +2 G_{2,2} \left[ \begin{matrix} 1 \\ 0 
\end{matrix} \right] \right) \\
V_{4} &=& 12 \left( -G_{2,2} \left[ \begin{matrix} 0 \\ 1 
\end{matrix} \right] +2 G_{2,2} \left[ \begin{matrix} 1 \\ 1 
\end{matrix} \right] - G_{2,2} \left[ \begin{matrix} 1 \\ 0 
\end{matrix} \right] \right) .
\end{eqnarray*}
This can also be written in terms of the Weierstrass $\wp$ function : 
\begin{eqnarray*}
V_{2} (\tau) &=& 3 \left( 2 \wp \left( \frac{1}{2} ; \tau \right) - \wp \left( \frac{\tau + 1}{2} ; \tau \right) - \wp \left( \frac{\tau}{2} ; \tau \right) \right) \\
V_{3} (\tau) &=& 3 \left( - \wp \left( \frac{1}{2} ; \tau \right) - \wp \left( \frac{\tau + 1}{2} ; \tau \right) +2 \wp \left( \frac{\tau}{2} ; \tau \right) \right) \\
V_{4} (\tau) &=& 3 \left( - \wp \left( \frac{1}{2} ; \tau \right) +2 \wp \left( \frac{\tau + 1}{2} ; \tau \right) - \wp \left( \frac{\tau}{2} ; \tau \right) \right) .
\end{eqnarray*}
These two ways of writing the potentials make the action of dualities manifest. 
For instance, the transformation properties (\ref{Weierstrass_Jacobi}) show  that under $S$-duality, $\wp (\frac{1}{2},\tau)$ becomes $\wp (\frac{1}{2},\frac{-1}{\tau})=\tau ^2 \wp (\frac{\tau}{2},\tau)$ while $\wp (\frac{\tau + 1}{2},\tau)$ becomes $\tau ^2 \wp (\frac{\tau + 1}{2},\tau)$, so that $V_2$ and $V_3$ are $S$-dual, et cetera.
The result can also be written using perhaps more familiar modular forms
\begin{eqnarray*}
V_{2}(q) &=& - 6 \pi^2 E_{2,2}(q)  \\
V_{3}(q) &=& \frac{3}{2} \pi^2 \left(2 E_{2,2}(q) - 3 \theta_2^4 (q) \right)  \\
V_{4}(q) &=& \frac{3}{2} \pi^2 \left(2 E_{2,2}(q) + 3 \theta_2^4 (q) \right)  \, .
\end{eqnarray*}
The action of  $T$-duality is again clear from these expressions. For $S$-duality it is slightly more intricate. Given that $E_{2,2}(q) = -\theta_2^4(q^2) - \theta_3^4(q^2)$, it relies on the  identities
\begin{eqnarray*}
 2 \theta_3^4 (2 \tau) +  2 \theta_2^4 (2 \tau) +  3 \theta_2^4 (\tau) &=& -  
\theta_2^4 (\tau /2) +  2 \theta_3^4 (\tau/2)\\
\theta_3^4 (\tau / 2) +  \theta_4^4 (\tau / 2) - 6 \theta_4^4 ( \tau) &=& - 4 \theta_3^4 (2 \tau) - 4 \theta_2^4 (2 \tau) + 6 \theta_2^4 ( \tau)  \, ,
\end{eqnarray*}
for $S$-duality between extrema 2 and 3, and self-$S$-duality for extremum 4, respectively.

\subsubsection{The quadruplet}
We move on to discuss  the extremal values of the potential in the quadruplet.
We can arrive at the following closed form for the potential in extremum 6: 
\begin{equation*}
V_{6}(q) = -24 \pi^2 (- \frac{1}{24} E_{2,3}(q) + (\eta(q)^3 + 9 \eta(q^9)^3) \eta(q^3)^2/\eta(q) + 
   3 (\eta(q^3)^3/\eta(q))^2) \, .
\end{equation*}
Note that this can alternatively  be written as
\begin{equation*}
V_{6}(q) = -24 \pi^2 (g_0(q) + q^{1/3} g_1(q) + 3 q^{2/3} g_2(q)) \, ,
\end{equation*}
where the $g_i$ are functions that can be expanded into series with only integer powers of $q$ (and the three summands in this expression correspond to the same summands in the expression above). 
Thus we know how the operation $\tau \rightarrow \tau+1$ acts on the extremum, and it generates two other extrema, whose  potential we also know exactly. These are extrema 7 and 8:
\begin{eqnarray*}
V_{7}(q) = -24 \pi^2 (g_0(q) + e^{2 i \pi /3} q^{1/3} g_1(q) + 3 e^{-2 i \pi /3} q^{2/3} g_2(q)) \\
V_{8}(q) = -24 \pi^2 (g_0(q) + e^{-2 i \pi /3} q^{1/3} g_1(q) + 3 e^{2 i \pi /3} q^{2/3} g_2(q))  \, .
\end{eqnarray*}
The potential for the extremum 5 is: 
\begin{equation*}
V_{5}(q) = -3 \pi^2 E_{2,3}(q) \, .
\end{equation*}
In the basis $\{5,6,7,8 \}$ the matrices for $S$- and 
$T$-dualities are : 
\begin{equation*}
T = 
\begin{pmatrix}
1 & 0 & 0 & 0 \\
0 & 0 & 0 & 1 \\
0 & 1 & 0 & 0 \\
0 & 0 & 1 & 0 
\end{pmatrix}
\qquad
\qquad
S = 
\begin{pmatrix}
0 & 1 & 0 & 0 \\
1 & 0 & 0 & 0 \\
0 & 0 & 0 & 1 \\
0 & 0 & 1 & 0 
\end{pmatrix} \, .
\end{equation*}
We can also apply the same method as above. The generators of $\Gamma(3)$ are all trivial 
in this basis. Thus the potentials are weight 2 modular forms of this congruence 
subgroup. The latter form a 3-dimensional space, generated by the zero-sum linear 
combinations of the 4 Eisenstein series associated to the order 3 vectors in 
$(\mathbb{Z}_3)^2$ (there are 8 such vectors, but the Eisenstein series are 
invariant under $v \rightarrow -v$, leaving only 4 distinct functions, see appendix). We 
find 
\begin{eqnarray*}
V_{5} &=&\frac{27}{2} \left(3 G_{2,3} \left[ \begin{matrix} 0 \\ 1 
\end{matrix} \right] - G_{2,3} \left[ \begin{matrix} 1 \\ 0 
\end{matrix} \right] - G_{2,3} \left[ \begin{matrix} 1 \\ 1 
\end{matrix} \right] - G_{2,3} \left[ \begin{matrix} 1 \\ 2 
\end{matrix} \right] \right)  \\
V_{6} &=& \frac{27}{2} \left(-G_{2,3} \left[ \begin{matrix} 0 \\ 1 
\end{matrix} \right] +3 G_{2,3} \left[ \begin{matrix} 1 \\ 0 
\end{matrix} \right] - G_{2,3} \left[ \begin{matrix} 1 \\ 1 
\end{matrix} \right] - G_{2,3} \left[ \begin{matrix} 1 \\ 2 
\end{matrix} \right] \right) \\
V_{7} &=& \frac{27}{2} \left(-G_{2,3} \left[ \begin{matrix} 0 \\ 1 
\end{matrix} \right] - G_{2,3} \left[ \begin{matrix} 1 \\ 0 
\end{matrix} \right] +3 G_{2,3} \left[ \begin{matrix} 1 \\ 1 
\end{matrix} \right] - G_{2,3} \left[ \begin{matrix} 1 \\ 2 
\end{matrix} \right] \right) \\
V_{8} &=& \frac{27}{2} \left(-G_{2,3} \left[ \begin{matrix} 0 \\ 1 
\end{matrix} \right] - G_{2,3} \left[ \begin{matrix} 1 \\ 0 
\end{matrix} \right] - G_{2,3} \left[ \begin{matrix} 1 \\ 1 
\end{matrix} \right] +3 G_{2,3} \left[ \begin{matrix} 1 \\ 2 
\end{matrix} \right] \right)  \, ,
\end{eqnarray*}
or alternatively,
\begin{eqnarray*}
V_{5}(\tau) &=&\frac{3}{2} \left(3 \wp  \left( \frac{1}{3} ; \tau \right) - \wp  \left( \frac{\tau}{3} ; \tau \right) - \wp  \left( \frac{\tau +1}{3} ; \tau \right) - \wp  \left( \frac{\tau+2}{3} ; \tau \right) \right) \\
V_{6}(\tau) &=&\frac{3}{2} \left(- \wp  \left( \frac{1}{3} ; \tau \right) +3 \wp  \left( \frac{\tau}{3} ; \tau \right) - \wp  \left( \frac{\tau +1}{3} ; \tau \right) - \wp  \left( \frac{\tau+2}{3} ; \tau \right) \right) \\
V_{7}(\tau) &=&\frac{3}{2} \left(- \wp  \left( \frac{1}{3} ; \tau \right) - \wp  \left( \frac{\tau}{3} ; \tau \right) +3 \wp  \left( \frac{\tau +1}{3} ; \tau \right) - \wp  \left( \frac{\tau+2}{3} ; \tau \right) \right) \\
V_{8}(\tau) &=&\frac{3}{2} \left(- \wp  \left( \frac{1}{3} ; \tau \right) - \wp  \left( \frac{\tau}{3} ; \tau \right) - \wp  \left( \frac{\tau +1}{3} ; \tau \right) +3 \wp  \left( \frac{\tau+2}{3} ; \tau \right) \right) \, . \\
\end{eqnarray*}
%
The dualities act on the vectors characterising the modular forms as follows
\begin{eqnarray*}
T & : & \left[ \begin{matrix} 0 \\ 1 \end{matrix} \right] \rightarrow \left[ 
\begin{matrix} 0 \\ 1 \end{matrix} \right] \\
 & & \left[ \begin{matrix} 1 \\ 0 \end{matrix} \right] \rightarrow 
 \left[ \begin{matrix} 1 \\ 1 \end{matrix} \right] \rightarrow 
 \left[ \begin{matrix} 1 \\ 2 \end{matrix} \right] \rightarrow 
 \left[ \begin{matrix} 1 \\ 0 \end{matrix} \right] \, , \\
 S & : & \left[ \begin{matrix} 0 \\ 1 \end{matrix} \right] \leftrightarrow 
\left[ \begin{matrix} 1 \\ 0 \end{matrix} \right] \\
  &   & \left[ \begin{matrix} 1 \\ 1 \end{matrix} \right] \leftrightarrow 
\left[ \begin{matrix} 1 \\ 2 \end{matrix} \right] \, .
\end{eqnarray*}
This reproduces the action of the dualities on the associated extrema. Thus, while the pattern of
the positions of the extrema is non-linear, the arguments of the values of the potential at certain extrema do provide a linear realisation of the duality group. 

Finally, we note that triality generates three copies of the triplet as well as of the quadruplet. Indeed, each of these extrema is left invariant by a $\mathbb{Z}_2$ subgroup of $S_3$ (as described in 
appendix \ref{listofextremaso8}).

Up to now, we have discussed the singlet, triplet and quadruplet whose duality diagrams are summarised in figure \ref{dualities_so8}.

\subsubsection{The duodecuplet and a point of monodromy}
In the multiplet of size twelve, also depicted in figure \ref{dualities_so8}, a new feature appears. We find that the extrema exhibit a monodromy around a point
in the interior of the fundamental domain of the parameter $\tau$. Thus, to be able to describe the multiplet structure in
this case we must first discuss the monodromy.

\subsubsection*{The point of monodromy}
We find a single point in the interior of the fundamental domain around which there is monodromy amongst extrema. It is possible to determine this point  numerically\footnote{The most immediate manifestation of the monodromy phenomenon can be seen as a symmetry breaking in the equilibrium positions for extrema 13 and 16 when moving on the imaginary axis across the point of monodromy $\tau_M$ (which is purely imaginary). Below this critical value, as can be seen in the diagrams drawn at $\tau=i$ (in appendix \ref{listofextremaso8}), the two extrema are exchanged by the $\mathbb{Z}_2$ action $X_i \leftrightarrow -\bar{X_i}$, while above the critical value, they are both invariant with respect to this action. This makes it possible to determine $2.41557 \leq 
\mathrm{Im} \tau_{M} \leq 2.41558$. } and its value is close to $\tau_M \sim 2.41558 i$. In particular,
the extrema 13 and 16 are exchanged when we follow a loop in the $\tau$-plane that closely circles the value
$\tau_M$. Moreover, using the geometry of the positions of the extrema 13 and 16, one can 
show that $\tau_M$ is a solution of the system of equations
\begin{equation}
\left\{
\begin{array}{l}
\wp (z; \tau)^2 + \wp (z - \omega_3; \tau)^2 + \wp (2z - \omega_3; \tau)^2 = \frac{\pi ^4}{3} E_4 (\tau)\\
2\wp' (z; \tau) + 2\wp' (z - \omega_3; \tau) + \wp' (2z - \omega_3; \tau)=0  \, ,
\end{array}
\right.
\end{equation}
where $\omega_3 = \omega_1 + \omega_2$, which gives the numerical result
\begin{equation*}
 \tau_M = 2.415576987549484510777262081474158860468152563579077460 ... i \, . 
\end{equation*}
Using the large accuracy of the value of the point of monodromy $\tau_M$, we find the corresponding
rational Klein invariant (with the normalisation (\ref{KleinInvariant})):
\begin{equation*}
j (\tau_M) = \frac{488095744}{125} = 1728 \times \frac{7626496}{3375} \, .
\end{equation*}
This can be considered as an exact statement -- the uncertainty is as low as $10^{-200}$.
Elliptic curves with rational Klein invariant have interesting arithmetic properties
(see e.g. \cite{Diamond}).

\subsubsection*{The extended duality group}
\begin{figure}
\centering
\includegraphics[width=0.5\textwidth]{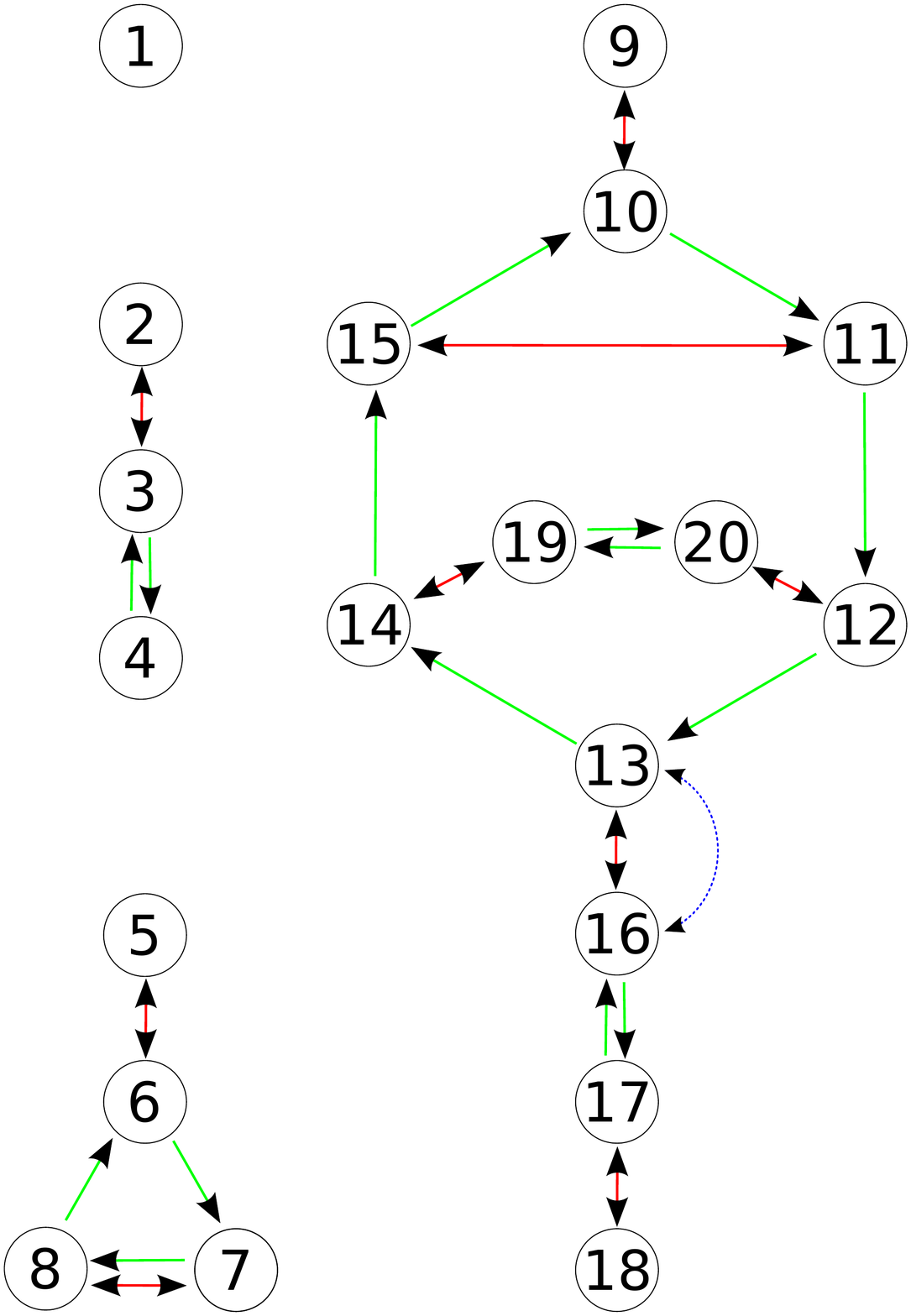}
\caption{The diagram of the action of dualities on the $D_4=so(8)$ extrema. In red we exhibit the action of $S$-duality, in green, $T$-duality, and in dotted blue, the monodromy. }
\label{dualities_so8}
\end{figure}

We can add the monodromy group to the set of generators $S$ and $T$ that act on our vector of extrema.
The resulting diagram of dualities then becomes the one in figure \ref{dualities_so8}.
The generators satisfy the relations: 
\begin{itemize}
 \item $S^2 = M^2 = 1$ and $T^6=1$, while $(TM)^8=1$
 \item $SM=MS$
 \item $(MST)^3 = 1$.
\end{itemize}
Once we are underneath the point of monodromy in the canonical fundamental domain, 
the matrix $MT$ plays the role usually taken by the matrix
$T$ in $SL(2,\mathbb{Z})$. In particular, relations like $(ST)^3=1$ implied by the geometry of the
fundamental domain of the modular group take on the form $(SMT)^3=1$, et cetera.
Triality leaves each extremum invariant.

In appendix \ref{listofextremaso8}, we give terms in the Fourier expansion of the extremal values
of the potential in the duodecuplet. We note that
a consistency and exhaustivity check on all multiplets is provided by the fact that the sum of all extrema in a given multiplet of $SL(2,\mathbb{Z})$ has to be a weight 2 modular form. The check works out: the sum equals zero in each multiplet separately, as it must. 
An analytic understanding of the duodecuplet extrema remains desirable.

\subsection{The Dual Cases $B_3=so(7)$ and $C_3=sp(6)$}

\label{thecaseC3}

\subsubsection{Exact multiplets}

For the twisted elliptic integrable models associated to the dual Lie algebra root
systems $so(7)$ and $sp(6)$, we present our results succinctly. 
We have found 17 isolated extrema for each, and they are Langlands dual.
We have therefore 34 extrema in total.
We identified two
quadruplets of the full duality group for which we found analytic expressions
for the potential at the extrema. 
The list of the corresponding extrema is given in appendix \ref{listofextremaso7}.
We find the following duality properties and analytic values for the extrema
of the potential. The extrema labelled
$\{ 1 , 2 
\}$ have extremal values for the $so(7)$ potential
equal to 
$V_1(\tau)$ and $V_{2} (\tau)$.
{From} the diagram of dualities (figure \ref{dualities_so7}), we read off that these extremal values 
are modular forms of $\Gamma_0 (4)$ with weight 2. 
Moreover, Langlands duality then implies that
$V_{1^\vee}(2 \tau)$ and $V_{2^{\vee}} (2 \tau)$ are also of that ilk.
The space $\mathcal{M}_2 (\Gamma_0 (4))$ of these
weight 2 forms has the two generators
\begin{eqnarray*}
-E_{2,2}(\tau)=\theta_2^4(2\tau) + \theta_3^4(2\tau) = 1/2 (\theta_3^4(\tau) + \theta_4^4 (\tau)) \\
-E_{2,4}(\tau) = 3 \theta_3^4 (2\tau) = 3 /4(\theta_3^2 (\tau) + \theta_4^2 (\tau))^2 \, .
\end{eqnarray*}
In terms of the generators, the extrema are:
\begin{eqnarray*}
V_{1}(\tau) &=&  \pi ^2 \left( -E_{2,2}(\tau) - 2 E_{2,4}(\tau) \right) \\
V_{2}(\tau) &=& \pi ^2 \left( -7 E_{2,2}(\tau) + 2 E_{2,4}(\tau) \right) \\
V_{1^\vee}(2 \tau) &=&  \pi ^2 \left(+ E_{2,2}(\tau) + 0 E_{2,4}(\tau) \right) \\
V_{2^\vee}(2 \tau) &=&  \pi ^2 \left( -2 E_{2,2}(\tau) + 1 E_{2,4}(\tau) \right) \, .
\end{eqnarray*}
For the other quadruplet under the full duality group, we have a similar story, with the happy ending:
\begin{eqnarray*}
V_{3}(2 \tau) &=&  \pi ^2/6 \left( -15 E_{2,2}(\tau) + 7 E_{2,4}(\tau) \right) \\
V_{4}(2 \tau) &=& \pi ^2/6 \left( +9 E_{2,2}(\tau) - 1 E_{2,4}(\tau) \right) \\
V_{3^\vee}(\tau) &=& 8 \pi ^2/3 \left( -3 E_{2,2}(\tau) + 1 E_{2,4}(\tau) \right) \\
V_{4^\vee}(\tau) &=& 8 \pi ^2/3 \left( 0 E_{2,2}(\tau) - 1 E_{2,4}(\tau) \right)  \, .
\end{eqnarray*}
The action of Langlands $S_2$ duality as well as T-duality can be found explicitly using these exact expressions, for instance by exploiting properties of $\theta$ functions.
As an example, we note that the action of 
$T$-duality is summarised in the equalities: 
\begin{eqnarray*}
E_{2,2} \left( \tau + \frac{1}{2} \right) &=& -2 E_{2,2}(\tau) +  E_{2,4}(\tau) \\
E_{2,4} \left( \tau + \frac{1}{2} \right) &=& -3 E_{2,2}(\tau) +  2E_{2,4}(\tau) \, .
\end{eqnarray*}
Moreover, on the extrema, the Langlands duality $S_2$
acts as
\begin{equation*}
\frac{1}{2\tau ^2} V_{1} \left( - \frac{1}{2\tau}\right) = V_{1^\vee}( \tau) + 3\pi ^2 E_{2,2} (\tau)
\, ,
\end{equation*}
and similar relations hold for the other $S_2$-dual couples, as predicted by the duality formula (\ref{BCduality}). 

\subsubsection{The duodecuplet, the quattuordecuplet and the points of monodromy}
\begin{figure}
\centering
\includegraphics[width=0.8\textwidth]{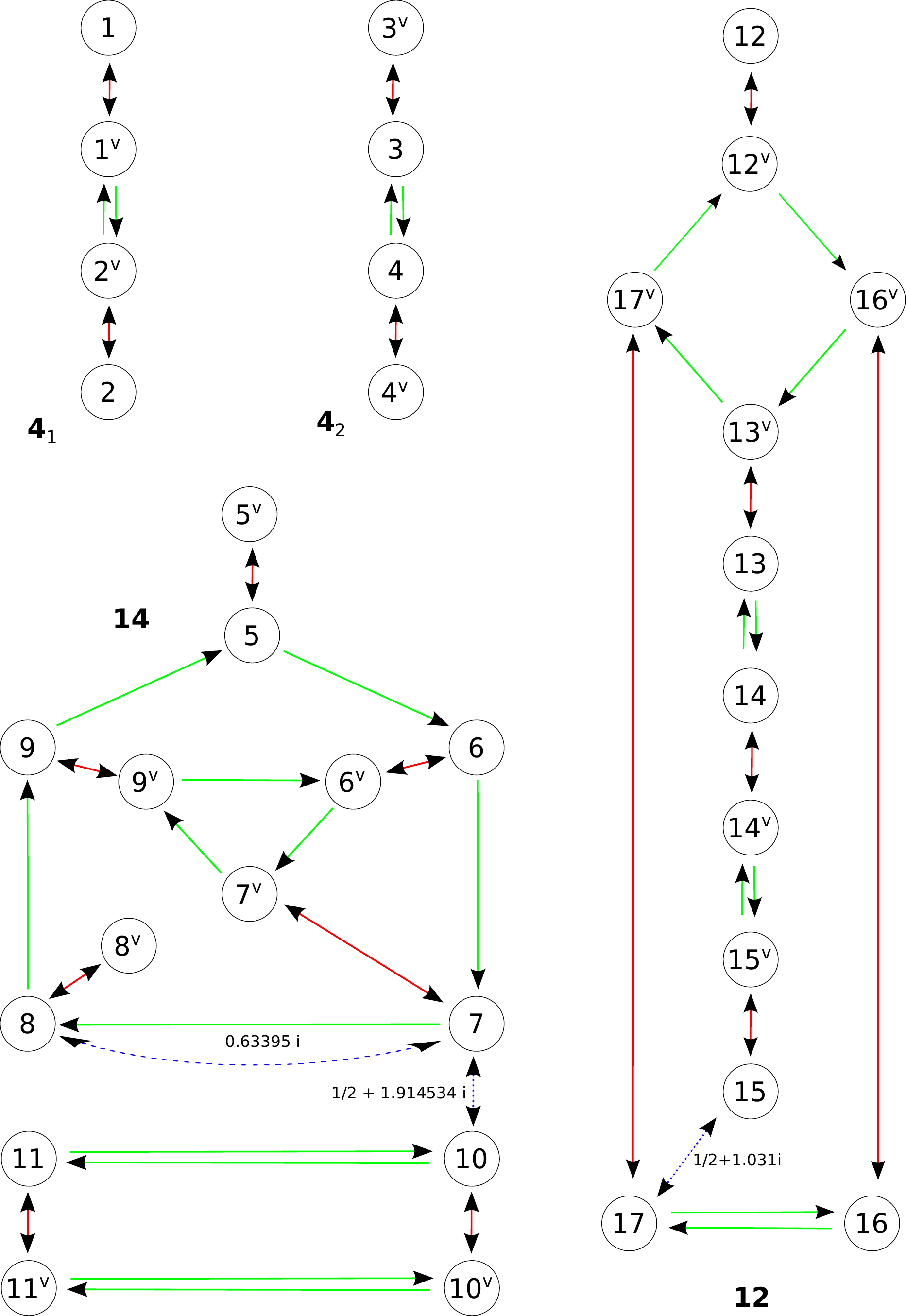}
\caption{The diagram of dualities for $so(7)$ and $sp(6)$ extrema. In red, we show the action of Langlands $S_2$-duality on the extrema, in green, $T$-duality, and in dotted blue, monodromies, with the corresponding approximate values of the points of monodromy $\tau$. As discussed in the text, monodromies relating $sp(6)$ extrema exist but are not represented here as they are equivalent to those already depicted.  }
\label{dualities_so7}
\end{figure}

We further identified a duodecuplet and a quattuordecuplet under the duality
group (for a total of $(4+4+12+14)/2=17$ extrema for $B_3=so(7)$). Sufficient data to reproduce them is provided in appendix \ref{listofextremaso7}. These multiplets exhibit  points of monodromy,
and the full duality diagram is captured in  figure \ref{dualities_so7}.
\begin{figure}
\centering
\includegraphics[width=0.3\textwidth]{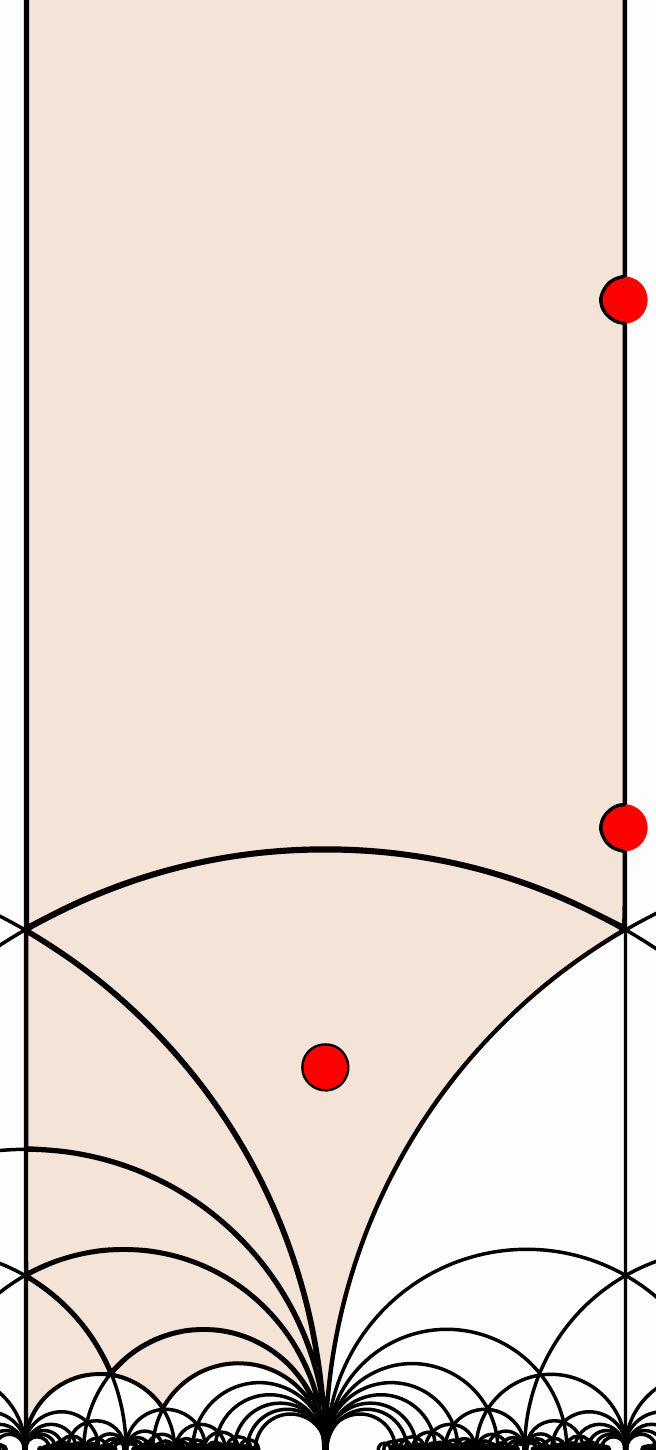}
\caption{The positions of the monodromies (red dots) inside the fundamental domain of $\Gamma_0 (4)$ (shaded)}
\label{fundamental_domain}
\end{figure}
It should be understood that we only represent points of monodromy that are inequivalent (where  two monodromies are taken to be equivalent when they are
equal up to conjugation by other elements of the duality group).
For instance $S_2 M_{\tau} S_2$ is the monodromy around $-1/(2\tau)$.

We draw attention to a few features of the diagram. 
There are 5 extrema that form a quintuplet under T-duality (around $\tau = i \infty$), labelled $5,6,7,8,9$. When we also turn around the point of monodromy,
the quintuplet enhances to a septuplet. This is reminiscent of a feature of the duality diagram for the duodecuplet of $so(8)$. 

Finally, we performed an exhaustivity check on the extrema by summing 
 the extremal values of the potential. We found\footnote{We evaluated the sum of the extrema numerically at two different values of $\tau$ to identify the linear combination  of $E_{2,2}$ and $E_{2,4}$ that equals the sum. We can then perform arbitrary many numerical checks at other values of $\tau$, and these work out. }
 \begin{eqnarray*}
 \sum\limits_{i \in \mathbf{4}_1} V_i (\tau) &=& -8 \pi^2 E_{2,2} (\tau) \\
  \sum\limits_{i \in \mathbf{4}_2} V_i (\tau) &=& 2 \pi^2 E_{2,2} (\tau) \\
   \sum\limits_{i \in \mathbf{12}} V_i (\tau) &=& -20 \pi^2 E_{2,2} (\tau) \\
    \sum\limits_{i \in \mathbf{14}} V_i (\tau) &=& 19 \pi^2 E_{2,2} (\tau) \, ,
 \end{eqnarray*}
 showing again that the sum of potentials over every multiplet is a modular form of $\Gamma_0(4)$. 

This concludes our systematic case-by-case discussion of the low rank $B,C,D$ isolated extrema of (twisted) elliptic Calogero-Moser models. We finish the section with a few further remarks on general features of the problem of identifying isolated extrema.

\subsection{Limiting Behaviour}
\label{limitingbehaviour}
We wish to make a remark on the limiting behaviour of the integrable models near an extremum.
As an example, consider 
extremum number 7 for $B_2=so(5)$ which has its extremal positions equal to a real number plus
$\tau/2$. We can take the limit of the potential as $\tau \rightarrow i \infty$ while keeping
the difference between the extremal positions and $\tau/2$ fixed. The limit of the integrable system is then
a (trigonometric) Sutherland system of type $D_2=so(4)$, and indeed, the real part of the extremal positions agrees
with those of the Sutherland system. This is but one example of the limiting behaviour of the models
near the extrema.

\subsection{Partial Results for Other Lie Algebras}
In this subsection, we discuss very partial results for some higher rank Lie algebras.
We think of the elliptic integrable model as a perturbation of the
Sutherland model, with trigonometric potential. The Sutherland model has a ground state
with all particles sprinkled on the real circle.  We can perturb this traditional ground
state by turning
on the elliptic deformation by powers of the small parameter $q$, and follow the ground
state under perturbation. In this
way, we can reconstruct the extremum of the complexified elliptic potential associated to the
Sutherland extremum on the real line. To take the limit from the elliptic integrable system towards
the Sutherland model, it is sufficient to use the expansion formula:
\begin{eqnarray}
\wp(x;\omega_1,\omega_2) &=&
-\frac{\pi^2}{12 \omega_1^2} E_2(q)
+ \frac{\pi^2}{4 \omega_1^2} \csc^2 \left( \frac{\pi x}{2 \omega_1} \right)
- \frac{2 \pi^2}{\omega_1^2} \sum_{n=1}^\infty \frac{n q^n}{1-q^n} \cos \frac{n \pi x}{\omega_1}
\, ,
\label{Weierstrasslargeqexpansion}
\end{eqnarray}
valid when the imaginary part of the modular parameter $\tau$ is sufficiently large.
The first term in the formula (\ref{Weierstrasslargeqexpansion}) is constant from
the perspective of the integrable system dynamics, while the second term gives rise
to the leading Sutherland potential. The minimum at the equilibrium of the Sutherland potential
on the real line can be computed analytically \cite{Corrigan:2002th} -- it is related to the
norm of the Weyl vector of the Lie algebra.
The positions of the equilibria are 
given in terms of zeroes of the Jacobi polynomials.
We can perform perturbation theory around these extrema (numerically),
and we find the following series in $q$ for the potential at perturbed Sutherland
extrema, for various gauge algebras:
\begin{eqnarray*}
\frac{V_{so(5)}}{\pi^2} &=& \frac{26}{3}+\frac{112 q}{3}+\frac{8656 q^2}{81}+\frac{392128 q^3}{2187}+\frac{5011568 q^4}{19683}+\frac{319117472
   q^5}{1594323} \\
&&   +\frac{12236034880 q^6}{43046721}+\frac{112088197760 q^7}{387420489}  + \dots\\
\frac{V_{so(6)}}{\pi^2} &=&8+64 q+192 q^2+256 q^3+192 q^4+384 q^5+768 q^6+512 q^7+192 q^8  + \dots\\
\frac{V_{so(7)}}{\pi^2} &=& 25+\frac{408 q}{5}+\frac{153816 q^2}{625}+\frac{23730528 q^3}{78125}+\frac{6103562136
   q^4}{9765625}+\frac{663346128528 q^5}{1220703125} \\
   && +\frac{129316813943136 q^6}{152587890625}+\frac{10819167546478272
   q^7}{19073486328125}  + \dots\\
 \frac{V_{so(8)}}{\pi^2} &=&  24+\frac{576 q}{5}+\frac{212544 q^2}{625}+\frac{39538944 q^3}{78125}+\frac{6618263616 q^4}{9765625}+\frac{909871629696
   q^5}{1220703125} \\
   && +\frac{171403608639744 q^6}{152587890625}+\frac{8112643818471936
   q^7}{19073486328125}+\frac{1087819119225488448 q^8}{2384185791015625}  + \dots\\
\frac{V_{so(9)}}{\pi^2} &=& \frac{164}{3}+\frac{992 q}{7}+\frac{5133728 q^2}{12005}+\frac{2305844608 q^3}{4117715}+\frac{168902799438112
   q^4}{176547030625} \\ 
   && +\frac{11307570247017024 q^5}{12111126300875}+\frac{640315787843154194816
   q^6}{370903242964296875}+\frac{1106383118191321793331968 q^7}{890538686357276796875} \\
   && +\frac{69929754265259380435436903968 q^8}{38181846177568242666015625}+\frac{17683503230173163609024329488224 q^9}{13096373238905907234443359375} + ...
\nonumber \\
\frac{V_{so(10)}}{\pi^2} &=& \frac{160}{3}+\frac{1280 q}{7}+\frac{1303808 q^2}{2401}+\frac{616518656
   q^3}{823543}+\frac{365560247552 q^4}{282475249} \\
   && +\frac{101140172889600
   q^5}{96889010407}+\frac{9869502718168064
   q^6}{4747561509943} \\
   && +\frac{18401127697466238976
   q^7}{11398895185373143}+\frac{6582207315175560008960
   q^8}{3909821048582988049}+...
\nonumber \\
\frac{V_{so(11)}}{\pi^2} &=& \frac{305}{3}+\frac{1960 q}{9}+\frac{30141880 q^2}{45927}+\frac{29034410080 q^3}{33480783}+\frac{4243088924219480
   q^4}{2790589782267} \\
   && +\frac{7560807432828504560 q^5}{6103019853817929}+\frac{4158609757083162994374880
   q^6}{1526041805387611692663} \\
   && +\frac{96348742286518866720674240
   q^7}{52975451244169948759587}+\frac{304885265038041162579660724924120
   q^8}{92724468600756742242419154123}+...
\nonumber \\
\frac{V_{so(12)}}{\pi^2}  &=&  100+\frac{800 q}{3}+\frac{4055200 q^2}{5103}+\frac{1335804800 q^3}{1240029}+\frac{63808646477600
   q^4}{34451725707} \\
   && +\frac{42945633858692800 q^5}{25115308040403}+\frac{6332155765834649948800
   q^6}{2093335809859549647}+...
\end{eqnarray*}
We see that at least for some extrema, it is fairly straightforward to generate interesting
data on the value of the potential at these extrema at higher rank. We note
a first pattern, valid at the order to which we have worked, in both the rank of the gauge group,
and the power of the modular parameter $q$.
Table \ref{N_and_k} gives the conjectured smallest integer $N$ such that for gauge algebra $g$, 
the potential $V_g (N q)/\pi ^2$ has a Fourier expansion with only integer coefficients 
in the following sense: the expansion can be written as $n_0 (r + n_1 q + n_2 q^2 + ...)$ 
where the $n_i$ are integers, and the first term $r$ is rational. 

As an example of this pattern, let us quote the formula: 
\begin{eqnarray*}
\frac{1}{66679200 \pi^2} V_{so(12)} \left(63^3 \times q \right) &=& \frac{1}{666792}+q+745143 q^2+252572301828 q^3+108583732036588599 q^4 \\ 
 && +25066769592690393853446   q^5+11087973934403204342320752348 q^6 \\
 && +1966652180387341854168182867614728 q^7+...
\end{eqnarray*}
As a final remark, we note that our numerical
searches in this and previous subsections
are far from exhausting the capabilities of present day computers.

\begin{table}[]
\centering
\begin{tabular}{|c|c|c|c|c|c|c|c|c|c|c|}
\hline
 $k$ & 5 & 6 & 7 & 8 & 9 & 10 & 11 & 12 & 13 & 14 \\
 \hline
 $N$ &  $3^3$ & $1$ & $5^3$ & $5^3$ & $7^3  5^2$ & $7^3$ & $3^6 7^2 $ & $3^6 7^3 $ & $3^1 7^1 11^3$ & 
 $3^2 7^1 11^3 $ \\
 \hline 
\end{tabular}
\caption{The integer $N$ for gauge algebra $so(k)$ rendering the Fourier expansion integral}
\label{N_and_k}
\end{table}

\section{The Massive Vacua of ${\cal N}=1^\ast$ gauge theories}
\label{gaugetheory}

In this section, we first briefly review properties of the infrared
physics of the ${\cal N}=1^\ast$ supersymmetric gauge theory,
and then show how the data we gathered  on elliptic integrable systems in section \ref{ellint}
elucidates the physics of this gauge theory further. 
We obtain the ${\cal N}=1^\ast$
gauge theory from ${\cal N}=4$ supersymmetric Yang-Mills theory with gauge group
$G$ by adding three masses $m_i$ for the three chiral ${\cal N}=1$ supermultiplets. We can then
go to the Coulomb branch of the gauge theory, and compactify the  theory on a circle \cite{Seiberg:1994rs,Seiberg:1996nz}.
Two massless scalars remain in the theory for each $U(1)$ in the unbroken $U(1)^r$ gauge
group, namely the Wilson lines $\phi= \int_{S^1} A_{\mu} dx^{\mu}$ and the scalar duals $\sigma$ of the
photons.   Since there are no fields in the theory which are charged
under the center of the gauge
group, we may choose the gauge group such that we 
allow for gauge transformations that twist around the circle by an
element of the center. This
lends a periodicity to the Wilson line under shifts taking values in the 
co-weight lattice $P^\vee$.
This reasoning corresponds to a choice of gauge group $G=\tilde{G}/C$ where
$\tilde{G}$ is the universal cover, and $C$ its center.\footnote{
Note that the dual theory to the one with gauge group $\tilde{G}/C$ has gauge group $\tilde{G}$, for a simply laced
group. The two scalars
are interchanged under S-duality. Thus, the duality symmetries mix various global choices of gauge groups. Duality
also acts on the twist direction of the twisted elliptic potential.} 

The gauge theory compactified on a circle gets non-perturbative  superpotential contributions 
from magnetic monopole configurations whose charges take values in the dual root lattice $Q^\vee$.
The scalar duals of the photons have as a result a smallest possible periodicity
equal to the weight lattice $P$. We choose to classify extrema of the superpotential with respect
to these identifications. We should mention that other choices will be physically relevant.  
 Since in deriving the effective superpotential
we compactified the theory on $\mathbb{R}^3 \times S^1$, the resulting effective theory
is influenced by the choice of the spectrum of line  operators that probe the
phases of our
four-dimensional theory  \cite{'tHooft:1981ht,Donagi:1995cf,Aharony:2013hda}. These
determine the set of allowed monopole operators in three dimensions, and this set may be larger
than the collection allowed by the minimal periodicity relation chosen above. Depending on the choice
of the spectrum of line operators, this can lead to an increase of the number of inequivalent solutions,
and therefore to an increase in the Witten index. This was analyzed carefully in \cite{Aharony:2013hda,Aharony:2013kma}.

We have identified the shift symmetries acting on the Wilson line and the dual photon.
We further divide out the configuration space by the Weyl group, which is the remnant
of gauge invariance. This classification of supersymmetric vacua agrees with the classification
we did in 
section \ref{ellint} in the elliptic integrable systems, on the condition that we identify the $\omega_2$ direction
with the Wilson line.

Our ${\cal N}=1^\ast$ theory is a deformation of ${\cal N}=4$ theory, and it inherits
some of its properties. In particular, the electric-magnetic duality group of
${\cal N}=4$ gauge theories in four dimensions 
\cite{Goddard:1976qe,Montonen:1977sn,Witten:1978mh} plays a crucial role. The duality symmetry
was determined 
to be the group $SL(2,\mathbb{Z})$ for simply laced gauge groups and $\Gamma_0(4)$ for
the $B$ and $C$ type gauge groups \cite{Girardello:1995gf,Dorey:1996hx,Kapustin:2005py}. Moreover, the $S_2$ generator of the Hecke group
exchanges the $B$ and $C$ type systems.
An infrared counterpart to these 
duality groups are present in our integrable systems, which allow for a (generalized) duality
group action on the infrared modular parameter $\tau$ \cite{Aharony:2000nt}, inherited after mass deformation from the
${\cal N}=4$ duality. Note in particular that the requirement of the 
$B$ type and $C$ type exchange is implemented 
in our integrable system by the Langlands duality we discussed in subsection \ref{dualityint}.
This duality 
provides a  further consistency check on the relative weight of the
short and long root contributions, fixed in  \cite{Kumar:2001iu} through consistency with 
the superpotential of the pure ${\cal N}=1$ super Yang-Mills theory.

In  \cite{Kumar:2001iu}, following the reasonings in 
\cite{Seiberg:1994rs,Seiberg:1996nz,D'Hoker:1998yi,Dorey:1999sj}, an exact effective superpotential
for ${\cal N}=1^\ast$ was proposed for any gauge group, equal to the potential of the twisted elliptic
Calogero-Moser model. The arguments were based on holomorphy,
uniqueness of the deformation from ${\cal N}=2^\ast$, the form of non-perturbative contributions,
and integrability. We have added to these reasonings the test of S-duality in
subsection \ref{dualityint}. We wish to further strengthen the arguments for the superpotential
by comparing the results for the exact quantum vacua for the theory on $\mathbb{R}^3 \times S^1$
with semi-classical results.

\subsection{Semi-classical Vacua}
A semi-classical analysis of the massive vacua of ${\cal N}=1^\ast$ on $\mathbb{R}^4$ 
proceeds in several steps.
First one solves the
equations of motion for constant scalar field configurations which are equivalent to the statement
that the three complex scalars satisfy a $su(2)$ algebra.
The enumeration
of inequivalent embeddings of $su(2)$ in the gauge algebra then provides the set
of classical solutions. In a second step, one analyzes the unbroken gauge group for each classical
vacuum,
and then counts the number of vacua that the corresponding pure ${\cal N}=1$
quantum theory gives rise to in the infrared (using e.g. the index calculation \cite{Witten:2000nv}).
For gauge algebra $su(n)$ the number of classical vacua was thus argued
to be equal to  the sum of the divisors of $n$ \cite{Donagi:1995cf}, and this number coincides precisely with the number
obtained from the exact superpotential \cite{Donagi:1995cf,Dorey:1999sj} for the theory on $\mathbb{R}^3 \times S^1$ (where
one 
classifies vacua in the manner described above).
For other gauge algebras, the semi-classical counting of vacua was performed in
\cite{Naculich:2001us}. For gauge algebra $so(n)$ it was argued to be:
\begin{eqnarray}
Z^{semi-class}(x,y) &=& 1 + x + x^2 y + 3 x^3 + 6 x^4 + x^5 (6 + y)   + x^6 (7 + 3 y) +
 x^7 (15 + 2 y) \nonumber \\ 
 & & + x^8 (26 + y^2)
+ x^9 (31 + 5 y) 
 +\dots
 \label{semiclassicalnumbers}
\end{eqnarray}
where the power of $x$ is equal to $n$ and the power of $y$ is the number of
massless $U(1)$'s in a given massless branch of vacua. Although we will concentrate on massive vacua, let us remark
that the semi-classical counting of massless vacua may well 
be futile in the full quantum theory, where there may be a single manifold of massless vacua of given rank \cite{Cachazo:2003yc} (albeit with different branches). Thus, we will only further consider the semi-classical formula (\ref{semiclassicalnumbers}) for $y=0$.  For low rank then, the
formula gives the following number of massive vacua, semi-classically:
\begin{equation}
so(5)  :_{sc}  6
\qquad
so(6)  :_{sc}  7
\qquad
so(7)  :_{sc} 15
\qquad
so(8)  :_{sc}  26 \, .
\end{equation}
We know the result for $so(6)=su(4)$ to be in agreement with the number of massive vacua of
the exact superpotential for the theory on $\mathbb{R}^3 \times S^1$. We moreover have that this counting
of $so(2r+1)$ vacua agrees with the semi-classical counting of vacua of $sp(2r)$
\cite{Wyllard:2007qw}, both on $\mathbb{R}^4$. In the following, we compare the predictions for
the number of vacua
some low rank gauge theories on $\mathbb{R}^4$ to the results we obtained for the massive vacua coded in the 
superpotential on $\mathbb{R}^3 \times S^1$. To make the comparison, we need to go into a little more detail of the semi-classical analysis.

\subsection{Low Rank Case Studies of Quantum Vacua}
In this subsection, we compare the analysis of integrable system extrema to the semi-classical analysis of massive 
vacua of ${\cal N}=1^\ast$ gauge theory on $\mathbb{R}^4$ case by case. We will moreover refine the counting at some
stages by taking into account the transformation properties of the vacua under remaining global symmetries.
This will also be the occasion to interpret the many duality 
properties that we found for the integrable systems in section \ref{ellint}.
We also briefly comment on the monodromies.

To wrap up a loose end first, let us note that the minimal mass  $M_{i}$ of a given vacuum $i$ can be computed  
using
the equation
\begin{equation}
\label{mass}
M^2_{k} = \min \left[ \operatorname{Spec} (\mathcal{M}_{k}^T \mathcal{M}_{k}) \right]  \, ,
\end{equation}
where $\mathcal{M}_k$ is the matrix of second derivatives of the potential
in vacuum $k$:
\begin{equation*}
\left(\mathcal{M}_{k}\right)_{ij} = \frac{\partial^2 V_k (X)}{\partial X_i \partial X_j}
\, .
\end{equation*} 
This clarifies the logic
behind our definition of isolated extrema of the integrable system (see equation (\ref{nonflat})
and the corresponding footnote).

\subsubsection*{The case $so(5)$}
Semi-classically, we expect six vacua for the gauge theory on $\mathbb{R}^4$.
Let's recall in a little more detail how this counting arises. We allow for various five-dimensional representations
of $su(2)$ as vacuum expectation values for the three complex scalars 
of ${\cal N}=1^\ast$. Even-dimensional representations
must appear in even numbers.
They need to take values in the gauge Lie algebra, and we classify them up to gauge equivalence.
One then finds the following allowed representations \cite{Naculich:2001us} -- we indicate
the dimensions of the $su(2)$ representations, the unbroken part of the gauge group, and then the number of
massive vacua they give rise to in the infrared:
\begin{eqnarray}
5 & : & \, \, \, \, \, 1 \quad : 1
\nonumber \\
3+1+1 & : & so(2) : 0
\nonumber \\
2+2+1 & : & sp(2) : 2 
\nonumber \\
1+1+1+1+1 & : & so(5) : 3 \, .
\label{scso5}
\end{eqnarray}
For instance, the $2+2+1$ dimensional representation breaks the gauge algebra down to
$sp(2)$. Classically, this gives rise to a pure ${\cal N}=1$ theory with $sp(2)$ gauge
algebra at low energies, which gives rise to two quantum vacua. Summing all the resulting numbers
of semi-classical vacua, we find six massive vacua in total.

When we compare this analysis to the exact quantum vacua that we found for the $so(5)$ gauge theory
on $\mathbb{R}^3 \times S^1$,
we remark that we have a neat correspondence. In particular, there is one vacuum, 
on the real axis, that we can identify in the exact quantum regime as 
the fully Higgsed vacuum (corresponding to the $5$-dimensional irreducible representation of $su(2)$ in the list (\ref{scso5})).
Its S-dual we interpret as a confining vacuum, and it is a triplet under
the T-transformation, agreeing neatly with the $so(5)$ confining vacua (corresponding to the trivial
representation of $su(2)$ in (\ref{scso5})). We moreover found a doublet under T-transformation, again in agreement with the
two vacua corresponding to the $sp(2)$ classical vacuum. Thus, at this level, we find excellent agreement.
We note that the analysis of section \ref{ellint} demonstrates that the six vacua are in a single
$SL(2,\mathbb{Z})$ sextuplet and that their transformation properties are in correspondence with the
transformation properties of sublattices of the torus lattice. Their (generalized) S-duality and T-duality properties
are now entirely known.

The exact analysis has revealed a seventh vacuum on $\mathbb{R}^3 \times S^1$. Its origin lies in the massless vacuum on $\mathbb{R}^4$. After compactication, we can render the massless vacuum massive. 
Indeed, we can turn on a Wilson line that commutes with the semi-classical
configuration for the adjoint scalars, and that simultaneously breaks the abelian gauge group. The necessary
Wilson line is precisely the one we found in the seventh quantum vacuum. We
have thus found its semi-classical origin.

One can wonder whether our identification used for the dual of the photon (mentioned in the introduction to section \ref{gaugetheory}), and therefore of the parameterization of the
Coulomb branch moduli space reduced the number of physical vacua on $\mathbb{R}^3 \times S^1$.  
Indeed, identifying our model as the one corresponding to gauge group $SO(5)_+$ (in the nomenclature of \cite{Aharony:2013hda,Aharony:2013kma}), leads to a doubling
of the triplet in the semi-classical analysis, while the other multiplets remain unchanged. For the 
$1+2+2$ semi-classical split, this is the case because the commutant is
 a $SU(2) \subset SO(5)$ gauge group (corresponding to a long root in $SO(5)$), and thus the pure ${\cal N}=1$ gauge theory gives rise to only a doublet of vacua upon compactification.

In the integrable system, this more careful analysis
corresponds to the rule that solutions can only be identified under shifts by $2 \omega_1$ (and not $\omega_1$). 
A look at the $so(5)$ extrema in the diagrams in subsection \ref{BCDmodels} shows that this relaxed equivalence 
relation adds precisely three vacua, namely each of the confining vacua (labelled $2,3$ and $4$) obtains a partner, as expected from the analysis of 
pure ${\cal N}=1$ \cite{Aharony:2013hda,Aharony:2013kma}.  Thus, in this more careful treatment, we increase the number
of vacua by three on both sides of the analysis.


We  have computed the masses of the vacua. They  are all roughly of the same order, and
much above the accuracy of our numerical approximations, thus guaranteeing that our vacua are indeed massive. Moreover, for a given
massive vacuum, the values of the masses are all approximately within a factor of 100 from each other. Interesting patterns in the (ratios) of masses (of various vacua) exist --
it should be fruitful to study them systematically.

\subsubsection*{The case $so(8)$}

In the case of the gauge algebra $so(8)$, we find a further set of subtleties. 
First, let's compare the quantum vacua on
$\mathbb{R}^3 \times S^1$ to the semi-classical analysis on $\mathbb{R}^4$.
The semi-classical analysis yields \cite{Naculich:2001us}:
\begin{eqnarray}
7+1 & : & H : 1_s
\nonumber \\
5+3 & : & H : 1_s
\nonumber \\
5+1+1+1 & : & so(3) : 2 
\nonumber \\
4+4 & : & sp(2) : 2^\ast
\nonumber \\
3+3 + 1+1 & : & so(2)\times so(2) : 0
\nonumber \\
3+2+2+1 & : & sp(2) : 2_s
\nonumber \\
3+1+1+1+1+1 & : & so(5) : 3
\nonumber \\
2+2+2+2 & : & sp(4) : 3^\ast
\nonumber \\
2+2+1+1+1+1 & : & sp(2) \times so(4) : 6_s
\nonumber \\
1+1+1+1+1+1+1+1 & : & so(8)   : 6_s
\end{eqnarray}
for a total of $26$ massive vacua. The semi-classical analysis was done under the assumption that the $\mathbb{Z}_2$
outer automorphism of $so(2N)$ is a gauge symmetry \cite{Naculich:2001us}, in contrast to our analysis in section \ref{ellint}. 
If we adopt this point of view, we are left with a single $\mathbb{Z}_2$ global symmetry, and we have indicated in the counting above whether
a set of vacua are a singlet ($s$) or are conjugate (${}^\ast$) under that remaining $\mathbb{Z}_2$.

Using this analysis, and the T-duality transformation properties of the integrable system extrema, we can partially match
the list of semi-classical and quantum vacua on $\mathbb{R}^4$ and $\mathbb{R}^3 \times S^1$ respectively.
The $6$ under the T-duality group makes for a match between 
extrema $10,11,12,13,14,15$ and the $su(2)$ representation $1+1+1+1+1+1+1+1$. These correspond to the confining
vacua. The doublets which are conjugate
under the remaining global $\mathbb{Z}_2$ match extrema $3$ and $4$ (as well as their $\mathbb{Z}_2$ reflections) to the representations $4+4$ and $5+1+1+1$. The conjugate triplets match $6,7,8$ (as well as their $\mathbb{Z}_2$ reflections) onto $3+1+1+1+1+1$ and $2+2+2+2$. The smaller representations of the T-duality group are harder to match.
We can still identify the Higgs vacuum with the extremum number 9, which lies on the real axis and which we can
therefore follow all the way to weak
coupling.
For other extrema, it is harder to follow the change of effective description of the gauge theory dynamics from the ultraviolet to the infrared. There is again a seeming mismatch of one in the total number of vacua.
 The origin is the same as before: one extra quantum vacuum arises from the massless vacuum in $\mathbb{R}^4$ by turning on the appropriate Wilson line after compactification on $S^1$.

Of course, our modular analysis of extrema again obtains a gauge theory interpretation. Recall that we found a singlet, triplet
and quadruplet under the modular group. The modular group plays the role of a generalized duality group \cite{Aharony:2000nt}, acting on the 
effective gauge coupling in the infrared.

Note that we also found a more surprising feature: a new duality group, with a generator corresponding
to a monodromy around a point in the fundamental domain of the effective coupling that we used to describe our theory.
We found a duodecuplet of vacua transforming under this new duality group. It could be very interesting to understand
this group better in terms of gauge theory physics, or, as associated to the choice of parameterisation in the infrared.

Again, the masses all lie very amply above our numerical accuracy, such that we can claim that we indeed
identified massive vacua. Masses are again within a factor of 100 or so from each other, and  exhibit
interesting patterns that could be explored.

\subsubsection*{The cases $so(7)$ and $sp(6)$}
For gauge algebra $so(7)$, the semi-classical analysis on $\mathbb{R}^4$ predicts
fifteen massive vacua, that arise as follows \cite{Naculich:2001us,Wyllard:2007qw}:
\begin{eqnarray}
7 & : & 1 : 1
\nonumber \\
5+1+1 & : & so(2) : 0
\nonumber \\
3+3+1 & : & so(2) : 0
\nonumber \\
3+2+2 & : & sp(2) : 2
\nonumber \\
3+1+1+1+1 & : & so(4) : 3
\nonumber \\
2+2+1+1+1 & : & sp(2) \times so(3) : 4
\nonumber \\
1+1+1+1+1+1+1 & : & so(7) : 5 \, .
\end{eqnarray}
In the quantum theory on $\mathbb{R}^3 \times S^1$, we do find a quintuplet under T-duality associated to the confining vacuum on the imaginary axis,
dual to the Higgs vacuum on the real axis, near weak effective coupling. It enhances to a septuplet under T-duality
at stronger effective coupling. From our analysis of the quantum theory on $\mathbb{R}^3 \times S^1$ we see
that the theory permits two more quantum vacua, labelled $1$ and $2$. 
Again, we checked explicitly that these arise from turning on Wilson lines in the 
massless vacua on $\mathbb{R}^4$. On the $sp(6)$ side of the duality, the two extra vacua $1^{\vee}$ and $2^{\vee}$ arise from an unbroken
$sp(2)$ gauge group (after breaking the abelian group that commutes with the $2+2+1+1$ representation). We note again that the multiplet structure under T-duality is blurred
at strong effective coupling. 

\subsection{Tensionless Domain Walls, Colliding Quantum Vacua and Masslessness}
The point in the fundamental domain around which we have found a monodromy in the case of the $so(8)$
gauge algebra, corresponds to a point at
which two massive vacua have equal superpotential. At this point, a supersymmetric
domain wall between the vacua  becomes tensionless
\cite{Abraham:1990nz,Dvali:1996xe}.
The physics associated to such a situation is hard to discuss in detail, because of the difficulty of controlling
the K\"ahler potential in gauge theories with ${\cal N}=1$ supersymmetry only. Explorations of the physics
in this regime can be found in \cite{Ferrari:2002kq,Cachazo:2002zk,Ritz:2003qq}. We note in particular 
that in a mass and cubically deformed $\mathcal{N}=1$ $U(N)$ theory in \cite{Ferrari:2002kq,Ritz:2003qq}, an 
extension of the $\mathbb{Z}_N$ action associated to shifts in the $\theta$ angle of the gauge theory 
to $\mathbb{Z}_{2N}$ was observed 
due to the presence of a point of monodromy in an effective coupling. 
The T-operation (shifting the $\theta$ angle of the gauge theory)
in our situation is also crucially influenced by the presence of the point of monodromy : above the point of monodromy (at weak effective coupling), we find a 
$\mathbb{Z}_{N-2}$ action, while below (at strong effective coupling), we find a $\mathbb{Z}_{N}$ action (for the case $N=8$ as well as for the case of $N=7$). We also note that the 
collision of the extrema of the superpotential indicates the existence of an effectively massless
excitation since there will be a zero mode for the matrix of second derivatives.
The physics, or at least the properties of the effective description, seem close to the discussion in  e.g. \cite{Ferrari:2002kq}. It would be interesting
to elucidate this point further.

\section{Conclusions and Open Problems}
\label{conclusions}

We studied the isolated extrema of complexified elliptic Calogero-Moser models, and encountered a plethora
of beautiful phenomena. The values of the integrable interparticle potential at the extrema are true vector-valued
modular forms in some cases, allowing for an analytic determination of the extrema in terms of modular forms of 
congruence subgroups of the modular group. This gives rise to webs of extrema that form representations under the duality
group of the model. The latter can either be a modular or a Hecke group. A more intricate phenomenon is the appearance of
monodromies amongst a second class of extrema as we loop around a point in the fundamental domain of the modular group. The duality group 
is then enlarged to include the monodromy generator. We determined the action of these generators 
on extrema. Moreover, we provided a  wealth of Fourier coefficients
of the extremal potential. These analyses can be viewed as a
considerable widening of the observation of the integrality of observables in
equilibria of integrable systems.

Secondly, we interpreted the results on extrema of Calogero-Moser systems in terms of mass-deformed ${\cal N}=4$ supersymmetric gauge theory
in four dimensions. 
We compared our results based on a low-energy effective action for the quantum theory on $\mathbb{R}^3 \times S^1$ to semi-classical predictions for the theory.
The total number of quantum vacua  matched the number resulting
from the semi-classical analysis, provided we took into account massive vacua that originate in massless
vacua on $\mathbb{R}^4$. A Wilson line on the circle can commute with semi-classical expectation values for the
adjoint scalars, yet break the remaining abelian factors in the gauge group to give rise to massive quantum vacua,
of either Higgs or confining type. We note that the appearance of extra vacua after compactification
is typical of the $B,C$ and $D$ series. The compactified theory manifestly differs from the theory on $\mathbb{R}^4$.
We also performed a more refined matching of quantum vacua in certain cases, thus providing further evidence that the 
superpotential gives a correct description of the physics of ${\cal N}=1^\ast$ theory on $\mathbb{R}^3 \times S^1$.
Furthermore, 
we noted that the precise multiplet structures in the quantum theory showed surprising features, including monodromy properties of the quantum vacua.

It should be clear that we only scratched the 
surface of this intriguing domain at the intersection of integrable systems, modularity and gauge theory. The new features of the extrema that we laid 
bare in the $B,C,D$-type  integrable models (compared to the 
 $A$-type theories) prompts the question of the generic counting of the extrema, the relevant duality group and modular 
structure (including monodromies) as well as their representation and number theoretic content. Clearly, our analysis begs to be extended to exceptional
algebras of low rank, to higher
rank root systems, to models with different choices of coupling constants, as well as to the integrable generalizations of the elliptic Calogero-Moser models,
including for instance the Ruijsenaars model with spin. Moreover, our study can be extended to other
observables, like the ratio of the frequencies of fluctuations.
It would also be interesting to attempt to characterize the positions of the extrema through
e.g. a generalization of 
zeroes of orthogonal polynomials \cite{Szego}.
All indications are that similarly intriguing phenomena as the ones we uncovered will appear in this broader field.

In gauge theory, one would like to analyze more closely the Seiberg-Witten curves of the
${\cal N}=2^\ast$ theories that underlie our models, and in particular, locate the 
points in Coulomb moduli space where the curve develops a number of nodes equal to the rank
of the gauge group, and where the vanishing cycles are mutually local (indicating the existence of 
massive vacua after mass deformation).
The Seiberg-Witten curves are defined by equations of  higher order, rendering this analysis
harder than in the cases treated in detail so far \cite{Donagi:1995cf}.

One would also like to have access to the large rank generalization of our results, to connect to holographic dual backgrounds with
orientifold planes \cite{Polchinski:2000uf,Aharony:2000nt,Naculich:2001us}. For these purposes it might suffice to have access
to the large rank generalization of a Higgs and confining vacuum, which one may hope to characterize analytically. It would also be useful to perform a more careful analysis of 
the discrete choices of gauge groups and line operators in our model \cite{Aharony:2013hda,Aharony:2013kma}, and to classify 
various supersymmetric vacua further \cite{'tHooft:1981ht,Aharony:2013hda}, e.g.  by understanding a single phase, then chasing it through the duality chains.

Finally, we already noted in passing that the branches of massless vacua predicted by the semi-classical analysis may 
turn out to be connected in the quantum theory, giving rise to a single massless vacuum manifold, consisting
of branches that can be characterized by differing algebraic equations  \cite{Cachazo:2003yc}. 
Our numerical explorations up to now are consistent with the fact
that all massless vacua have the same value of the superpotential. It would be interesting to clarify the structure of these vacuum manifolds further for the models at hand.

\section*{Acknowledgments}
We thank Antonio Amariti as well as the JHEP referee for encouraging comments.
We would like to acknowledge support from the grant ANR-13-BS05-0001-02, and from the \'Ecole Polytechnique and the \'Ecole Nationale Sup\'erieure des Mines de Paris.


\vspace{0.3cm}

\appendix

\section{Lie Algebra}
\label{Lie}
We briefly review Lie algebra concepts that are useful to us in discussing the symmetries of both the
integrable systems and gauge theories we discuss in the bulk of the paper. See e.g. \cite{OV} for a detailed
exposition of the following facts. Let us consider a (compact
simple) Lie group $G$ with maximal torus $T$. They have  corresponding tangent algebras $\mathfrak{g}$  and $\mathfrak{t}$.
We can then identify $T$ as a linear group, and its space of characters $\chi(T)$ is in bijection with a lattice in the
space $\mathfrak{t}^\ast (\mathbb{R})$ dual to the tangent algebra $\mathfrak{t}$, and defined over the real numbers $\mathbb{R}$.
To the Lie algebra, we can associate its space of weights in the adjoint representation, which is the set
of roots $\Delta$. Again, these roots are elements of the Euclidean space $\mathfrak{t}^\ast (\mathbb{R})$. The space
$\mathfrak{t} (\mathbb{R})$ comes
equipped with a non-degenerate scalar product, which we will denote $(\cdot,\cdot)$. 
This scalar product allows
us to identify a function $\lambda$ on the space $\mathfrak{t}(\mathbb{R})$ with an element $u_\lambda$ of the space $\mathfrak{t}(\mathbb{R})$ through the relation:
\begin{eqnarray}
\lambda (x) &=& (u_\lambda,x) \, ,
\end{eqnarray}
valid for all elements $x$ of $\mathfrak{t}(\mathbb{R})$.
We will occasionally abuse notation and write $\lambda(x)=(\lambda,x)$, and also $(u_{\lambda},u_{\lambda '})=(\lambda , \lambda ')$, which defines a dual scalar product on $\mathfrak{t}^\ast (\mathbb{R})$. The bijection between the space generated by the roots and its dual allows us to define the dual roots (i.e. the co-roots) through the relation:
\begin{equation}
  \alpha^\vee = \frac{2 u_\alpha}{( \alpha , \alpha )} \, . 
\end{equation}
The root lattice $Q$ is the lattice generated by the roots.   Any set of simple roots $\alpha_i$ spans
the space $\mathfrak{t}^{\ast}$. 
The weight lattice $P$ also sits inside $\mathfrak{t}^{\ast}$ and is defined to be generated by a 
basis $\pi_j$ such that:
\begin{eqnarray}
2 \frac{ ( \alpha_i , \pi_j ) }{ ( \alpha_i , \alpha_i ) }
&=& \delta_{ij} \, ,
\end{eqnarray}
for all $i$ and $j$ that run from $1$ to the rank of the group $G$. 
We moreover define the dual root lattice $Q^\vee$ to be the lattice generated by the dual roots, and the dual weight lattice $P^\vee$ to be the weight lattice corresponding to the dual root lattice. The dual of the lattice generated by the characters of a given group $G$ will be denoted $\mathfrak{t}(\mathbb{Z})$. We have the following properties. The center $C(G)$ of the group $G$ is given by:
\begin{eqnarray}
C(G)  \equiv  P^\vee /  \mathfrak{t}(\mathbb{Z}) \equiv \chi(T) / Q \, .
\end{eqnarray}
Moreover, when $G$ is simply connected it is equal to its universal cover $\tilde{G}$.
We then have that the space of characters is bijective to the whole of the weight lattice $\chi(T) = P $,
and that
 $\mathfrak{t}(\mathbb{Z})=Q^\vee$, such  that $C(G)$ is maximal and  $C(\tilde{G})=P/Q=P^\vee/Q^\vee$.
The group with minimal center $C(G)=1$
is the universal cover $\tilde{G}$ divided by its
center $C(\tilde{G})$. In this case we have that the 
set of weights is the set of roots $\chi(T)=Q$ and that $\mathfrak{t}(\mathbb{Z})=P^\vee$. 

Our definitions imply that the fundamental
weights $\pi^\vee_j$ that generate the dual weight lattice $P^\vee$ satisfy:
\begin{eqnarray}
(\pi^\vee_j , u_{\alpha_i}) = \delta_{ij} \, ,
\end{eqnarray}
and therefore that:
\begin{eqnarray}
\alpha(X) = (u_\alpha,X) = (\alpha,X)
\end{eqnarray}
is integer for $X$ in the dual weight lattice, i.e. for $X$ a co-weight.

We summarize inclusions and dualities in the diagram below. The arrows indicate that the lattices are dual, i.e. that the contractions give integers.  
\begin{center}
\begin{tabular}{ccccccccc}
$\mathfrak{t}^\ast  (\mathbb{R})$ & $\supset$ & $P$ & $\supset$ & $\chi(T)$ & $\supset$ & $Q$ & & \\
 & & $\updownarrow$ & & $\updownarrow$ & & $\updownarrow$ & & \\
 & & $Q^{\vee} = P^{\ast}$ & $\subset$ & $\mathfrak{t}(\mathbb{Z})$ & $\subset$ & $P^{\vee}= Q^{\ast}$ & $\subset$ & $\mathfrak{t} (\mathbb{R})$ 
\end{tabular}
\end{center}

We end this review on Lie group and Lie algebra theory with table \ref{liealgebradata} which exhibits
useful data on the Weyl group,  the outer automorphisms, the dual
Coxeter number  and the center of the universal covering group 
corresponding to the  classical Lie algebras: 
\vspace{.35cm}

\begin{minipage}{\linewidth}
 \begin{tabular}{|c|c|c|c|c|
 }
 \hline
Algebra & Weyl group & Outer Automorphisms 
& Dual Coxeter number & Center Univ. Cover 
\\
\hline 
$A_r$, $r>1$ & $S_{r+1}$ & $\mathbb{Z}_2$ 
& $r+1$ & $\mathbb{Z}_{r+1}$
\\
$A_1$ & $\mathbb{Z}_{2}$ & $1$ 
& $2$ &  $\mathbb{Z}_2$ 
\\
$B_r$ & $S_{r} \ltimes \mathbb{Z}_2^r$ & $1$ 
& $2r-1$ &  $\mathbb{Z}_2$
\\
$C_r$ & $S_{r} \ltimes \mathbb{Z}_2^r$  & $1$ 
& $r+1$ &  $\mathbb{Z}_2$ 
\\
$D_r$, odd $r$ & $S_{r} \ltimes
\mathbb{Z}_2^{r-1}$  & $\mathbb{Z}_2$ 
& $2r-2$ &   $\mathbb{Z}_4$ 
\\
$D_r$, even $r > 4$ & $S_{r} \ltimes
\mathbb{Z}_2^{r-1}$  & $\mathbb{Z}_2 $ 
& $2r-2$ & $\mathbb{Z}_2 \times \mathbb{Z}_2$  
\\
$D_4$ & $S_{4} \ltimes
\mathbb{Z}_2^3$  & $S_3$ 
& $6$ &  $\mathbb{Z}_2 \times \mathbb{Z}_2$ 
\\
\hline
\end{tabular}
\captionof{table}{Lie Algebra Data} 
\label{liealgebradata} 
\end{minipage}

\vspace{0.35cm}

\section{Elliptic Functions}
\label{ellipticfunctions}
Our conventions for the elliptic Weierstrass function are:
\begin{eqnarray}
\wp(x;\omega_1,\omega_2) &=&
\frac{1}{x^2} + \sum_{(m,n) \neq (0,0)}
\left( \frac{1}{(x+ 2m \omega_1 + 2n \omega_2)^2} -\frac{1}{(2 m \omega_1+2n \omega_2)^2} \right)
\nonumber \\
\wp(z;\tau) &=& \frac{1}{z^2} + \sum_{(m,n) \neq (0,0)}
\left( \frac{1}{(z+ m +n \tau)^2} -\frac{1}{( m+n \tau)^2} \right)
\end{eqnarray}
which entails the equality
\begin{eqnarray}
\wp(z;\tau=\omega_2/\omega_1)&=& 4 \omega_1^2 \, \wp (2 \omega_1 z ; \omega_1, \omega_2) \, .
\label{periodstotau} 
\end{eqnarray}
We impose the convention that $\Im (\omega_2/\omega_1)=\Im(\tau) >0$.
The Weierstrass function is a Jacobi form of level 2 and index 0 : 
\begin{eqnarray}\label{Weierstrass_Jacobi}
\wp \left(\frac{z}{c \tau +d};\frac{a \tau+b}{c \tau+d} \right) &=& (c \tau+d)^2 \wp(z;\tau) \, .
\label{wpmod}
\end{eqnarray}
It has the following expansion for large imaginary part of $\tau$:
\begin{eqnarray}
\wp(x;\omega_1,\omega_2) &=&
-\frac{\pi^2}{12 \omega_1^2} E_2(q)
+ \frac{\pi^2}{4 \omega_1^2} \csc^2 \left( \frac{\pi x}{2 \omega_1} \right)
- \frac{2 \pi^2}{\omega_1^2} \sum_{n=1}^\infty \frac{n q^n}{1-q^n} \cos \frac{n \pi x}{\omega_1}
\, .
\end{eqnarray}
For the twisted Weierstrass functions, we can derive the equalities:
\begin{eqnarray}
\wp (x; \omega_1 , \omega_2) + \wp (x + \omega_1 ; \omega_1 , \omega_2) &=& \wp 
(x; \frac{\omega_1}{2}, \omega_2) + \frac{ \pi^2}{6 \omega_1^2} (2 E_2
(2\frac{\omega_2}{\omega_1})-E_2 (\frac{\omega_2}{\omega_1})) \nonumber \\
\wp (x; \omega_1 , \omega_2) + \wp (x + \omega_2 ; \omega_1 , \omega_2) &=& \wp 
(x; \omega_1, \frac{\omega_2}{2}) - \frac{ \pi^2}{6 \omega_1^2} ( E_2
(\frac{\omega_2}{ \omega_1})- \frac{1}{2} E_2 (\frac{\omega_2}{2 \omega_1})) \, .
\label{halfperiod}
\end{eqnarray}
These can be proven using the definition of the Weierstrass function $\wp$,
as well as the definition of the second Eisenstein series $E_2$.

\section{Modular Forms}
\label{modularforms}
We present a compendium of modular forms that we put
to use in the bulk of our paper. 
\subsection{Theta and Eta Functions}
We first fix our conventions for the theta-functions with characteristics:
\begin{equation*}
\theta \left[ \begin{matrix} \alpha \\ \beta \end{matrix} \right] (\tau) = \sum\limits_{n \in \mathbb{Z}} \exp \left[ i \pi (n+ \alpha)^2 \tau + 2 \pi i \beta (n+\alpha)  \right] \, .
\end{equation*}
Particular examples of these theta-functions include:
\begin{eqnarray*}
\theta_2(q) &=& \theta \left[ \begin{matrix} \frac{1}{2}  \\ 0 \end{matrix} \right]  (q) = 2 q^{1/8}+2 q^{9/8}+2 q^{25/8}+2 q^{49/8} + ... \\
\theta_3 (q) &=& \theta \left[ \begin{matrix} 0 \\ 0 \end{matrix} \right] (q) = 1 + 2 q^{1/2}+ 2 q^2 + 2 q^{9/2} + 2 q^8 + ... \\
\theta_4 (q) &=& \theta \left[ \begin{matrix} 0 \\ \frac{1}{2} \end{matrix} \right] (q)  = 1 - 2 q^{1/2}+ 2 q^2 - 2 q^{9/2} + 2 q^8 + ... 
\end{eqnarray*}
We also make use of the Dedekind eta-function:
\begin{eqnarray*}
\eta(q) = q^{1/24} \prod\limits_{n=1}^{\infty} (1-q^n) \, , 
\end{eqnarray*}
and the  Klein invariant
\begin{equation}
\label{KleinInvariant}
j(q) = 1728 \frac{E_4^3(q)}{E_4^3(q)-E_6^2(q)} = \frac{1}{q}+744+196884 q+  ...
\end{equation}

\subsection{Modular Forms and Sublattices}

In this subsection we recall how to find a basis of the space of modular forms of weight $k$ for the congruence subgroup (\cite{Diamond}) :
\begin{equation*}
\Gamma (N) = \left\{  \begin{pmatrix} a & b \\ c & d \end{pmatrix} \in SL_2(\mathbb{Z}) : \begin{pmatrix} a & b \\ c & d \end{pmatrix} \equiv \begin{pmatrix} 1 & 0 \\ 0 & 1 \end{pmatrix} \, \mathrm{mod} \, N\right\}. 
\end{equation*}

First we note that the cusps of $\Gamma (N)$ can be identified with the pairs $\pm v \in (\mathbb{Z}/N\mathbb{Z})^2$ of order $N$. This makes it possible to count the number of such cusps : 
\begin{equation*}
\epsilon_{\infty} (\Gamma (N)) = 
\begin{cases}
3 & N=2 \\
\frac{N^2}{2} \prod\limits_{p|N} \left( 1- \frac{1}{p^2}\right) & N\geq 3 \, .
\end{cases}
\end{equation*}
For any congruence subgroup $\Gamma$ the space of modular forms $\mathcal{M}_k(\Gamma)$ of weight $k$ decomposes into the subspace of cusp forms and the Eisenstein space: $\mathcal{M}_k(\Gamma) = \mathcal{S}_k(\Gamma) \oplus \mathcal{E}_k(\Gamma)$. For $N=2$ we have 
\begin{equation*}
\dim \mathcal{S}_2(\Gamma(2))=0
\end{equation*}
and for $N \geq 3$, 
\begin{equation*}
\dim \mathcal{S}_2(\Gamma(N))= 1 + \frac{N^2 (N-6)}{24} \prod\limits_{p|N} \left( 1- \frac{1}{p^2}\right). 
\end{equation*}
In particular, $\dim \mathcal{S}_2(\Gamma(3))=0$ and $\dim \mathcal{S}_2(\Gamma(6))=1$. 

We want an explicit basis of the Eisenstein space. For any vector $v = \left[ \begin{matrix} c \\ d 
\end{matrix} \right] \in (\mathbb{Z}/N\mathbb{Z})^2$ of order $N$, and for $k \geq 3$, we define the (non-normalized) Eisenstein series 
\begin{equation*}
G_{k,N} \left[ v \right] (\tau) = G_{k,N}  \left[ \begin{matrix} c \\ d 
\end{matrix} \right] (\tau) = \sideset{}{'}\sum\limits_{v' \equiv v (N)} \frac{1}{(c' \tau + d')^k} \, , 
\end{equation*}
and for weight two
\begin{equation*}
G_{2,N} \left[ v \right] (\tau) = G_{2,N}  \left[ \begin{matrix} c \\ d 
\end{matrix} \right] (\tau) = \frac{1}{N^2} \left[ \wp \left( \frac{c \tau + d}{N} , \tau  \right) + G_2 (\tau ) \right] \, ,
\label{WPatlatticepoints}
\end{equation*}
where the primed sum runs over those non-vanishing vectors $v' = \left[ \begin{matrix} c' \\ d' 
\end{matrix} \right]$ that equal $v$ modulo $N$. 
One can show that the Fourier expansion of these functions in terms of $q = e^{2i \pi \tau}$ is: 
\begin{equation*}
G_{k,N}  \left[ \begin{matrix} c \\ d 
\end{matrix} \right] (q) = \delta (c) \zeta^d_N(k) + \frac{(-2\pi i)^k}{N^2 
(k-1)!} \sum\limits_{n=1}^{\infty} \sigma_{k-1,N} \left[ \begin{matrix} c \\ d 
\end{matrix} \right] (n) q^{n/N}
\end{equation*}
where
\begin{equation*}
\sigma_{k-1,N} \left[ \begin{matrix} c \\ d 
\end{matrix} \right] (n) =   \sum\limits_{m|n \, \mathrm{and} \, \frac{n}{m} 
\equiv c 
(N)} \mathrm{sgn}(m) m^{k-1} \exp \left( 2 \pi i \frac{d m}{N}\right)
\end{equation*}
and
\begin{equation*}
 \zeta^d_N(k) = \sideset{}{'}\sum\limits_{d' \equiv d \, (N)} \frac{1}{(d')^k} \, .
\end{equation*}
This Fourier expansion is valid for all $k \geq 2$, including $k=2$ which is the case we are mostly interested in. 

For $k \geq 3$, any set $\{ G_{k,N} [v] \}$ with one $v$ corresponding to each cusp of $\Gamma (N)$ represents a basis of the space $\mathcal{E}_k (\Gamma (N))$ of Eisenstein series of weight $k$ on $\Gamma (N)$ (and in particular $\dim \mathcal{E}_k (\Gamma (N)) = \epsilon_{\infty}$). For the case $k=2$ these statements have to be modified, because of the lack of modularity of the (ordinary) weight 2 Eisenstein series. It turns out that $\dim \mathcal{E}_2 (\Gamma (N)) = \epsilon_{\infty}-1$, and that $\mathcal{E}_2 (\Gamma (N))$ is the set of linear combinations of the $\{ G_{k,N} [v]\}$ (where $v \in (\mathbb{Z}/N\mathbb{Z})^2$ is of order $N$) whose coefficients sum to 0.\footnote{Theorem 4.6.1 in \cite{Diamond}}

 The Eisenstein series $G_{k,N} \left[v \right]$ have good transformation properties under 
$SL(2,\mathbb{Z})$ for $k \geq 3$ and $N \in \{ 2,3\}$ provided the vector $v$ is transformed 
accordingly:\footnote{For generic $N$ the relation between the normalized Eisenstein series, which enjoy these good transformation properties, and the series $G_{k,N} 
\left[v \right]$ is not simply a proportionality relation (see formula (4.5) in \cite{Diamond}), but it is
a simple rescaling for 
$N=2$ and $N=3$. }  
\begin{equation*}
\frac{1}{(c \tau + d)^k} G_{k,N}  \left[ v \right] \left( \frac{a \tau + b}{c \tau + d}  \right) = G_{k,N}  \left[ \begin{pmatrix} a & b \\ c & d \end{pmatrix} v \right] (\tau) \, .
\end{equation*}
For $k=2$,  we have to take into account a 
non-holomorphic term, except for linear combinations where the sum of the coefficients vanishes, as is 
the case for the potentials considered in the bulk of the paper. 

Finally, we also define
\begin{eqnarray}
E_{2,N}(\tau) &=& E_2(\tau)- N E_2(N \tau) \, . 
\end{eqnarray}
These are weight 2 modular forms of $\Gamma_0 (N)$. We use extensively the fact that $\mathcal{M}_2(\Gamma_0 (4))$ has dimension 2 and is generated by 
\begin{eqnarray*}
-E_{2,2}(q) &=& 1+24 q+24 q^2+96 q^3+24 q^4+144 q^5 + ... \\
-E_{2,4}(q) &=& 3+24 q+72 q^2+96 q^3+72 q^4+144 q^5 + ...
\end{eqnarray*}
We note the transformation property of the form $E_{2,2}$ under $\tau \rightarrow -1/(2 \tau)$:
\begin{eqnarray}
E_{2,2} (-1/(2 \tau)) &=& E_2(-1/(2 \tau)) - 2 E_2 (-1/\tau)
=-2 \tau^2 E_{2,2}(\tau) \, .  
\label{S2E2}
\end{eqnarray}

\section{The List of Extrema}
In this appendix, we list the extrema of the complexified (twisted) elliptic Calogero-Moser models
with root systems $D_4=so(8)$ and $B_3=so(7)$. We provide a few more details on how we obtained them, how to
relate them through dualities, as well as  Fourier expansions of the extremal potentials.

\subsection{The List of Extrema for $so(8)$}

\label{listofextremaso8}

The strategy we used to find extrema boils down to finding all the minima (which are also zeros) of the (auxiliary, gauge theory) potential (\ref{nonflat}) with non vanishing mass (\ref{mass}) using a simple gradient algorithm with random initial conditions. Then we identify those configurations that are related by one of the symmetries we quotient by. 
This procedure is executed at a given value of $\tau$. Once the complete list of extrema is known, we can follow a given extremum along any curve in the $\tau$ upper half plane, by adiabatically varying $\tau$. The $T$-dual extrema and the monodromies are obtained in this way, while the action of $S$-duality is known exactly. We thus unfold the whole web of dualities. 

In order to determine the potential at the extrema, we first make use of our knowledge of $T$-duality, which dictates the Fourier expansion variable  $q^{1/n}$, where $n$ is the smallest positive integer such that $T^n$ acts trivially on the extremum under consideration. Then we evaluate the extremal potential at many different values of $\tau$ and find recursively the rational Fourier coefficients.

\subsubsection{The diagrams of the extrema}

In the diagrams that follow, the black dots represent the values of the components $X_i$, $i=1,2,3,4$
at the extrema. The dark grey dots are images under the symmetries discussed in section \ref{BCDmodels}.

In some of the diagrams, there are five black dots instead of four, reflecting the fact that they represent  three extrema related by the global $S_3$ symmetry.
For every such extremum, one subgroup $\mathbb{Z}_2 \subset S_3$ acts trivially.
One of the three extrema is obtained by choosing one of the circled black dots and the three ordinary black dots, a second one is obtained by choosing the other circled black dot and the three black dots,
while the third is determined by the four small black dots. (The pale grey dots  show the possible translations of this extremum by half-periods.)

We note in passing that some exact information on the positioning of the extrema is available.
For instance, for extremum number 9, some exact information on the positions is the following. At $\tau \rightarrow i \infty$, the system reduces to the Sutherland system (with trigonometric potential). According to \cite{Corrigan:2002th}, the positions at equilibrium are related to the roots of a Jacobi polynomial. Explicitly in the case of $D_4$, the polynomial  is $P^{(1,1)}_2 (y)=\frac{15}{4} (y-1)^2+\frac{15 }{2}(y-1)+3$, from which we deduce the positions $X_1=0$, $X_{2,3}=\frac{1}{2\pi} \arccos (\pm 1 / \sqrt{5})$ and $X_4=\frac{1}{2}$. For $\tau \rightarrow 0$, the positions converge on $X_1=0$, $X_{2}=1/6$, $X_{2}=1/3$ and $X_4=\frac{1}{2}$. This  numerical convergence is slow.

S-duality guarantees that the situation is similar for the extremum on the imaginary axis, with the two limits exchanged. Moreover, T-duality then acts in the $\tau \rightarrow i \infty$ limit as $X_0 \rightarrow X_0$, $X_1 \rightarrow X_1 + 1/6$, $X_2 \rightarrow X_2 + 1/3$, $X_3 \rightarrow X_3 + 1/2$. (These transformations are exact within the precision of the numerics.) This generates the 6-cycle. Et cetera.

\begin{minipage}{\linewidth}
\begin{center}
Extrema at $\tau=i$ for $so(8)$
\end{center}
\vspace{1em}
\begin{tabular}{p{5cm}p{5cm}p{5cm}}

&

{%
\setlength{\fboxsep}{8pt}%
\setlength{\fboxrule}{0pt}%
\fbox{\includegraphics[width=3.5cm]{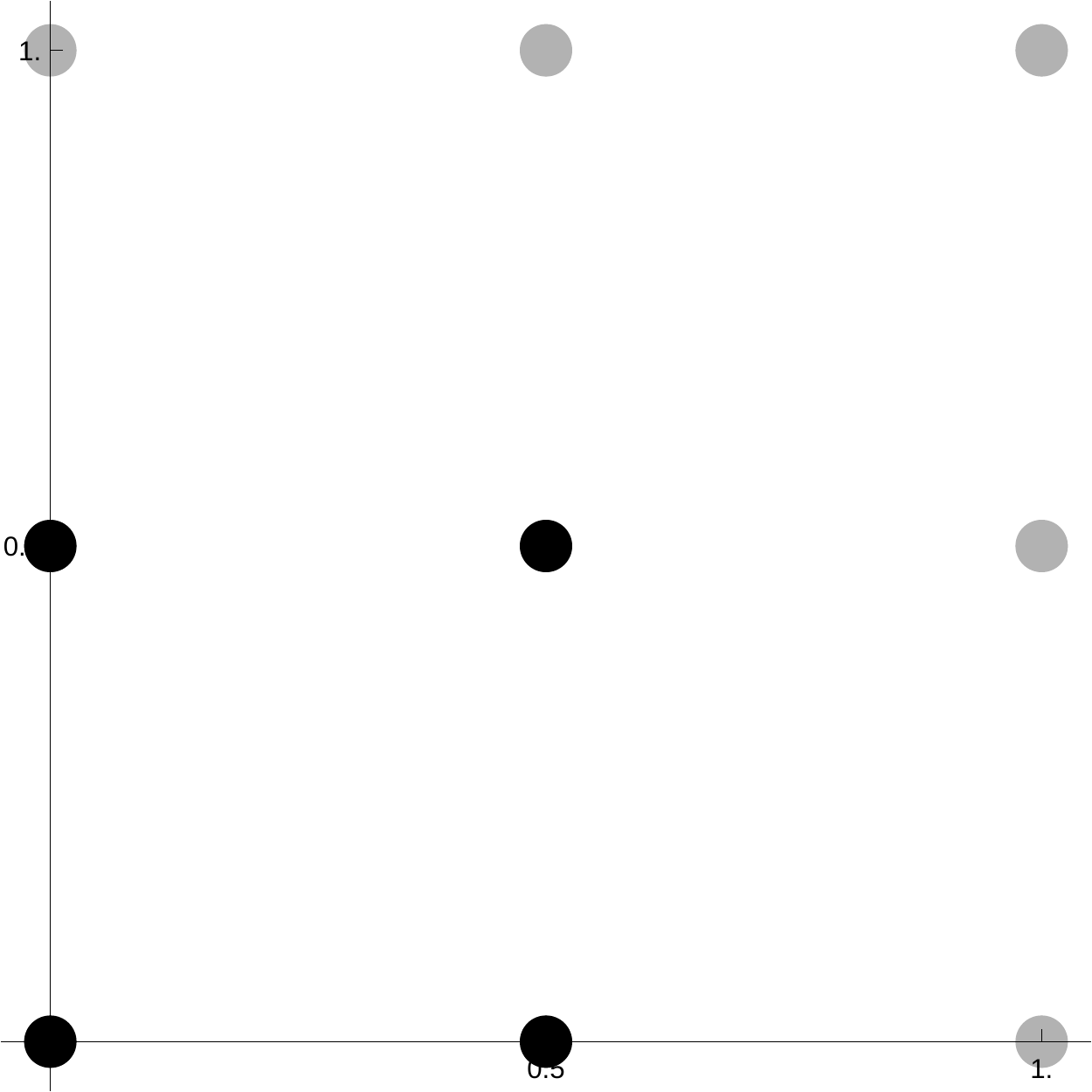}}%
}%
\begin{center}
{Extremum 1}
\end{center}

& \\

{%
\setlength{\fboxsep}{8pt}%
\setlength{\fboxrule}{0pt}%
\fbox{\includegraphics[width=3.5cm]{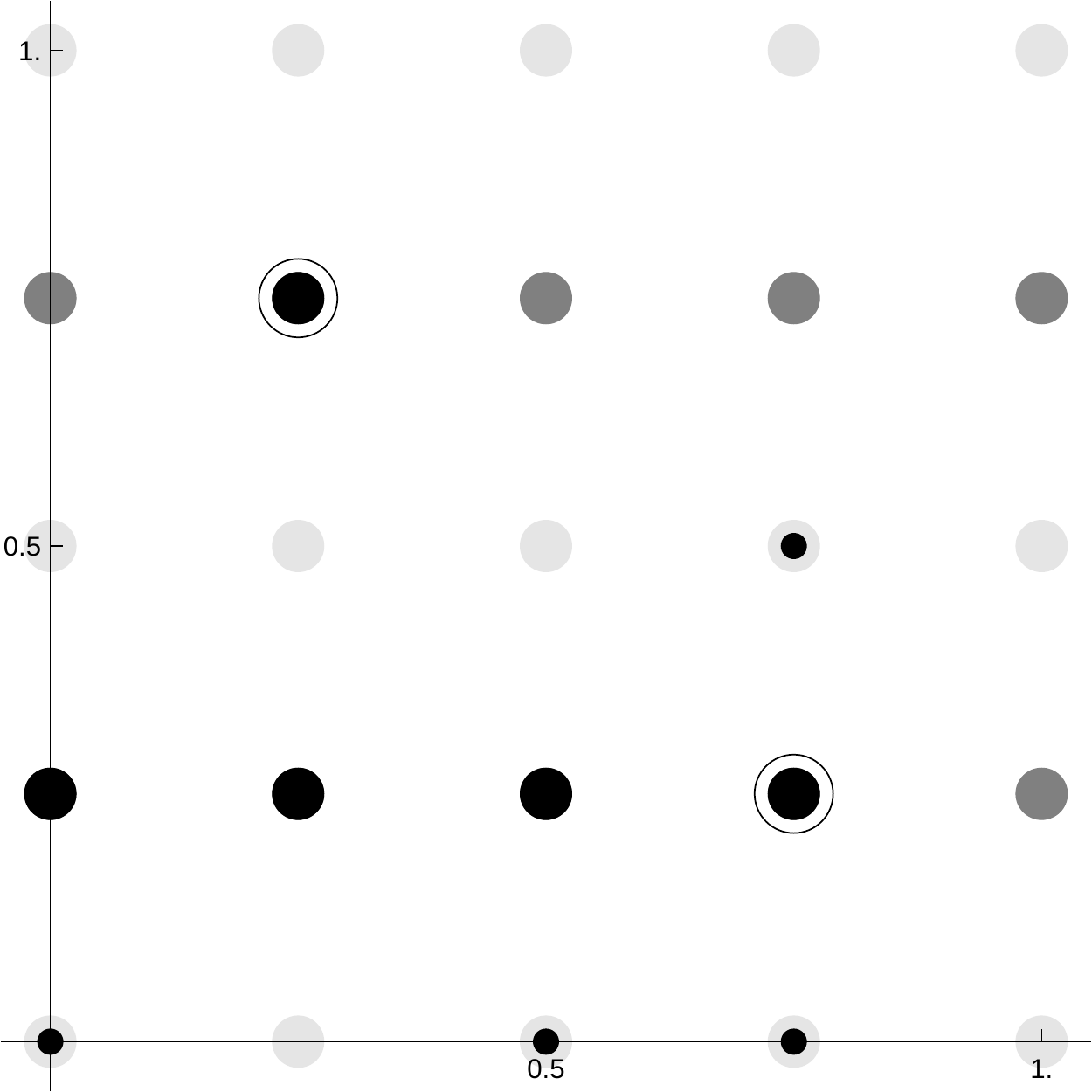}}%
}%
\begin{center}
{Extremum 2}
\end{center}

&

{%
\setlength{\fboxsep}{8pt}%
\setlength{\fboxrule}{0pt}%
\fbox{\includegraphics[width=3.5cm]{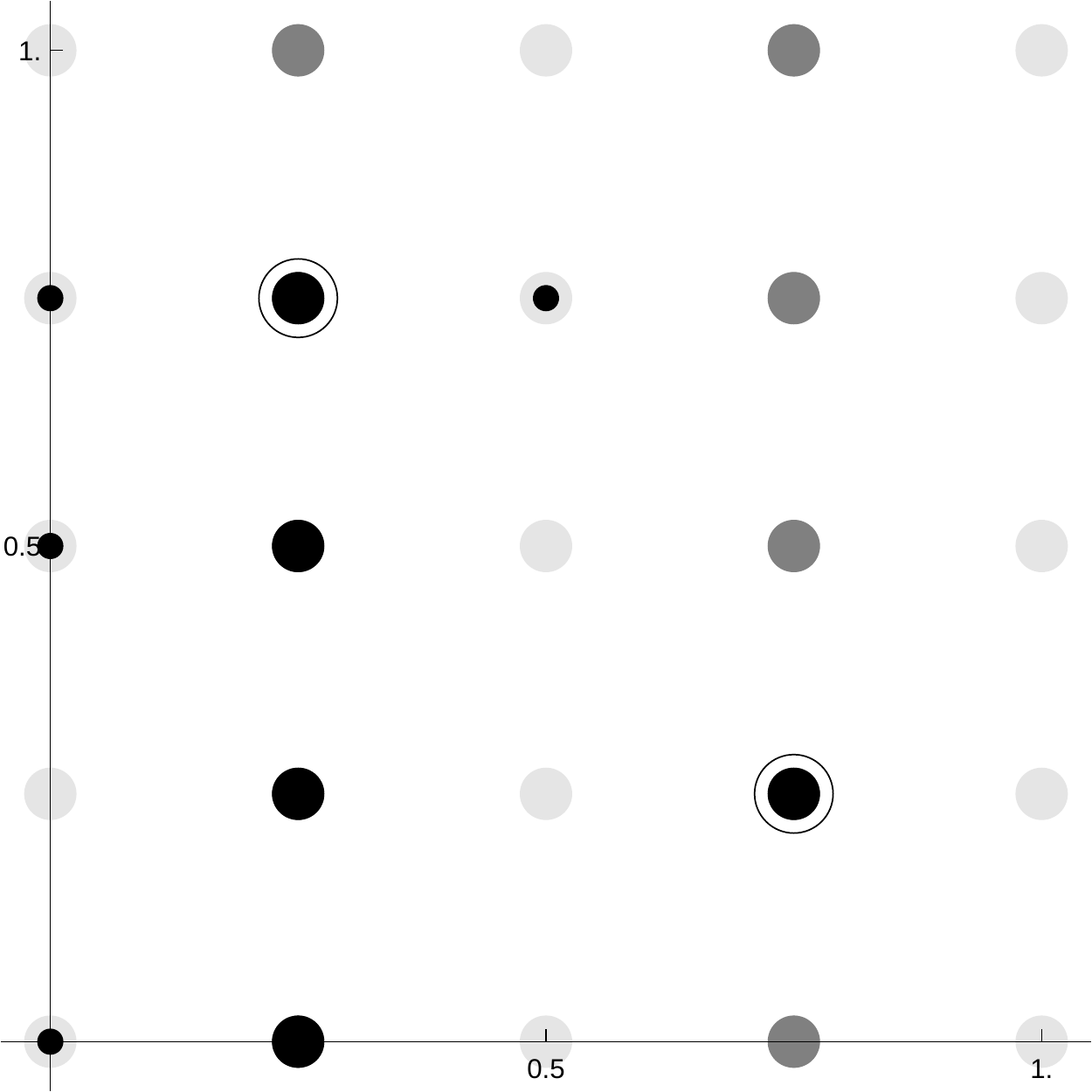}}%
}%
\begin{center}
{Extremum 3}
\end{center}

&

{%
\setlength{\fboxsep}{8pt}%
\setlength{\fboxrule}{0pt}%
\fbox{\includegraphics[width=3.5cm]{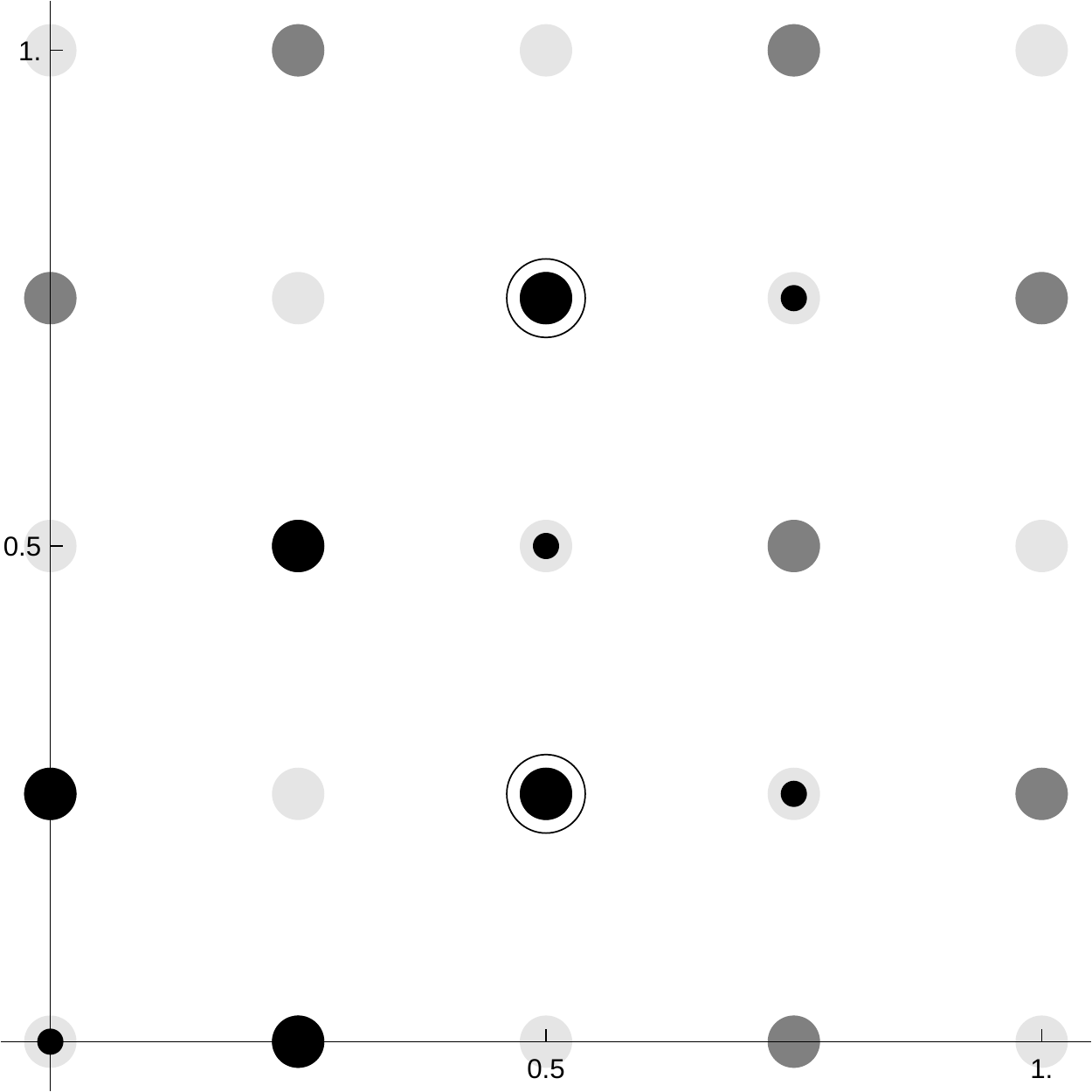}}%
}%
\begin{center}
{Extremum 4}
\end{center}

\\

\end{tabular}
\end{minipage}

\begin{minipage}{\linewidth}
\begin{tabular}{p{5cm}p{5cm}p{5cm}}

{%
\setlength{\fboxsep}{8pt}%
\setlength{\fboxrule}{0pt}%
\fbox{\includegraphics[width=3.5cm]{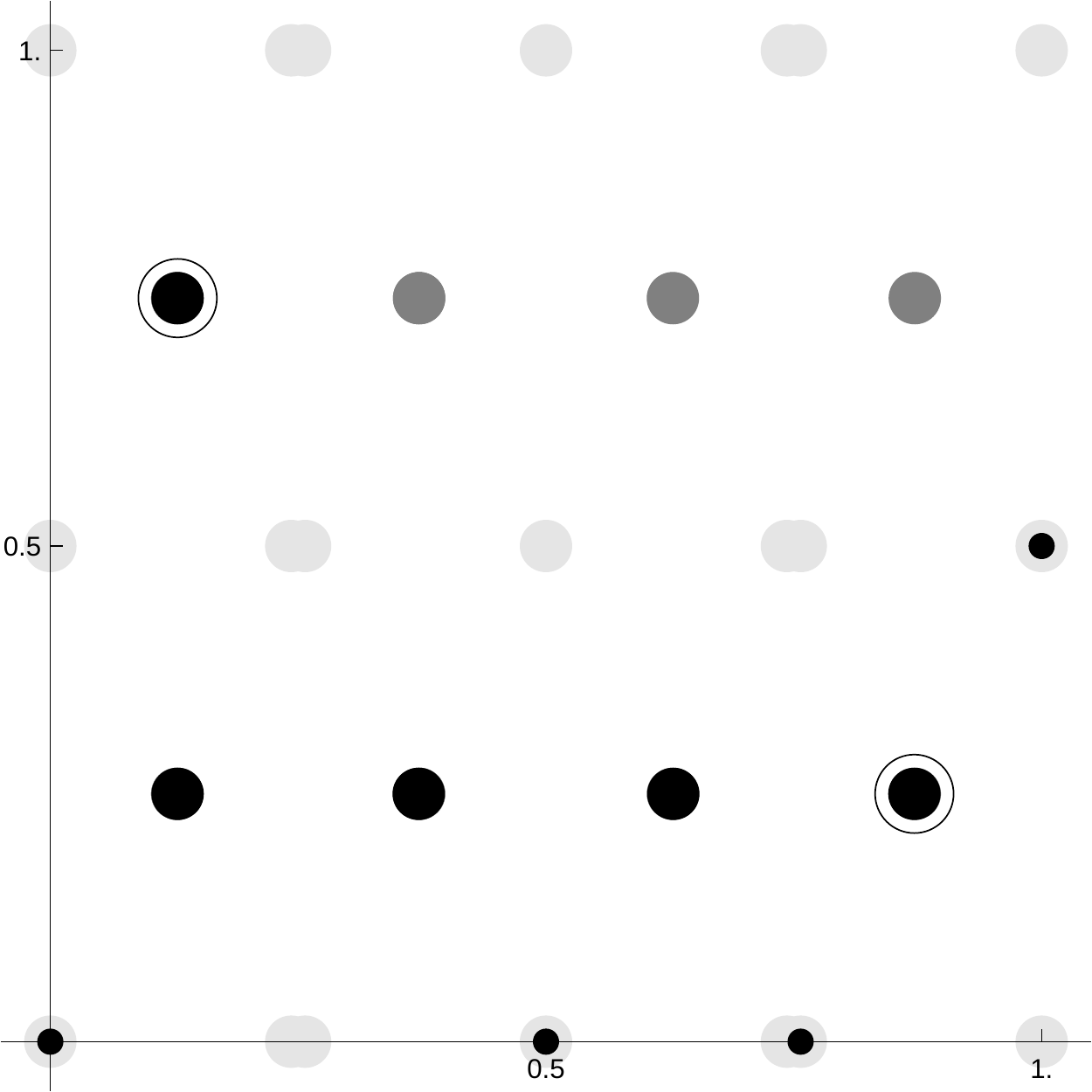}}%
}%
\begin{center}
{Extremum 5}
\end{center}

&

{%
\setlength{\fboxsep}{8pt}%
\setlength{\fboxrule}{0pt}%
\fbox{\includegraphics[width=3.5cm]{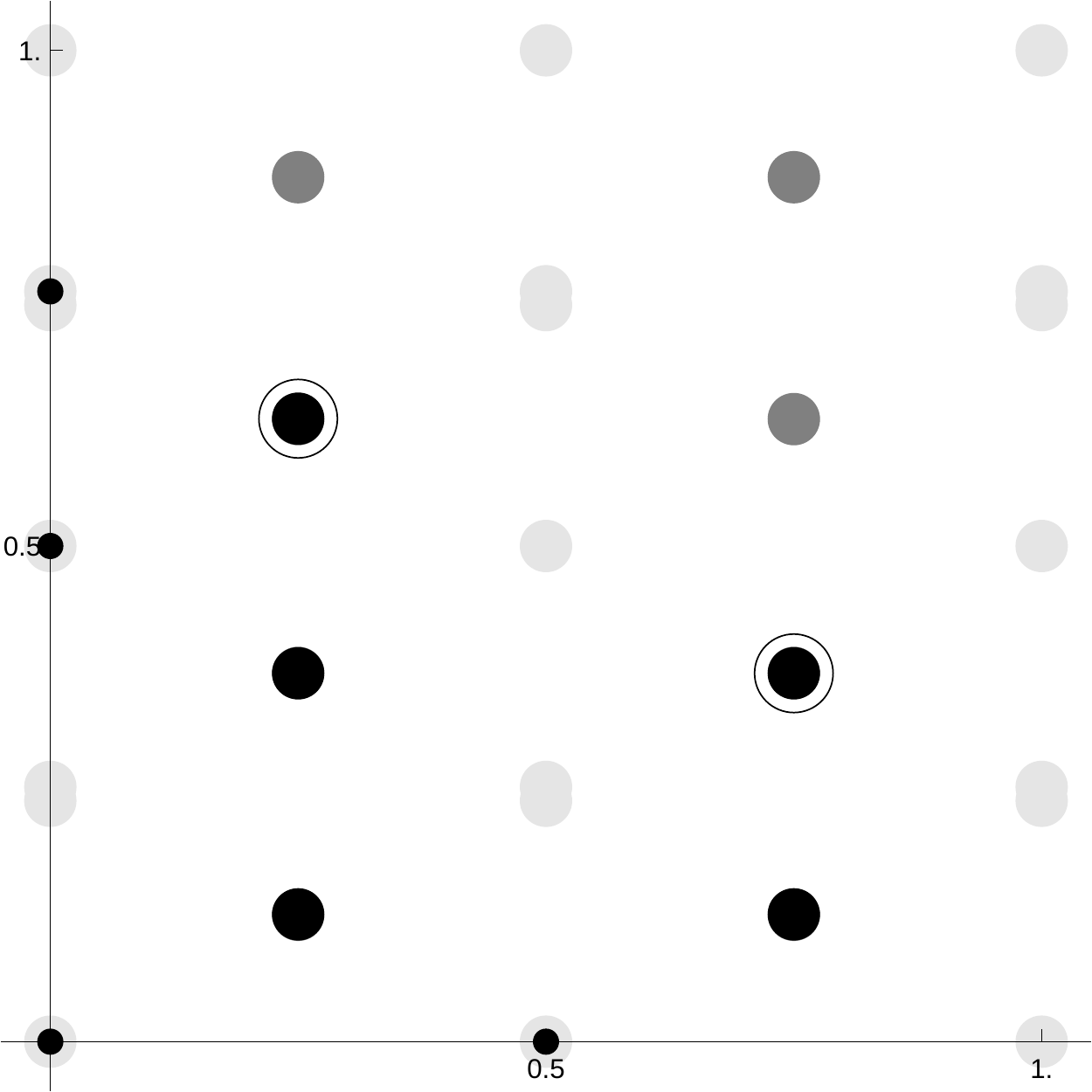}}%
}%
\begin{center}
{Extremum 6}
\end{center}

&

{%
\setlength{\fboxsep}{8pt}%
\setlength{\fboxrule}{0pt}%
\fbox{\includegraphics[width=3.5cm]{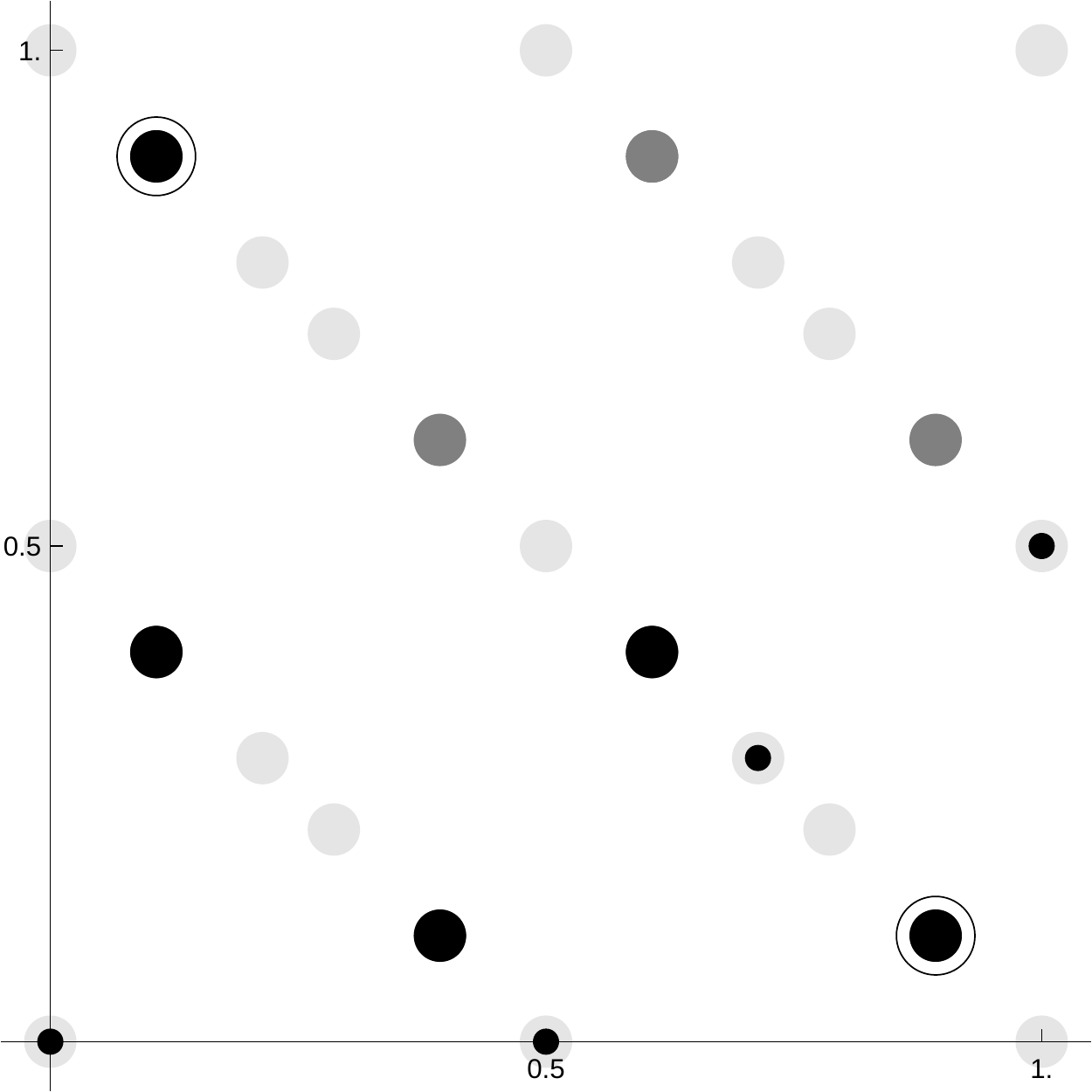}}%
}%
\begin{center}
{Extremum 7}
\end{center}

\\

&

{%
\setlength{\fboxsep}{8pt}%
\setlength{\fboxrule}{0pt}%
\fbox{\includegraphics[width=3.5cm]{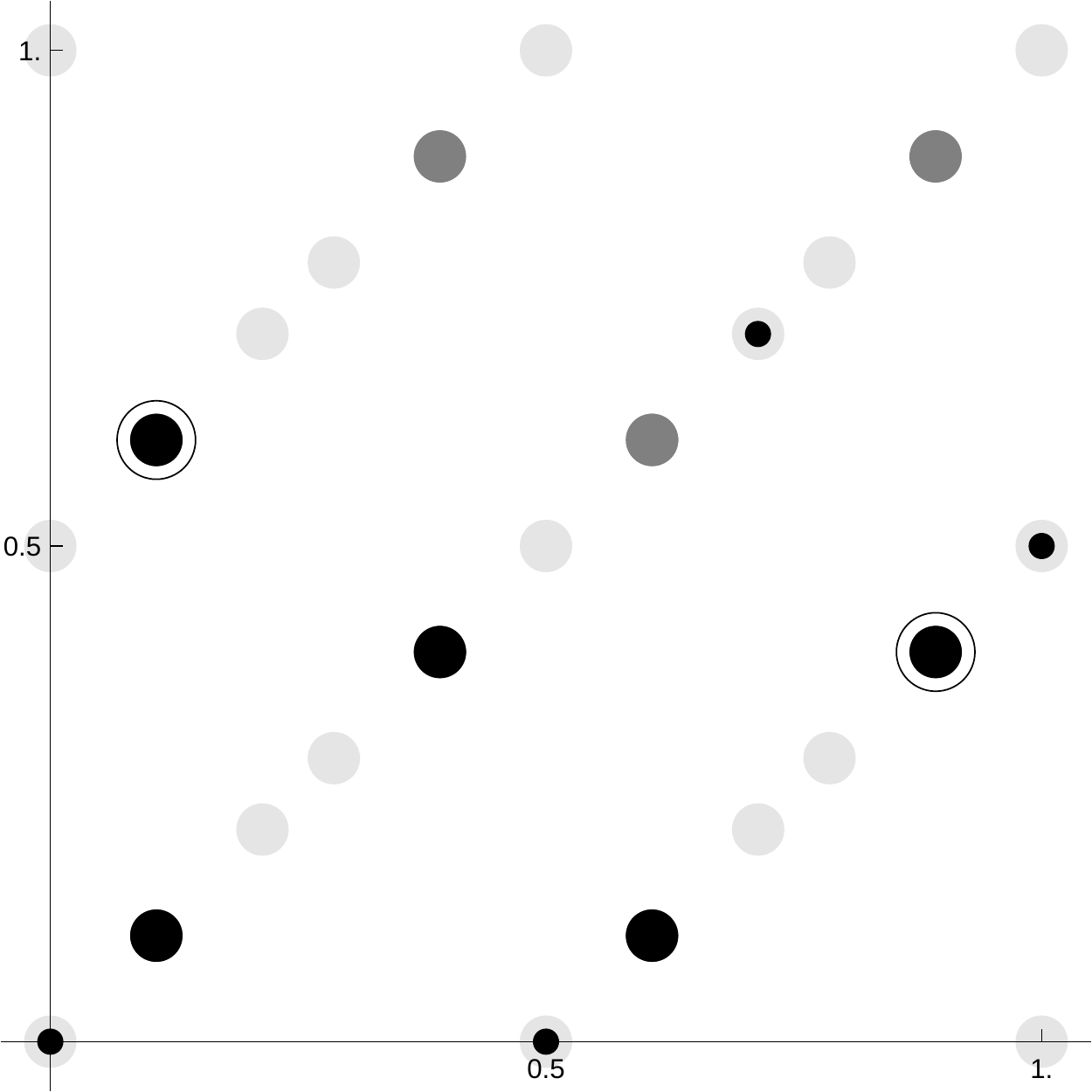}}%
}%
\begin{center}
{Extremum 8}
\end{center}

&

\end{tabular}
\end{minipage}

\begin{minipage}{\linewidth}
\begin{center}
Extrema at $\tau=i$
\end{center}
\vspace{1em}
\begin{tabular}{p{5cm}p{5cm}p{5cm}}

{%
\setlength{\fboxsep}{8pt}%
\setlength{\fboxrule}{0pt}%
\fbox{\includegraphics[width=3.5cm]{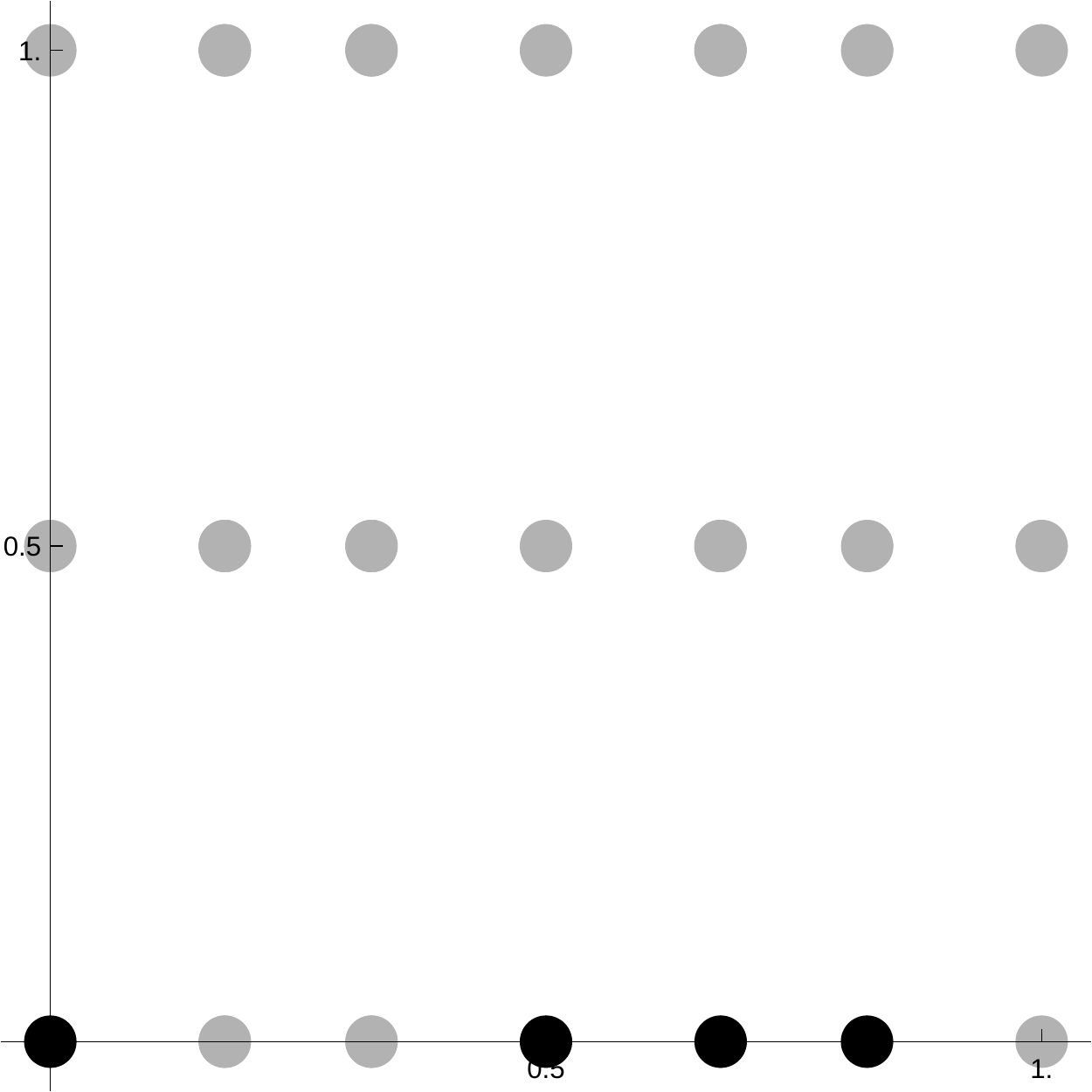}}%
}%
\begin{center}
{Extremum 9}
\end{center}

&

{%
\setlength{\fboxsep}{8pt}%
\setlength{\fboxrule}{0pt}%
\fbox{\includegraphics[width=3.5cm]{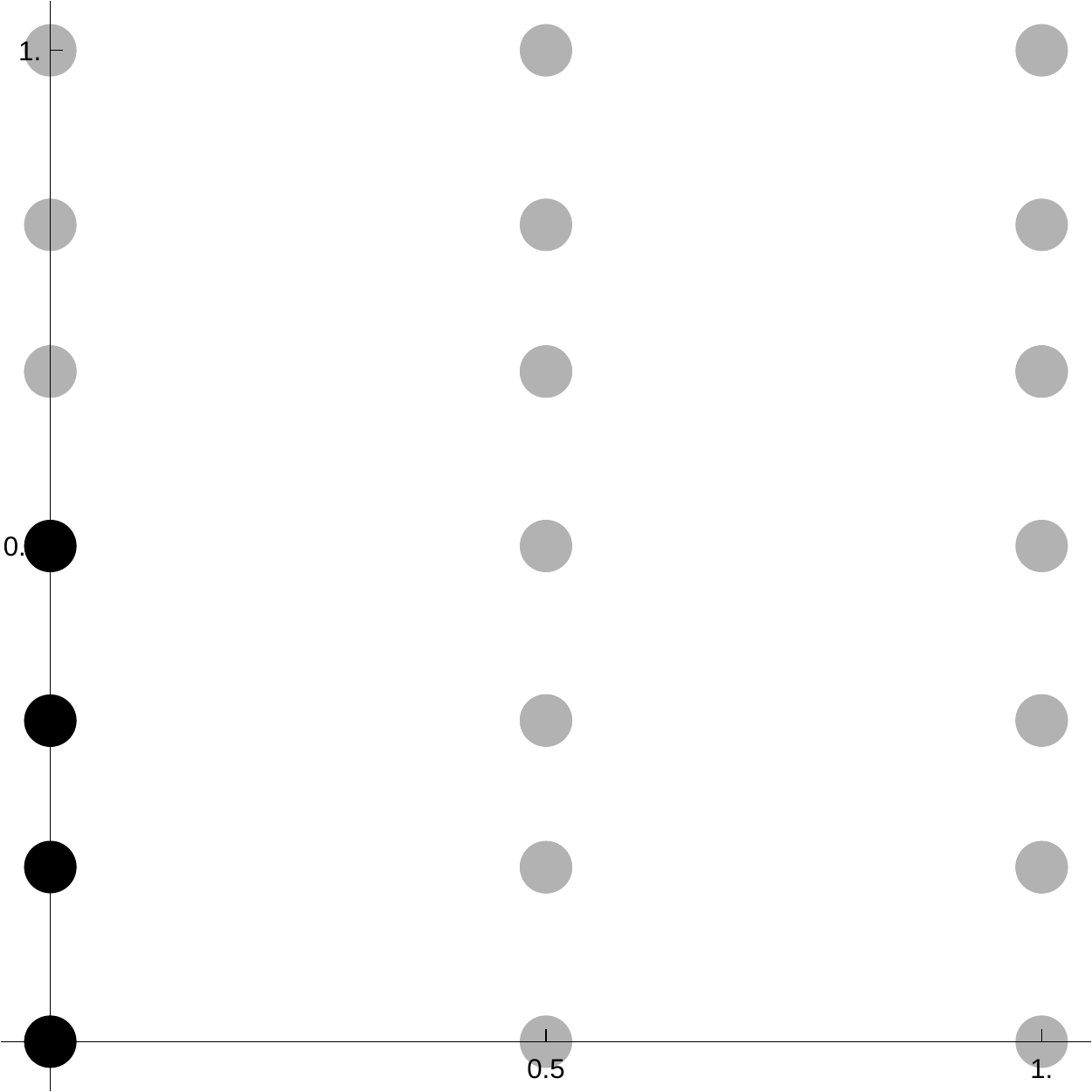}}%
}%
\begin{center}
{Extremum 10}
\end{center}

&

{%
\setlength{\fboxsep}{8pt}%
\setlength{\fboxrule}{0pt}%
\fbox{\includegraphics[width=3.5cm]{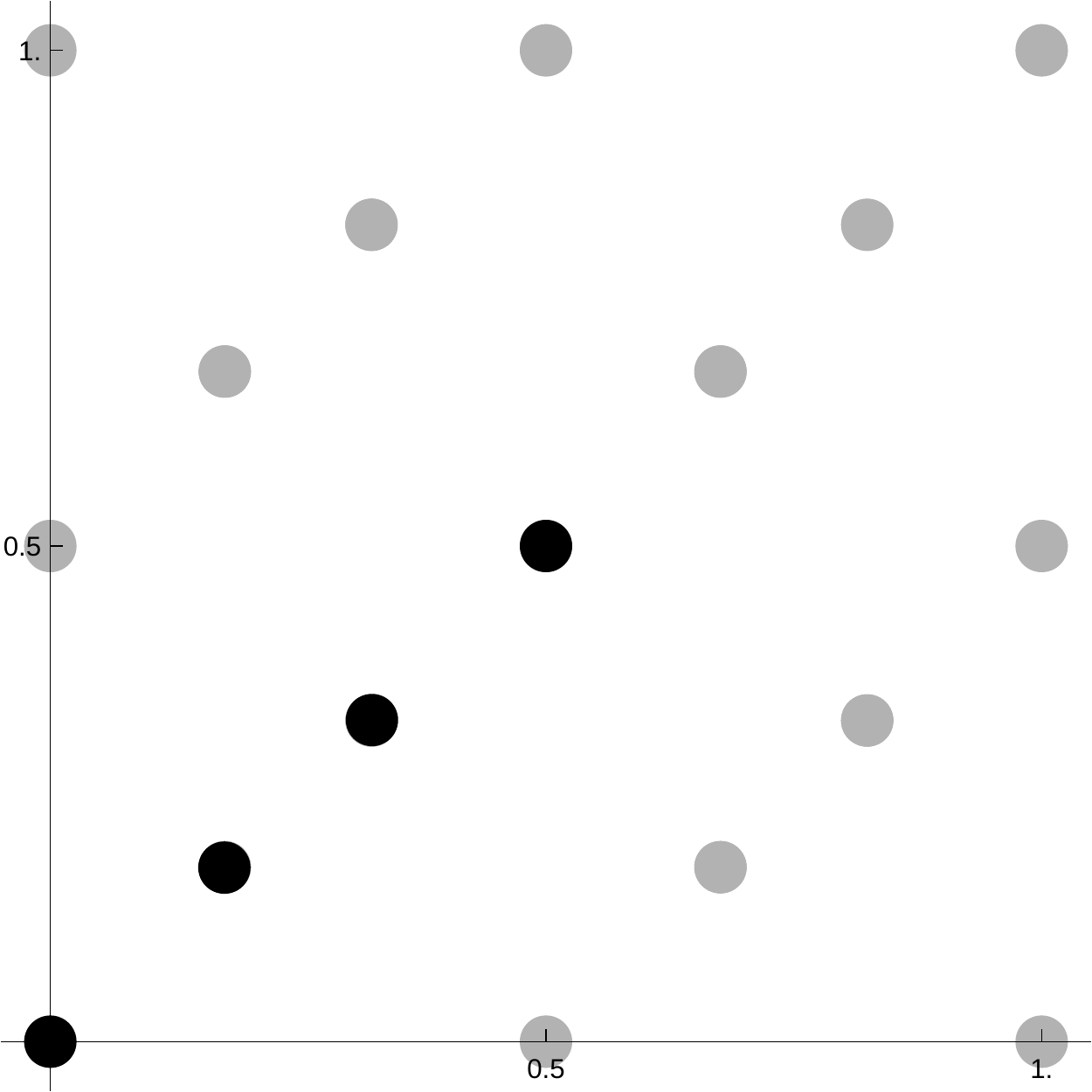}}%
}%
\begin{center}
{Extremum 11}
\end{center}

\\

{%
\setlength{\fboxsep}{8pt}%
\setlength{\fboxrule}{0pt}%
\fbox{\includegraphics[width=3.5cm]{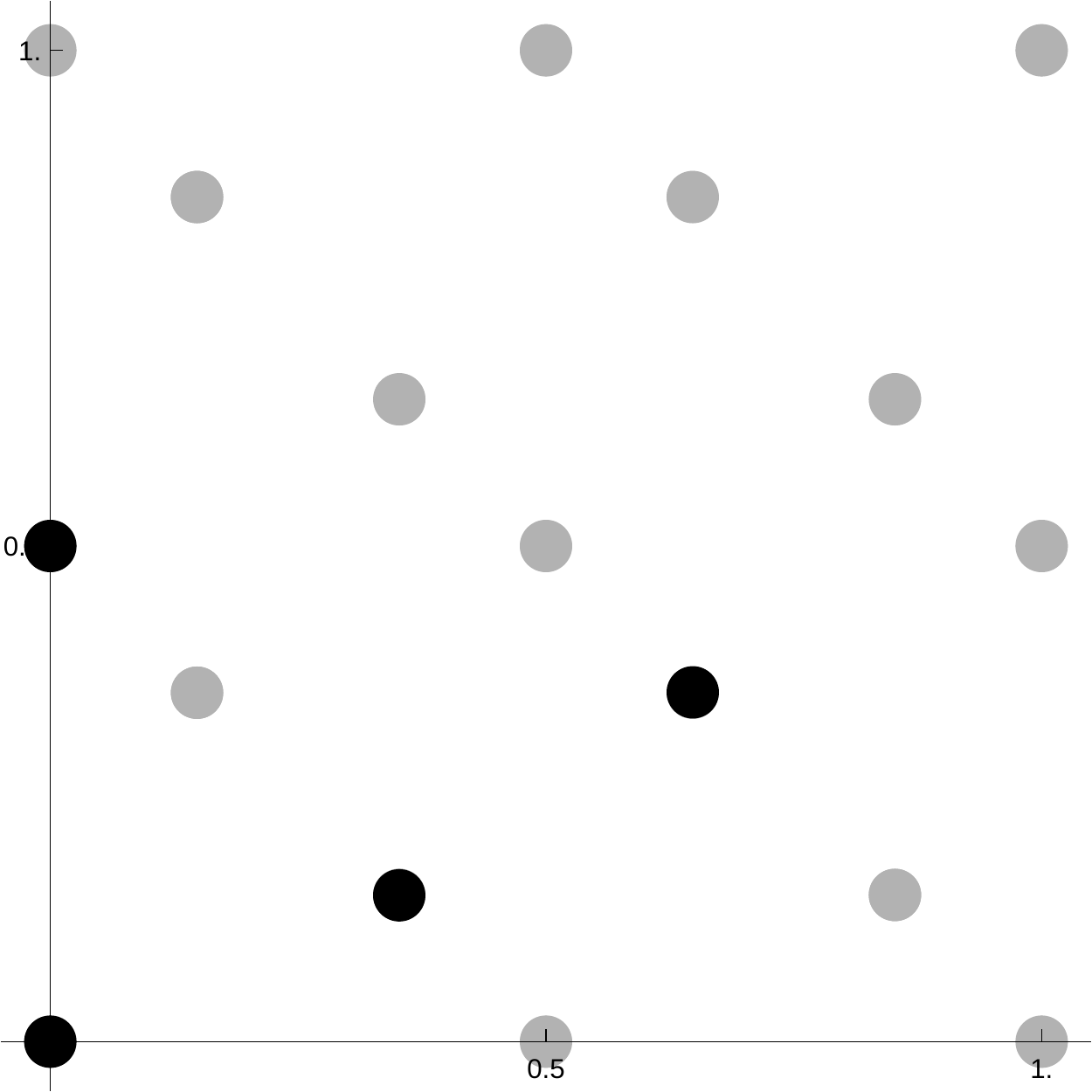}}%
}%
\begin{center}
{Extremum 12}
\end{center}

&

{%
\setlength{\fboxsep}{8pt}%
\setlength{\fboxrule}{0pt}%
\fbox{\includegraphics[width=3.5cm]{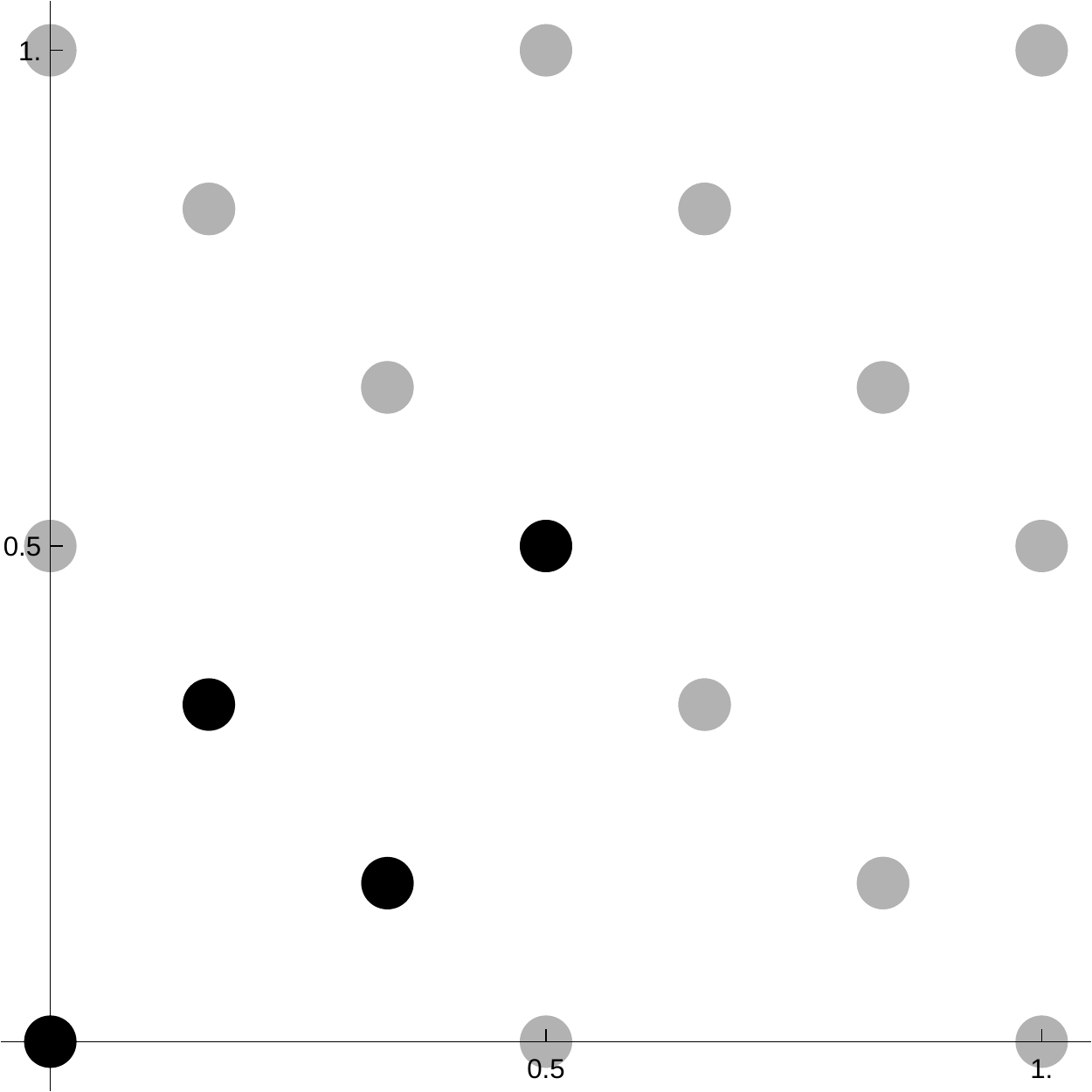}}%
}%
\begin{center}
{Extremum 13}
\end{center}

&

{%
\setlength{\fboxsep}{8pt}%
\setlength{\fboxrule}{0pt}%
\fbox{\includegraphics[width=3.5cm]{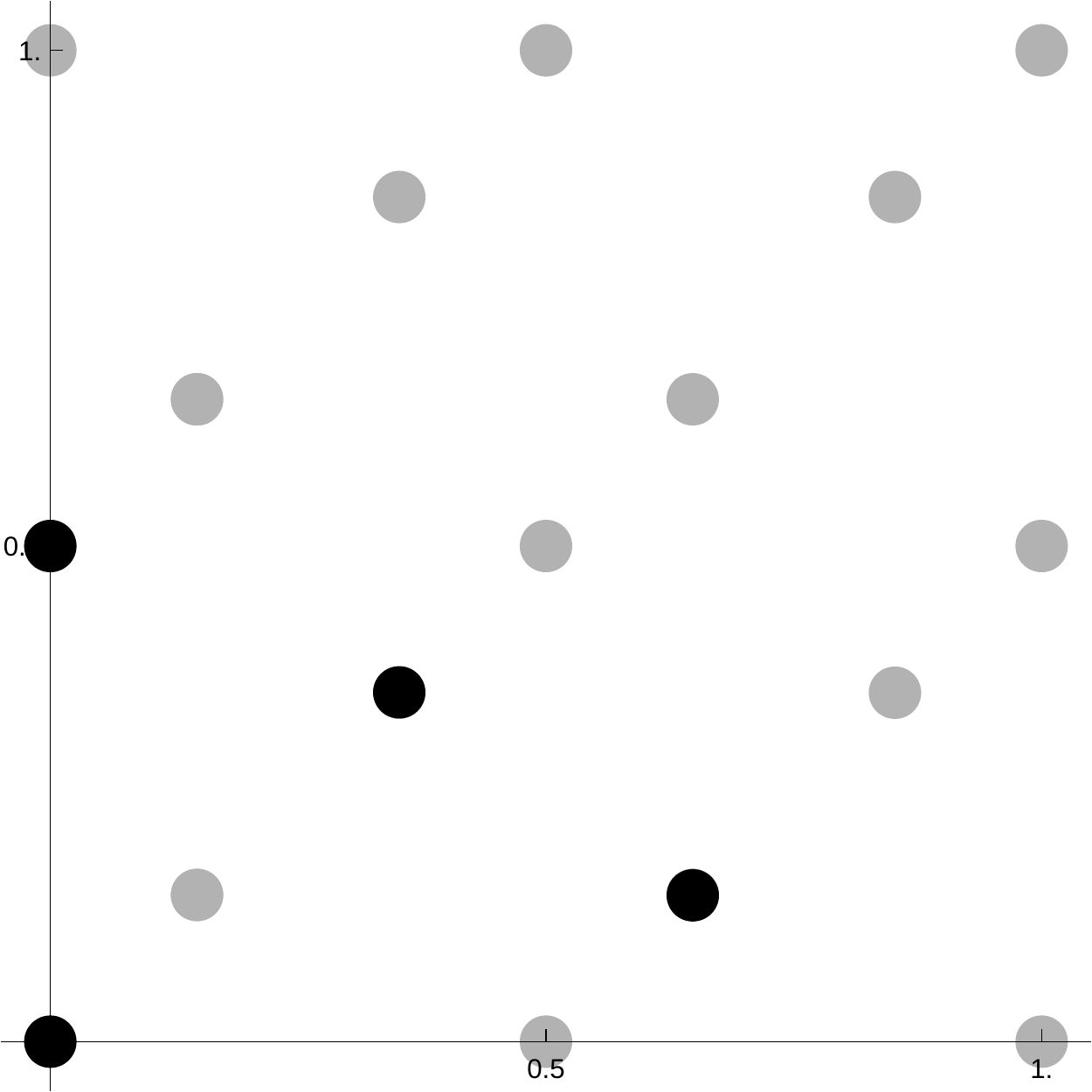}}%
}%
\begin{center}
{Extremum 14}
\end{center}

\\

{%
\setlength{\fboxsep}{8pt}%
\setlength{\fboxrule}{0pt}%
\fbox{\includegraphics[width=3.5cm]{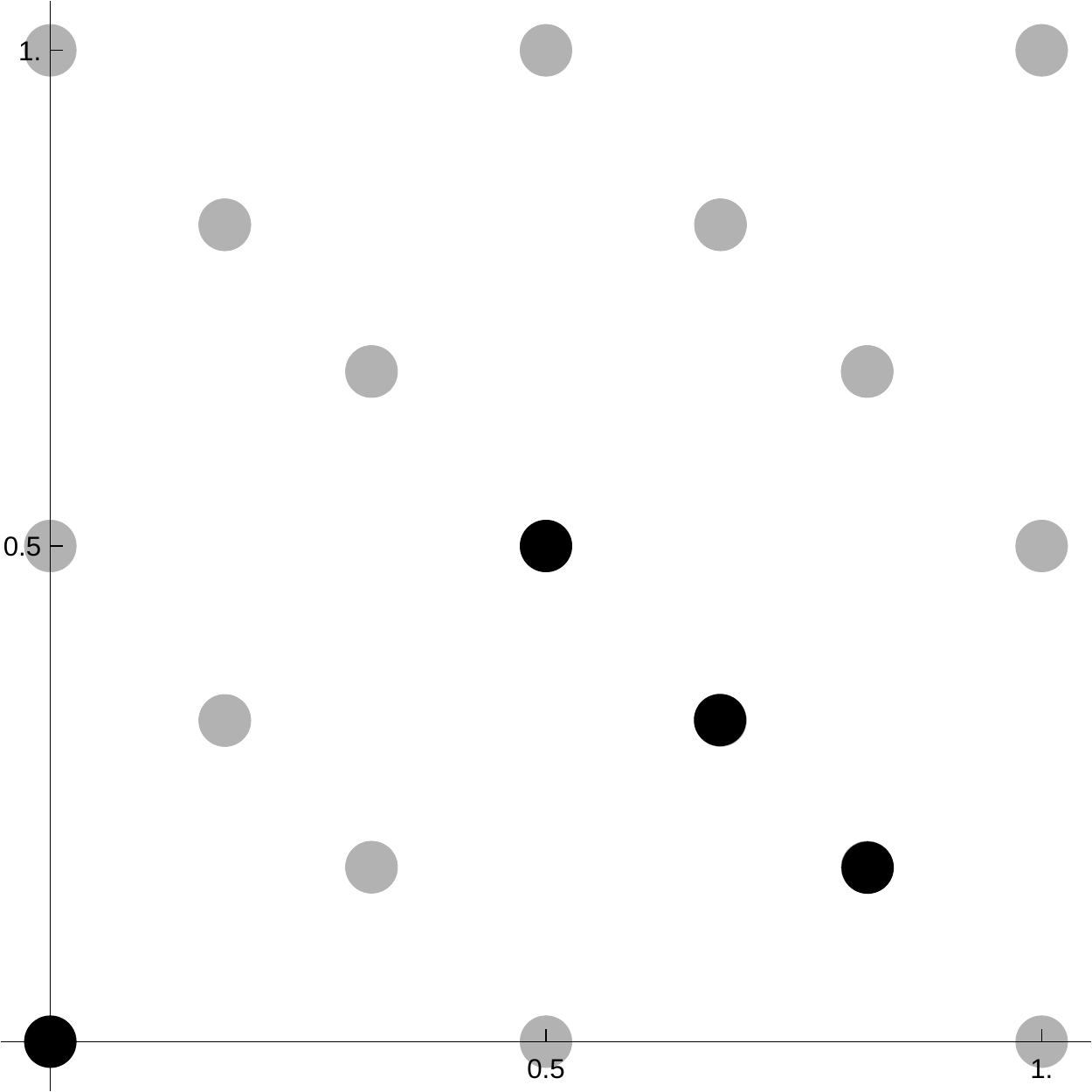}}%
}%
\begin{center}
{Extremum 15}
\end{center}

&

{%
\setlength{\fboxsep}{8pt}%
\setlength{\fboxrule}{0pt}%
\fbox{\includegraphics[width=3.5cm]{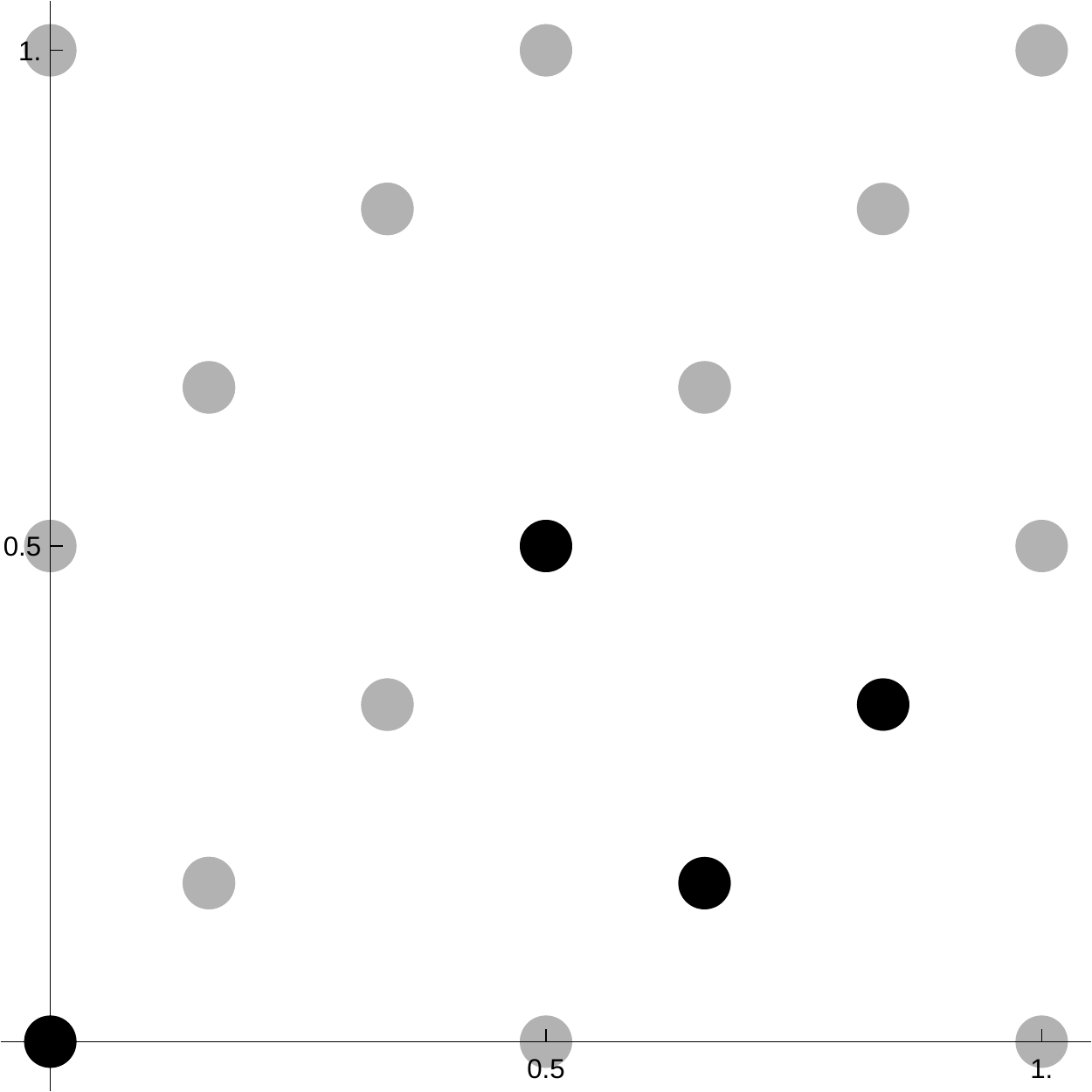}}%
}%
\begin{center}
{Extremum 16}
\end{center}

&

{%
\setlength{\fboxsep}{8pt}%
\setlength{\fboxrule}{0pt}%
\fbox{\includegraphics[width=3.5cm]{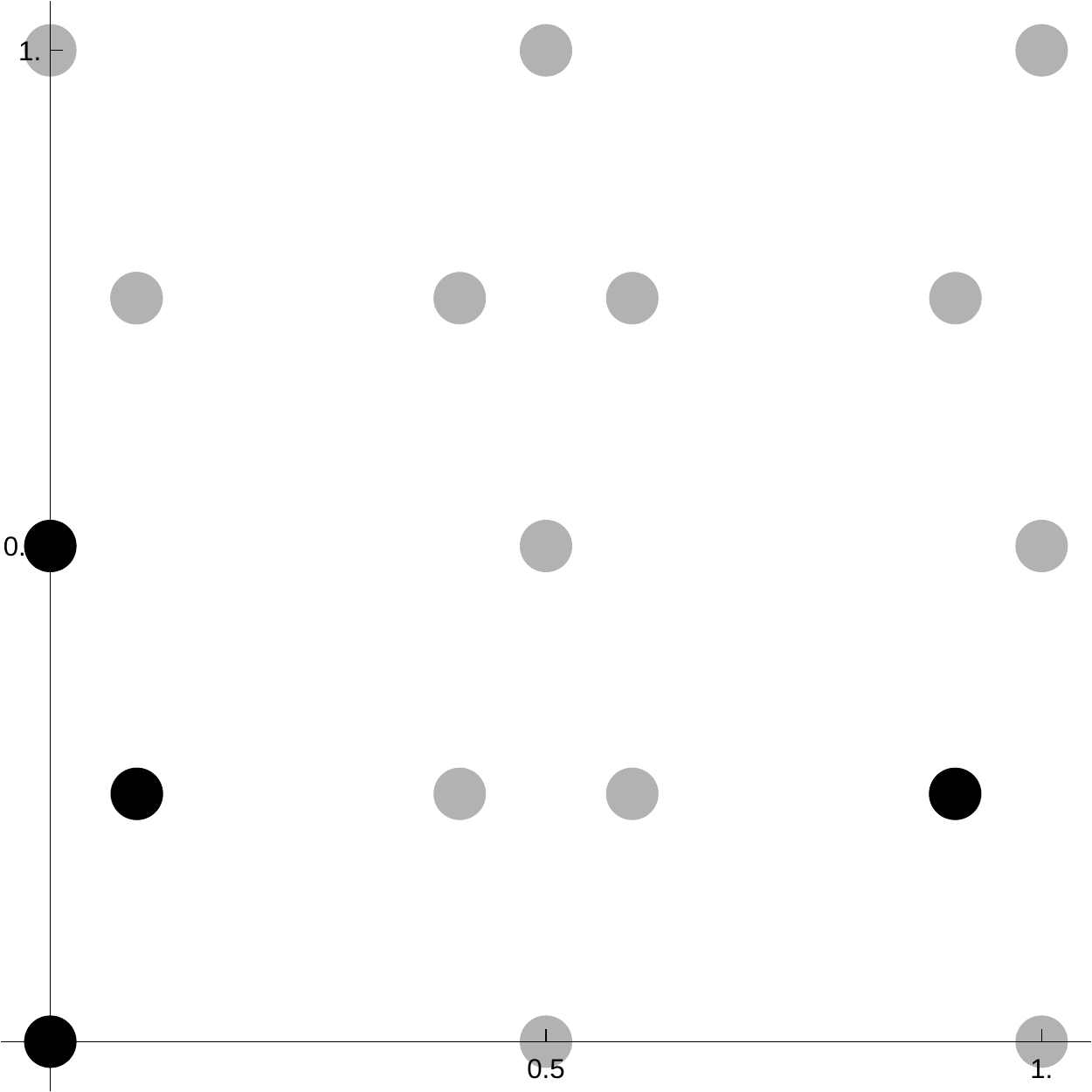}}%
}%
\begin{center}
{Extremum 17}
\end{center}

\\

{%
\setlength{\fboxsep}{8pt}%
\setlength{\fboxrule}{0pt}%
\fbox{\includegraphics[width=3.5cm]{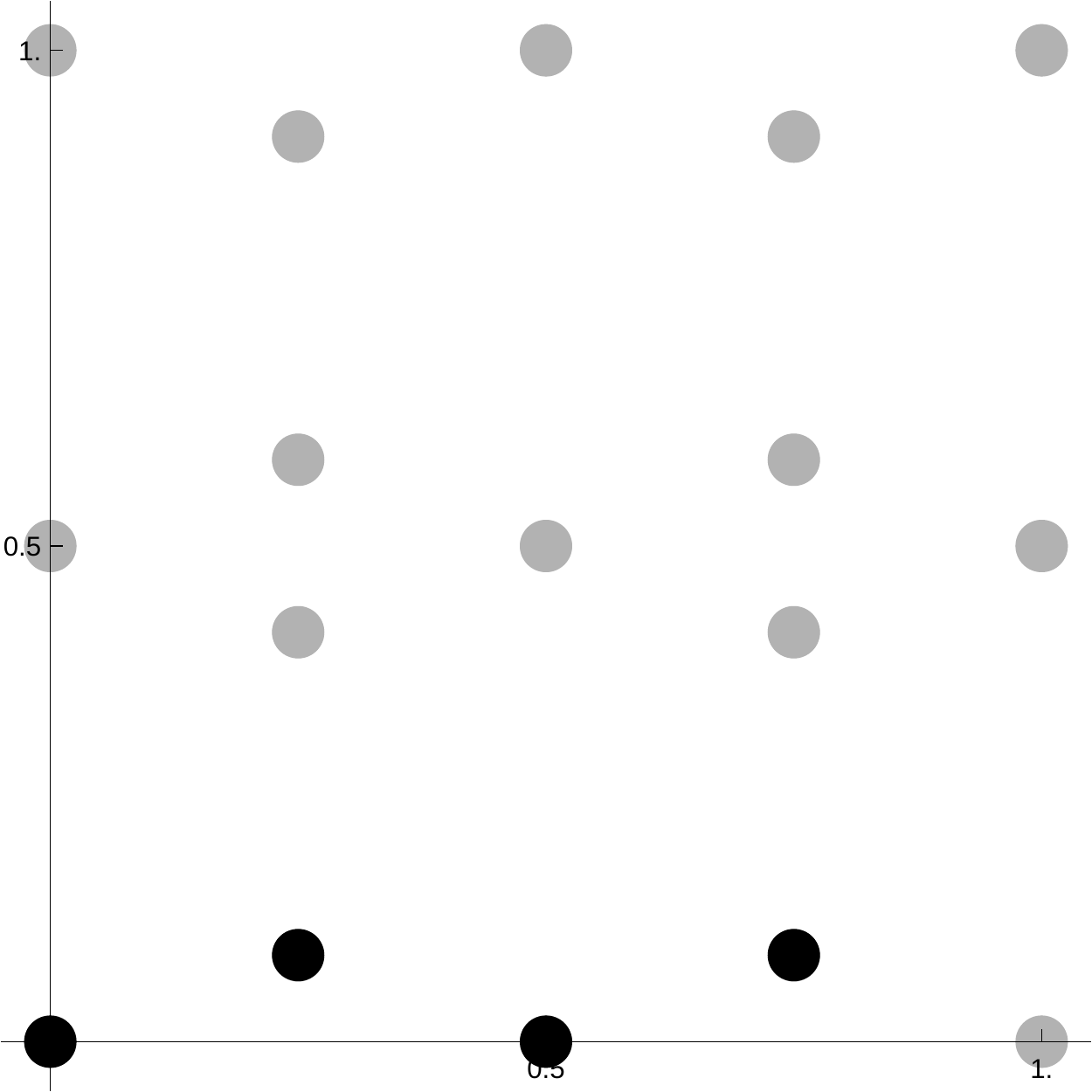}}%
}%
\begin{center}
{Extremum 18}
\end{center}

&

{%
\setlength{\fboxsep}{8pt}%
\setlength{\fboxrule}{0pt}%
\fbox{\includegraphics[width=3.5cm]{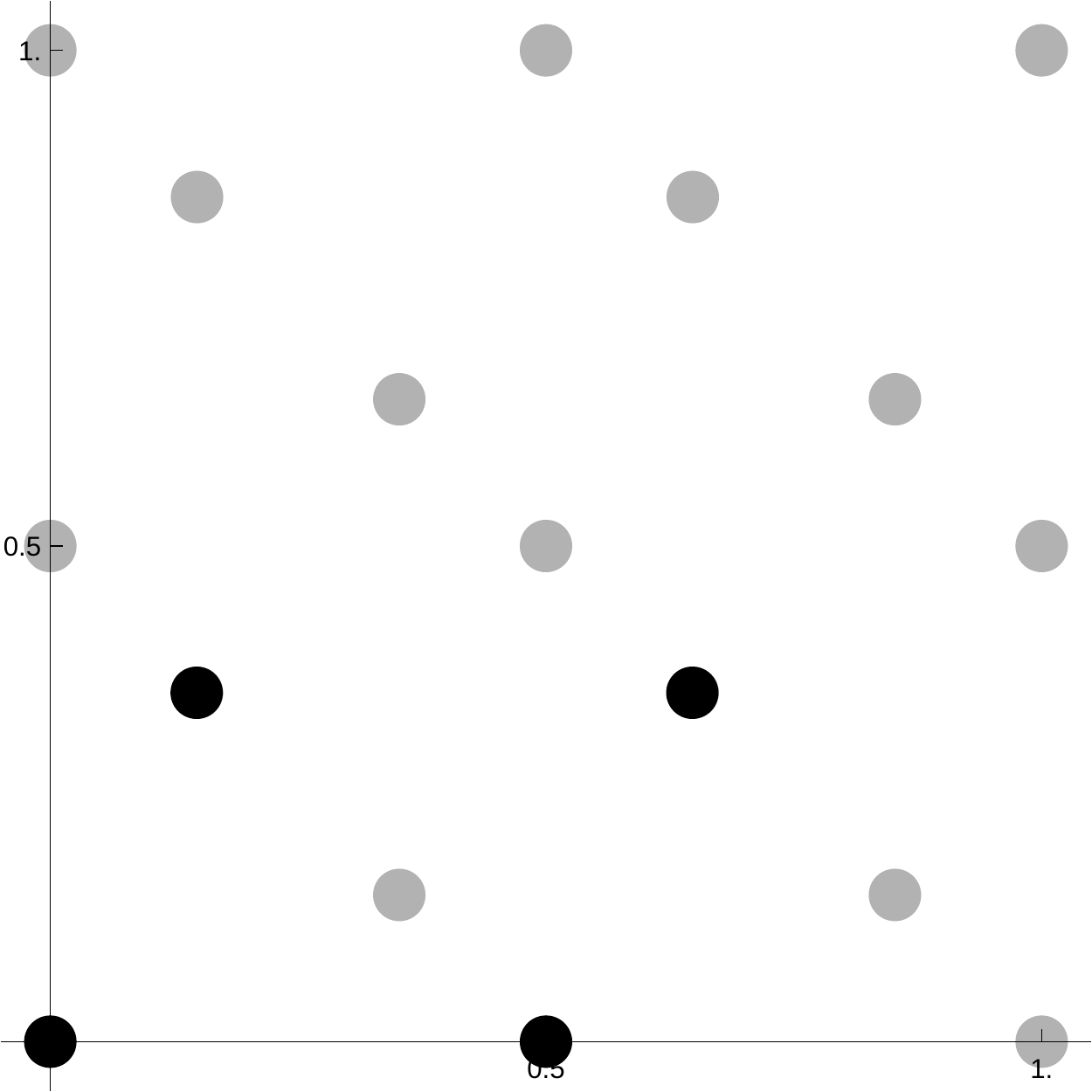}}%
}%
\begin{center}
{Extremum 19}
\end{center}

&

{%
\setlength{\fboxsep}{8pt}%
\setlength{\fboxrule}{0pt}%
\fbox{\includegraphics[width=3.5cm]{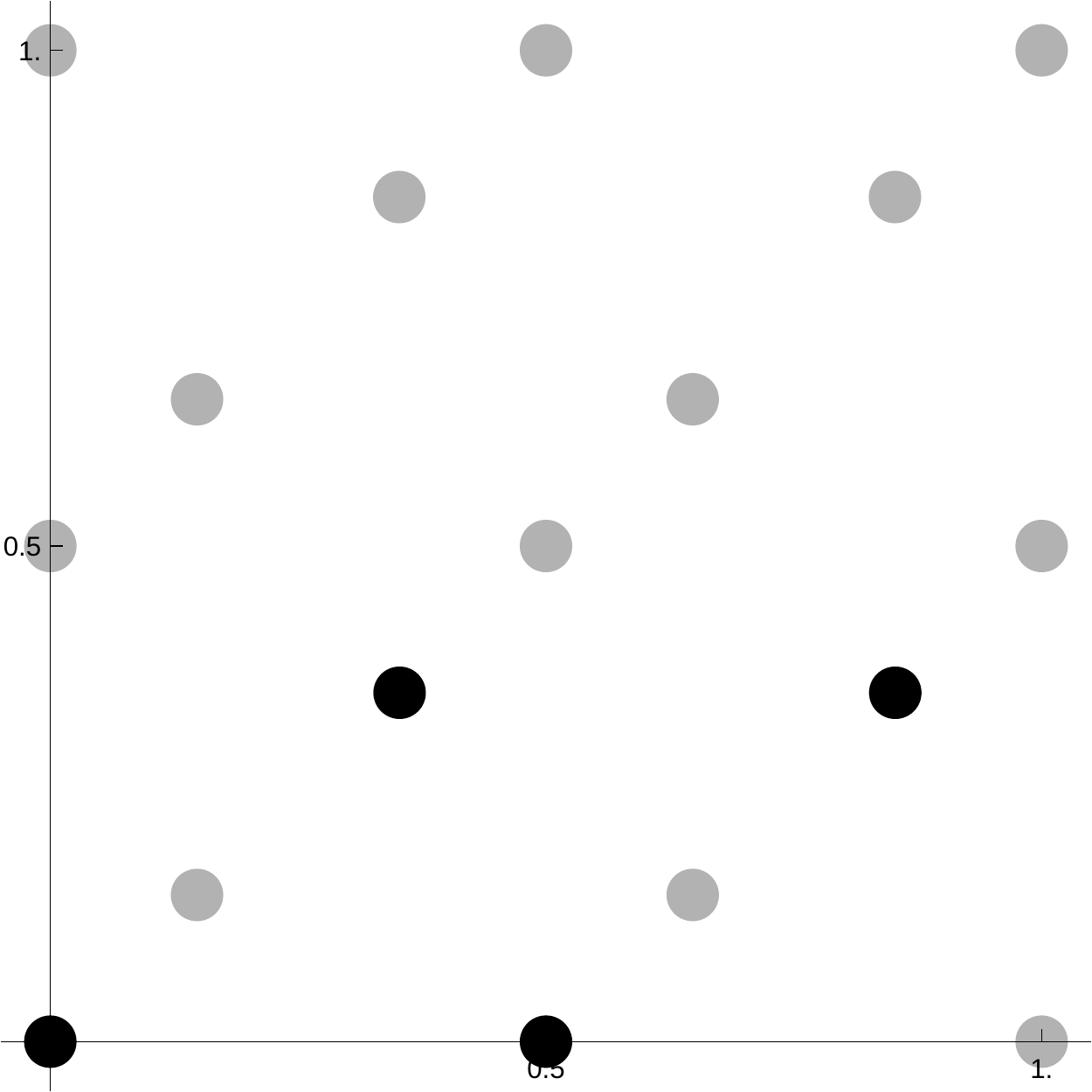}}%
}%
\begin{center}
{Extremum 20}
\end{center}

\end{tabular}
\end{minipage}

\subsubsection{The series for the $so(8)$ extremal potentials}
\begin{figure}
\centering
\includegraphics{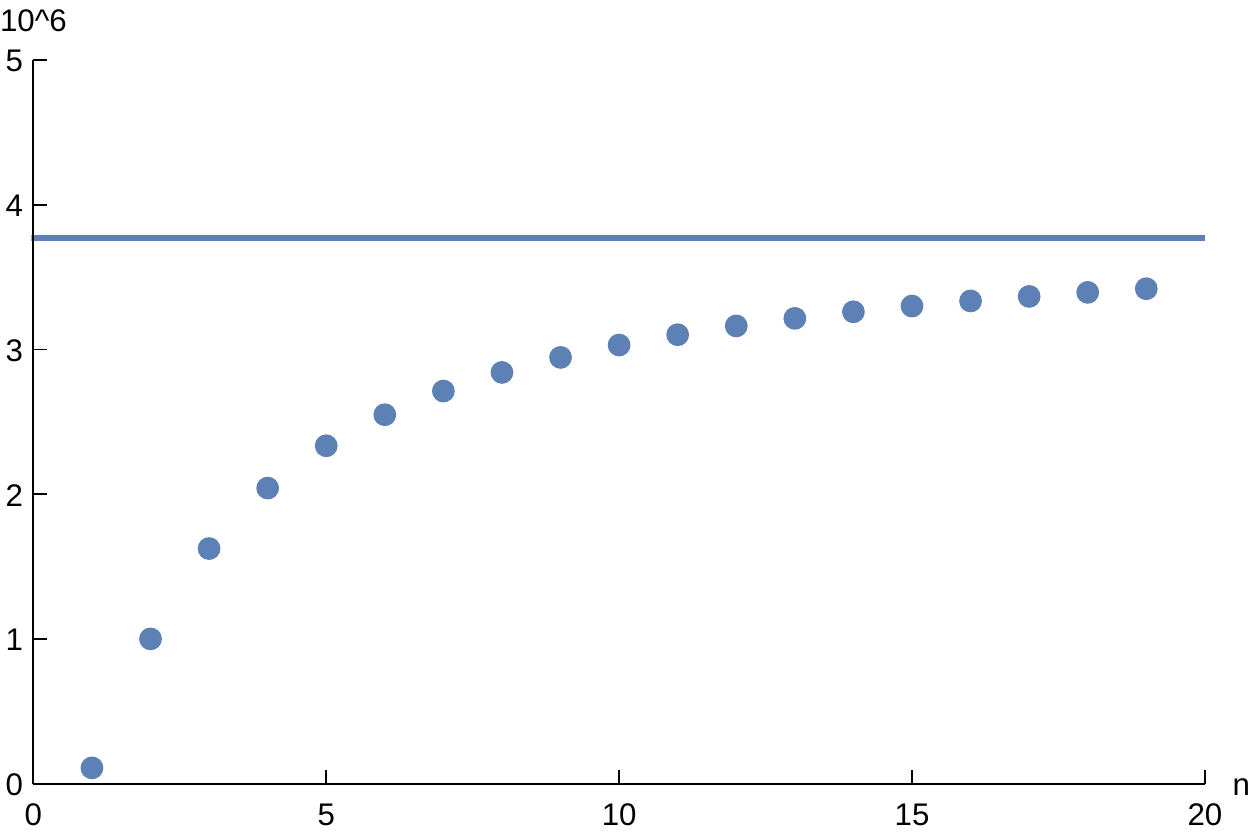}
\caption{The dots show the successive ratios of the coefficients of $f_{11}$, and a line has been drawn, for comparison, at the value $1/q_M = e^{- 2 \pi i \tau_M}$. }
\label{ratiosandmonodromy}
\end{figure}

We have been able to determine the $q$-expansions of the potentials in each extremum with great accuracy, in terms of functions with integer coefficients. For extrema 1 to 8, we gave the exact expression  in section \ref{Sectionso8}. To list the series for the remaining extrema, we  introduce 11 functions, for which we only reproduce the  first few coefficients -- more can be obtained -- : 
\begin{equation*}
\begin{array}{rcl}
f_1(q) &=& \frac{1}{1800}-467 q+45379 q^2-23993958092 q^3-44044347374301 q^4-711960536580667762 q^5 + ...  \\
f_2(q) &=& 1-15172 q+51582918 q^2-397077052296 q^3+5101142359347277 q^4+94300056917523369780 q^5 + ...  \\
f_3(q) &=& \frac{1}{600}+q+369 q^2+68644 q^3+11490041 q^4+1579638246 q^5 + ...  \\
f_4(q) &=& 1+3096 q+1818378264 q^2+2446348866170976 q^3+4535490919062930456600 q^4+  ...  \\
f_5(q) &=& 1-142284 q-2825331513294 q^2-110241726267588876840 q^3 + ...  \\
f_6(q) &=& 2+780960 q+18367562372664 q^2+762875530342634406144 q^3 + ...  \\
f_7(q) &=& 1-4478868 q-121113750523626 q^2-5314750232983801186536 q^3 \\
f_8(q) &=& \frac{1}{3} (14+79929712 q+2425403175787968 q^2+111756708524847535116096 q^3 + ... )\\
f_9(q) &=& \frac{1}{3} ( -37-489421748 q-16364614670173794 q^2-787663906596039662206584 q^3 +...) \\
f_{10}(q) &=& 1-12264 q-7273512936 q^2-9785395464683424 q^3-18141963676251721826280 q^4 + ... \\
f_{11}(q) &=& 1+110596 q+110757888006 q^2+180011523750912008 q^3+367762906594569664954381 q^4+ ...
\end{array}
\end{equation*}
The potentials then read 
\begin{equation*}
\begin{array}{rcl}
V_9 &=& 14400 \pi ^2 f_3 \left( \frac{q}{5^3}\right) \\
V_{10+k} &=& -4 \pi ^2 \sum\limits_{j=0}^5 (16q)^{j/6} \exp \left( 2 \pi i \frac{kj}{6}\right) f_{4+j} \left( \frac{q}{3^3}\right) \\
V_{16} &=& -3 \pi ^2 (f_{10}(q) - 72 \sqrt{q} f_{11}(q)) \\
V_{17} &=& -3 \pi ^2 (f_{10}(q) + 72 \sqrt{q} f_{11}(q)) \\
V_{19} &=& -24 \pi ^2 (75f_{1} (q/15^3) + i\sqrt{5q/3} f_{2}(q/15^3)) \\
V_{20} &=& -24 \pi ^2 (75f_{1} (q/15^3) - i\sqrt{5q/3} f_{2}(q/15^3)) \, ,
\end{array}
\end{equation*}
where $k=0,...,5$. The last series $V_{18}$ can then be deduced from the fact that the sum of all potentials in the duodecuplet vanishes. 
Note that the coefficients grow rapidly, preventing the functions above to be modular forms. 
The monodromy is responsible for this phenomenon, as can be confirmed by the estimation of the convergence radius given by the successive ratios of the coefficients (see figure \ref{ratiosandmonodromy}). 

\subsection{The List of Extrema for $so(7)$ and $sp(6)$}
\label{listofextremaso7}
Finally, in the case of the algebras $B_3=so(7)$ and $C_3=sp(6)$, we only present diagrams of the extremal positions for the $so(7)$ root system, since the corresponding extrema for $sp(6)$ can be found
by Langlands duality. We use the same conventions as for the $so(5)$ figures. Additional data, like the data
we presented for $so(8)$ in the previous section, can be found.

\begin{minipage}{\linewidth}
\begin{center}
Extrema at $\tau=i$ for $so(7)$
\end{center}
\vspace{1em}
\begin{tabular}{p{5cm}p{5cm}p{5cm}}

{%
\setlength{\fboxsep}{8pt}%
\setlength{\fboxrule}{0pt}%
\fbox{\includegraphics[width=3.5cm]{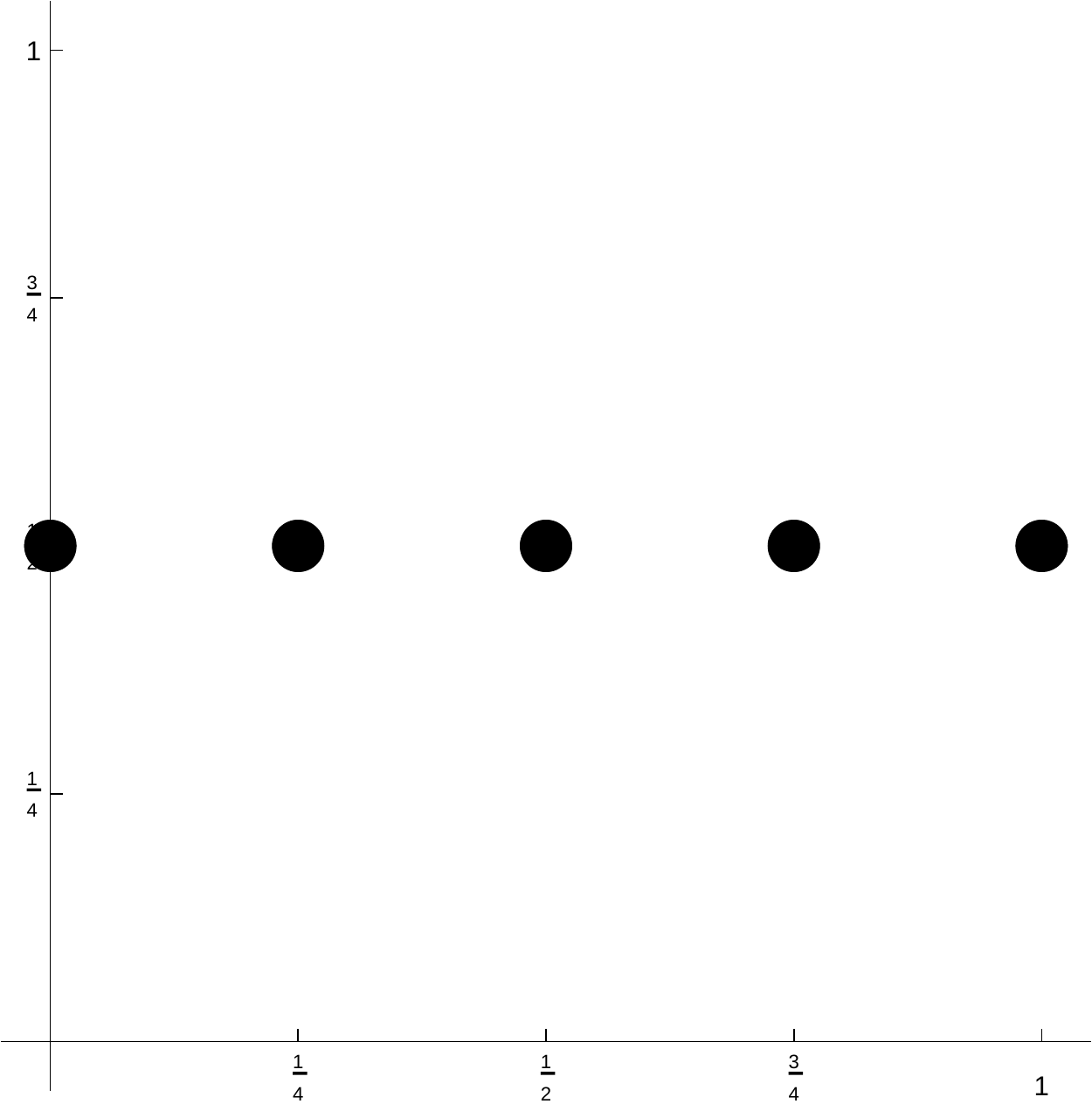}}%
}%
\begin{center}
{Extremum 1}
\end{center}

&

{%
\setlength{\fboxsep}{8pt}%
\setlength{\fboxrule}{0pt}%
\fbox{\includegraphics[width=3.5cm]{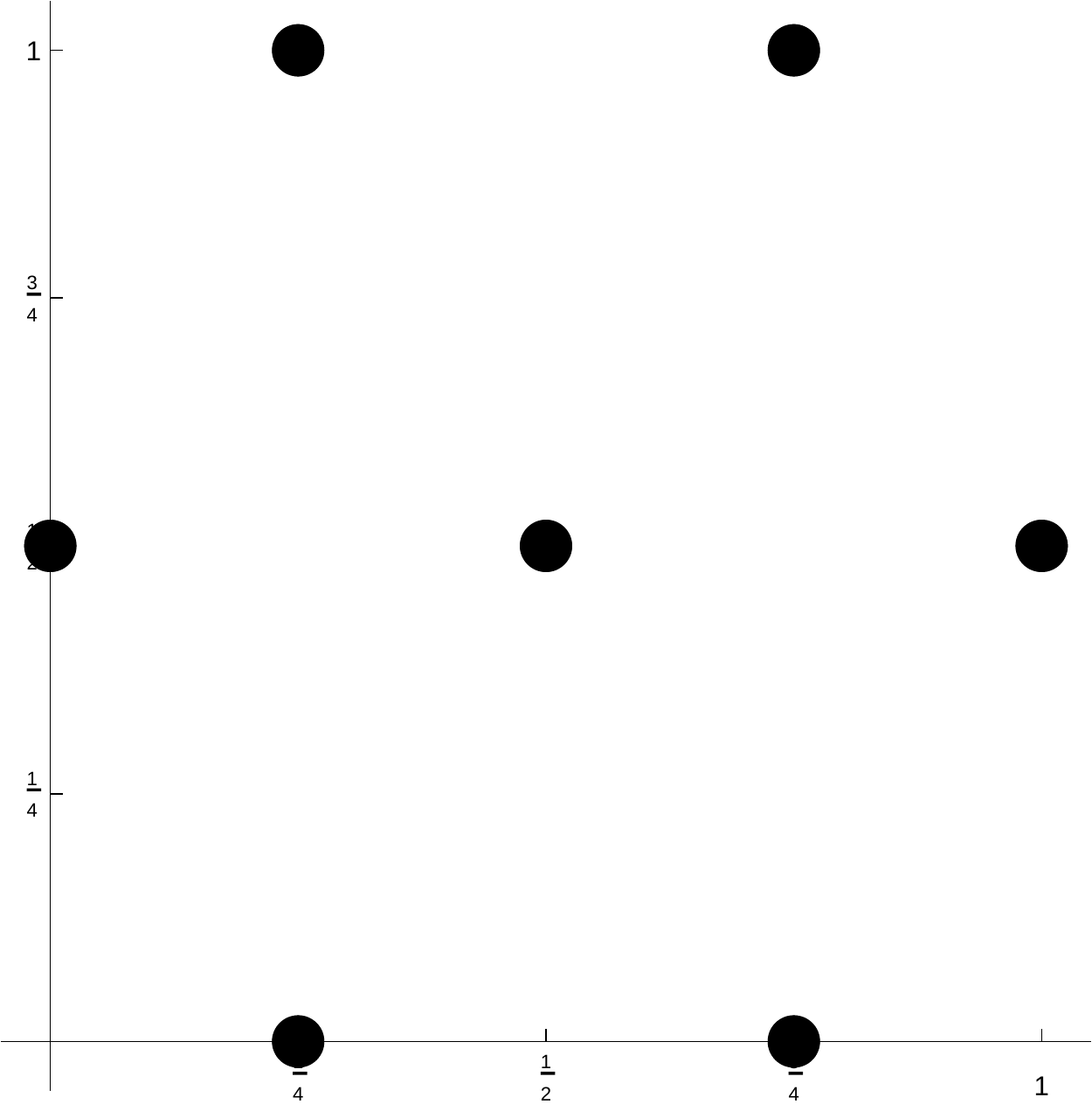}}%
}%
\begin{center}
{Extremum 2}
\end{center}

&

{%
\setlength{\fboxsep}{8pt}%
\setlength{\fboxrule}{0pt}%
\fbox{\includegraphics[width=3.5cm]{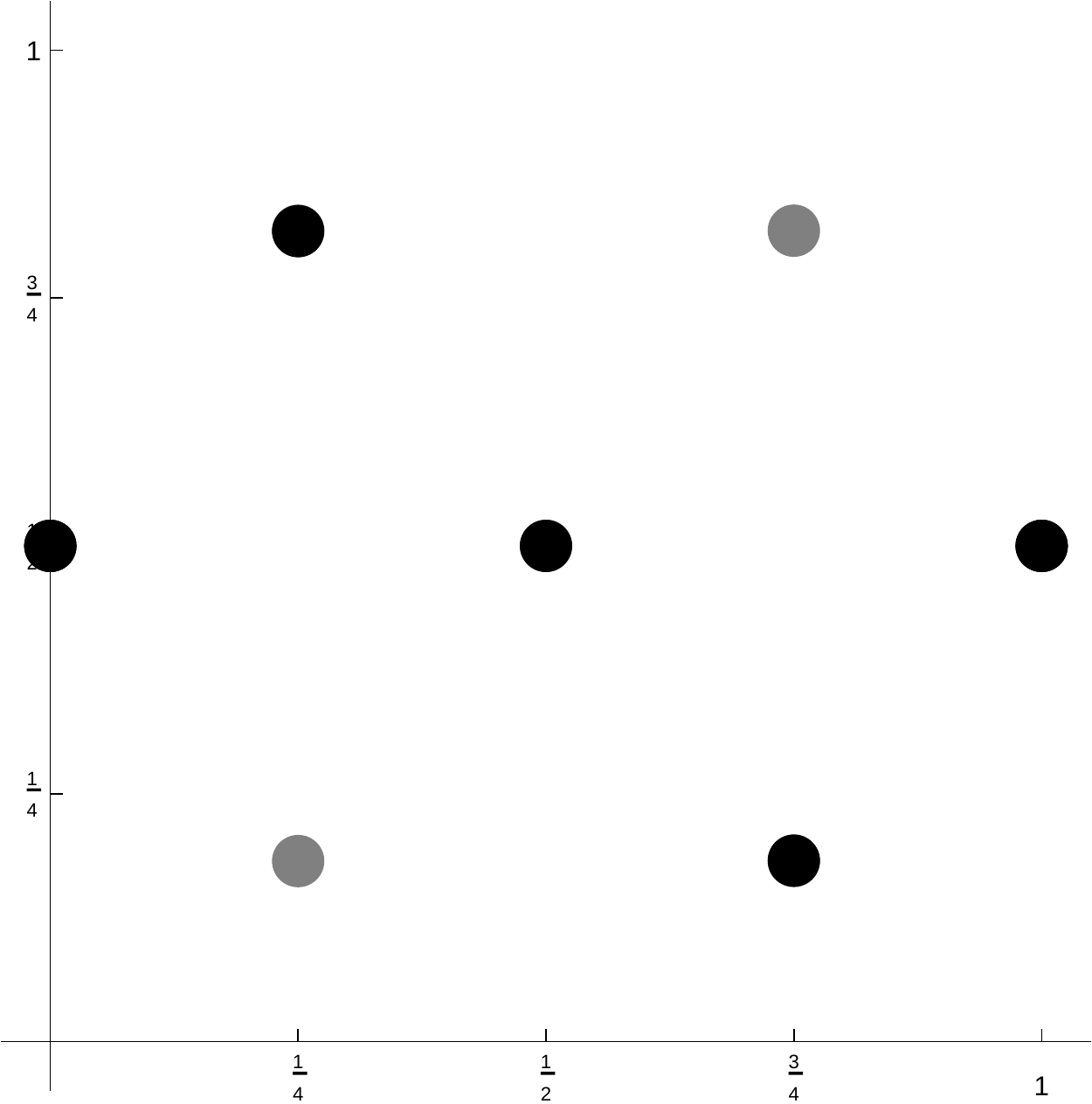}}%
}%
\begin{center}
{Extremum 3}
\end{center}
 
\\

{%
\setlength{\fboxsep}{8pt}%
\setlength{\fboxrule}{0pt}%
\fbox{\includegraphics[width=3.5cm]{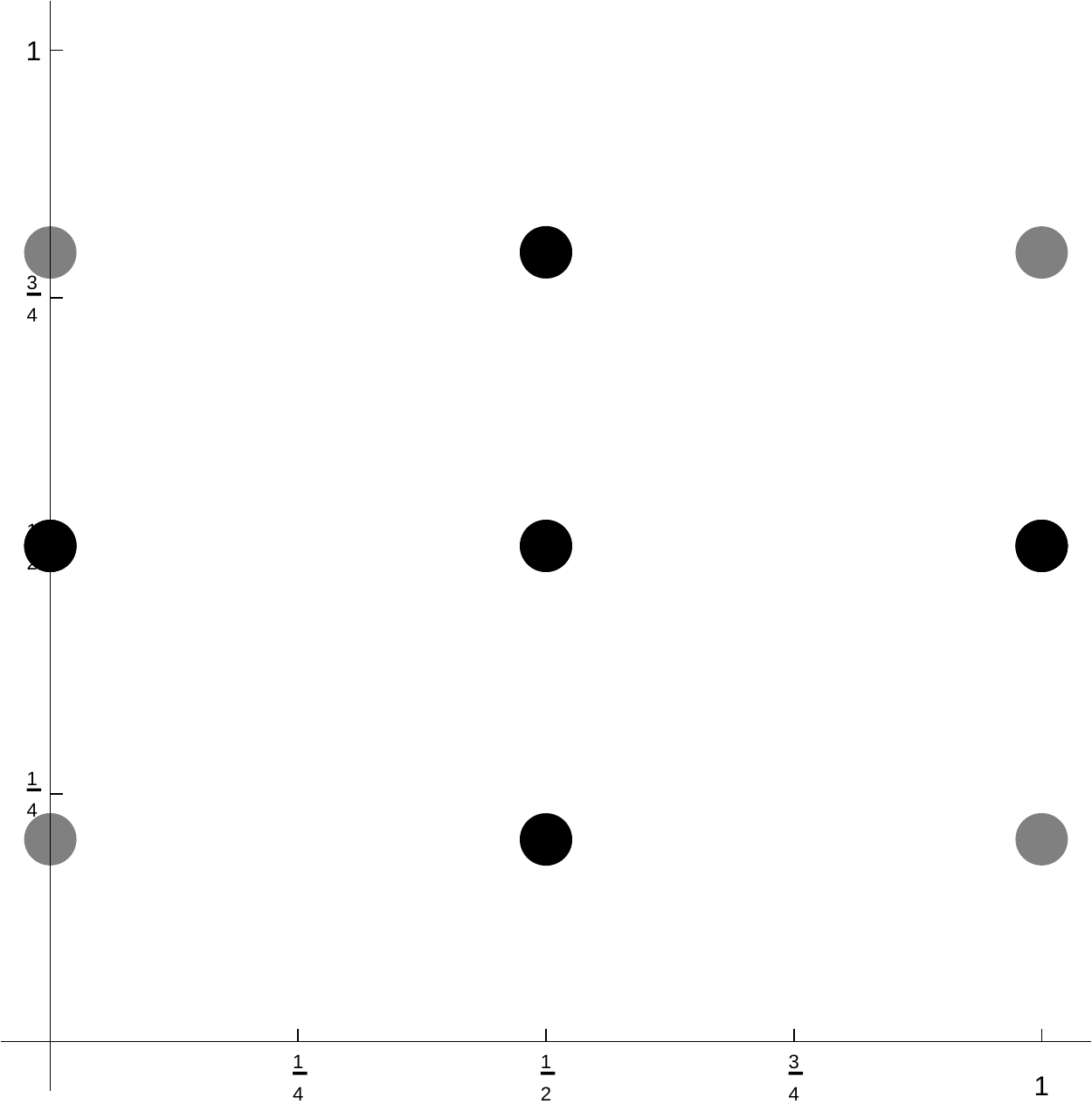}}%
}%
\begin{center}
{Extremum 4}
\end{center}

&

{%
\setlength{\fboxsep}{8pt}%
\setlength{\fboxrule}{0pt}%
\fbox{\includegraphics[width=3.5cm]{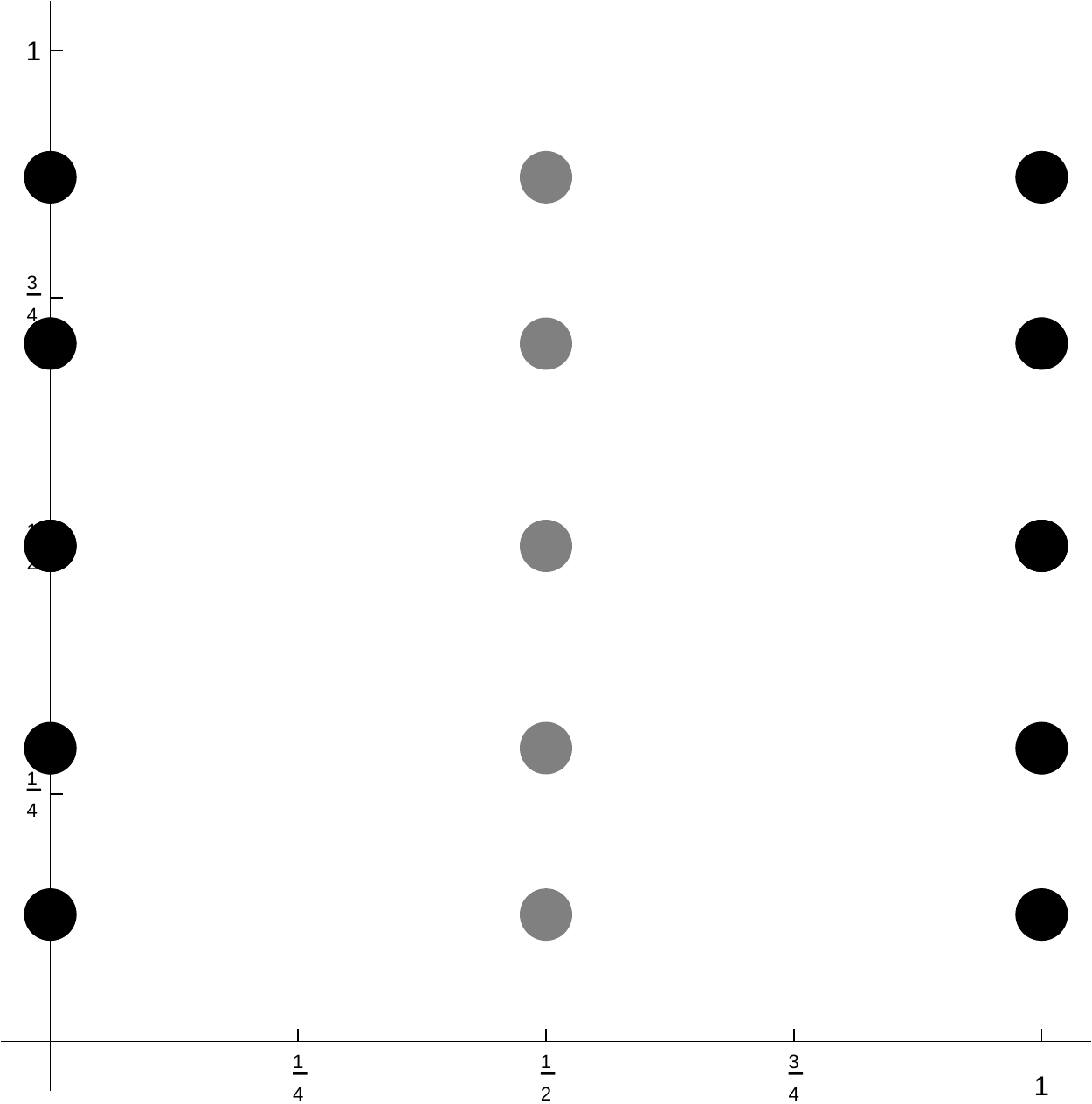}}%
}%
\begin{center}
{Extremum 5}
\end{center}

&

{%
\setlength{\fboxsep}{8pt}%
\setlength{\fboxrule}{0pt}%
\fbox{\includegraphics[width=3.5cm]{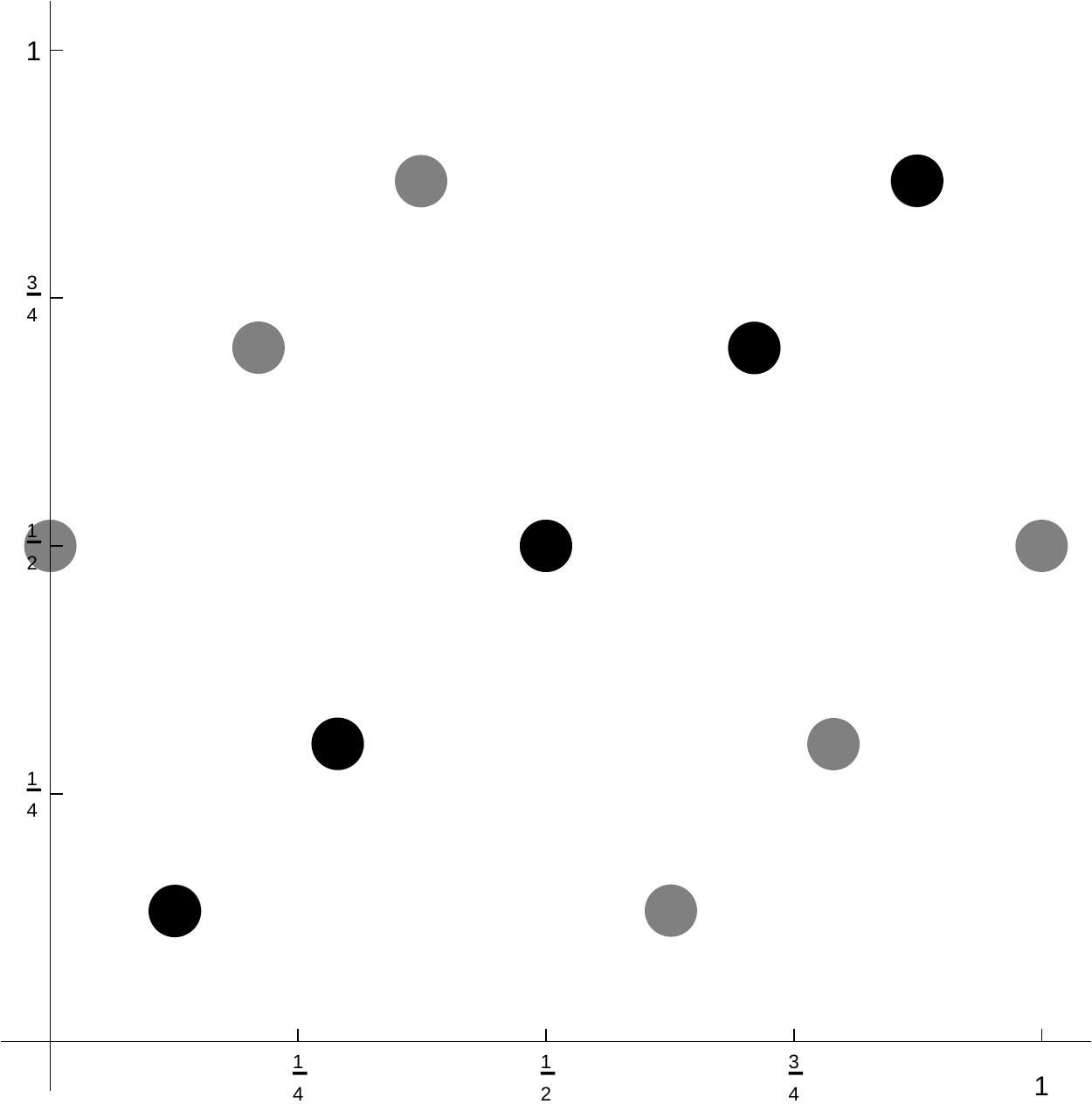}}%
}%
\begin{center}
{Extremum 6}
\end{center}
 
\\

{%
\setlength{\fboxsep}{8pt}%
\setlength{\fboxrule}{0pt}%
\fbox{\includegraphics[width=3.5cm]{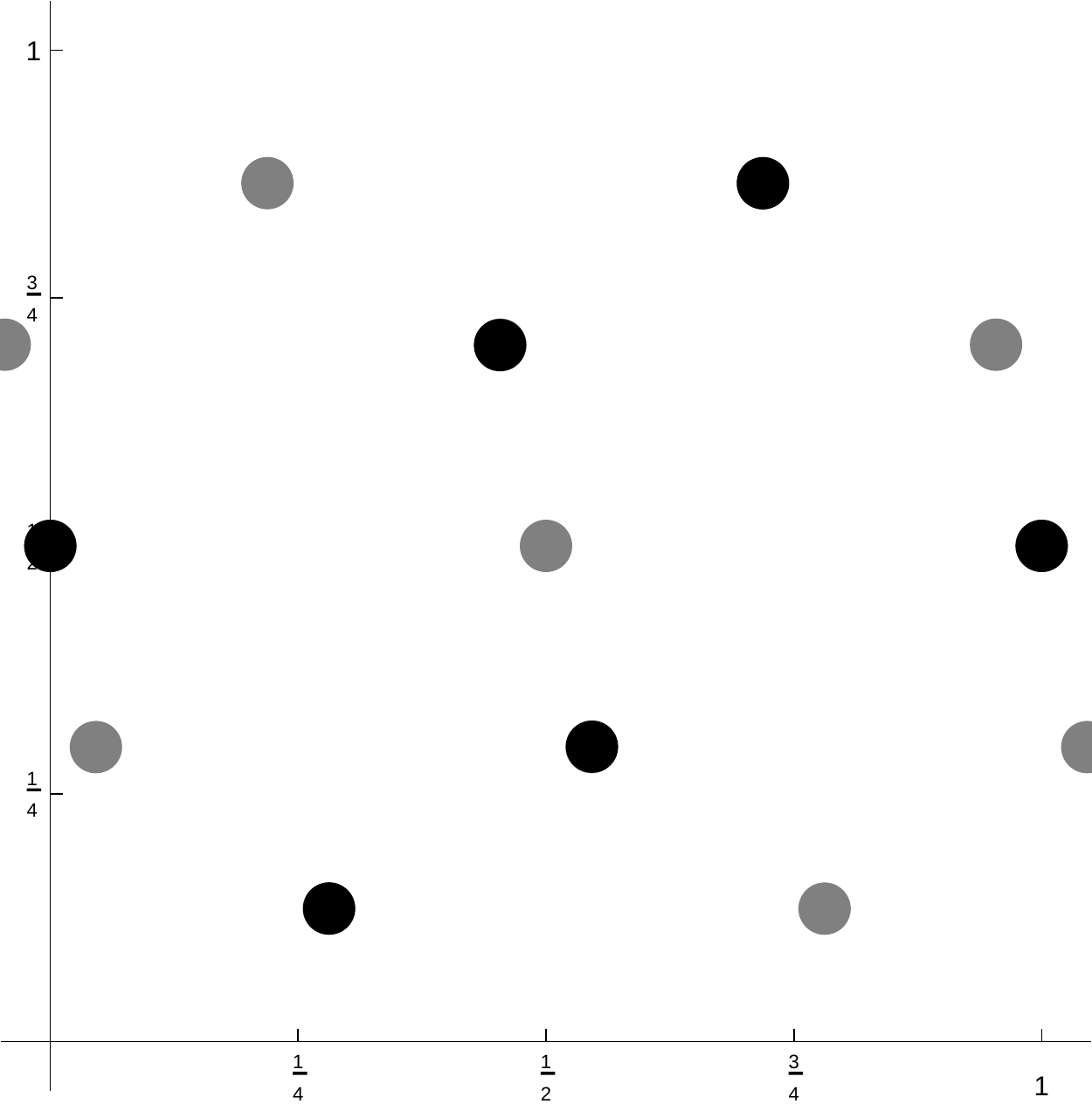}}%
}%
\begin{center}
{Extremum 7}
\end{center}

&

{%
\setlength{\fboxsep}{8pt}%
\setlength{\fboxrule}{0pt}%
\fbox{\includegraphics[width=3.5cm]{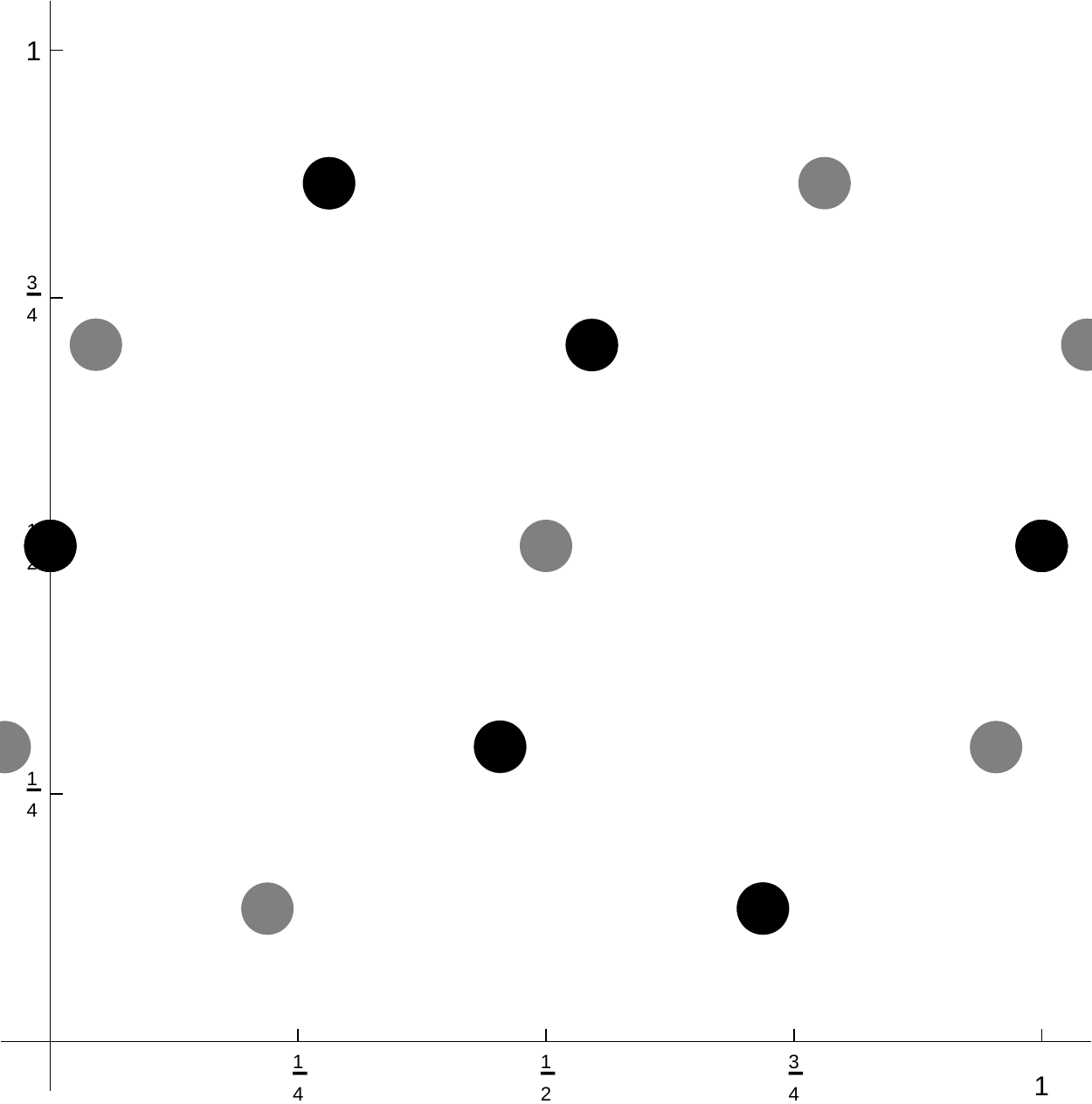}}%
}%
\begin{center}
{Extremum 8}
\end{center}

&

{%
\setlength{\fboxsep}{8pt}%
\setlength{\fboxrule}{0pt}%
\fbox{\includegraphics[width=3.5cm]{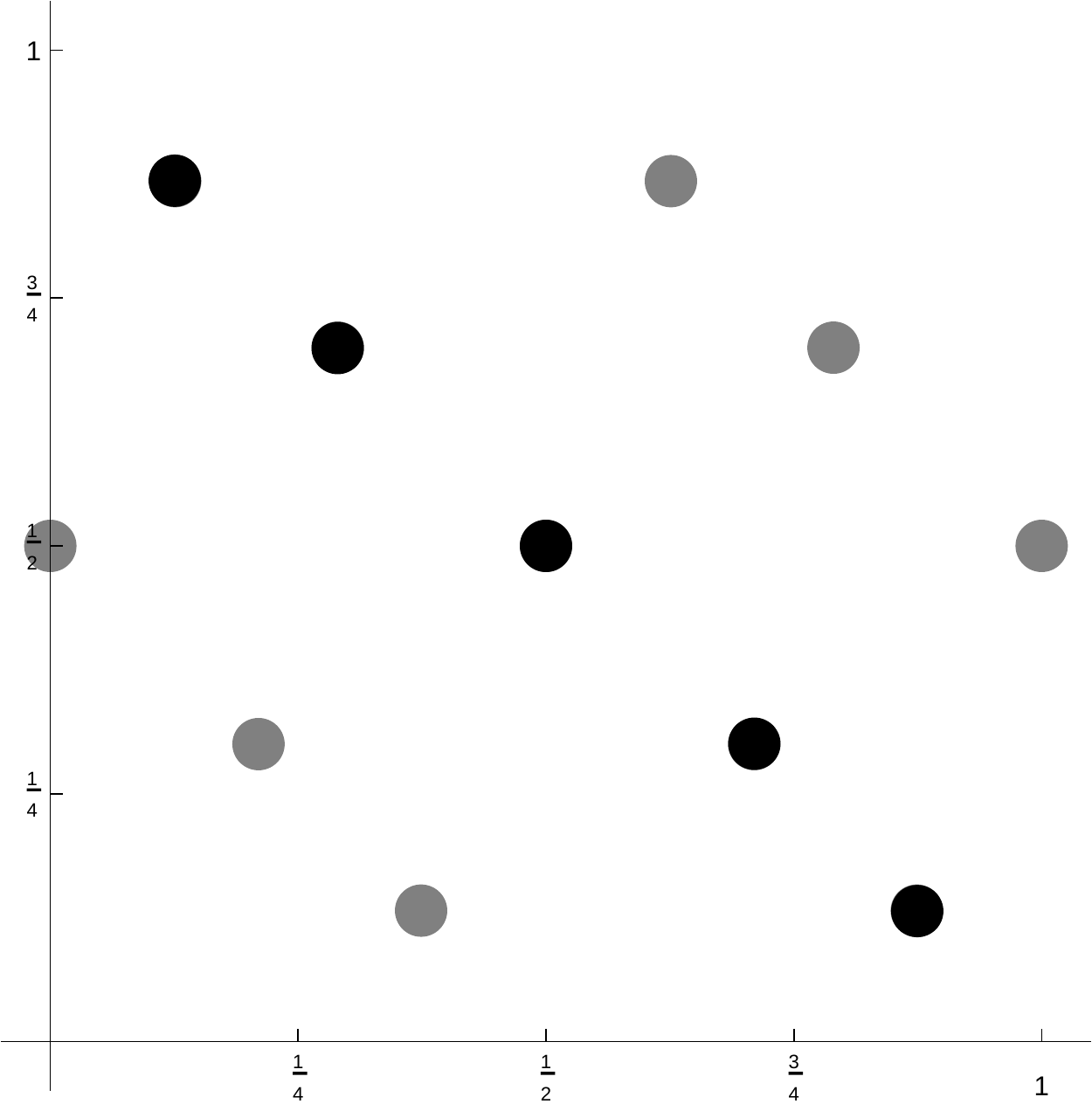}}%
}%
\begin{center}
{Extremum 9}
\end{center}
 
\\

{%
\setlength{\fboxsep}{8pt}%
\setlength{\fboxrule}{0pt}%
\fbox{\includegraphics[width=3.5cm]{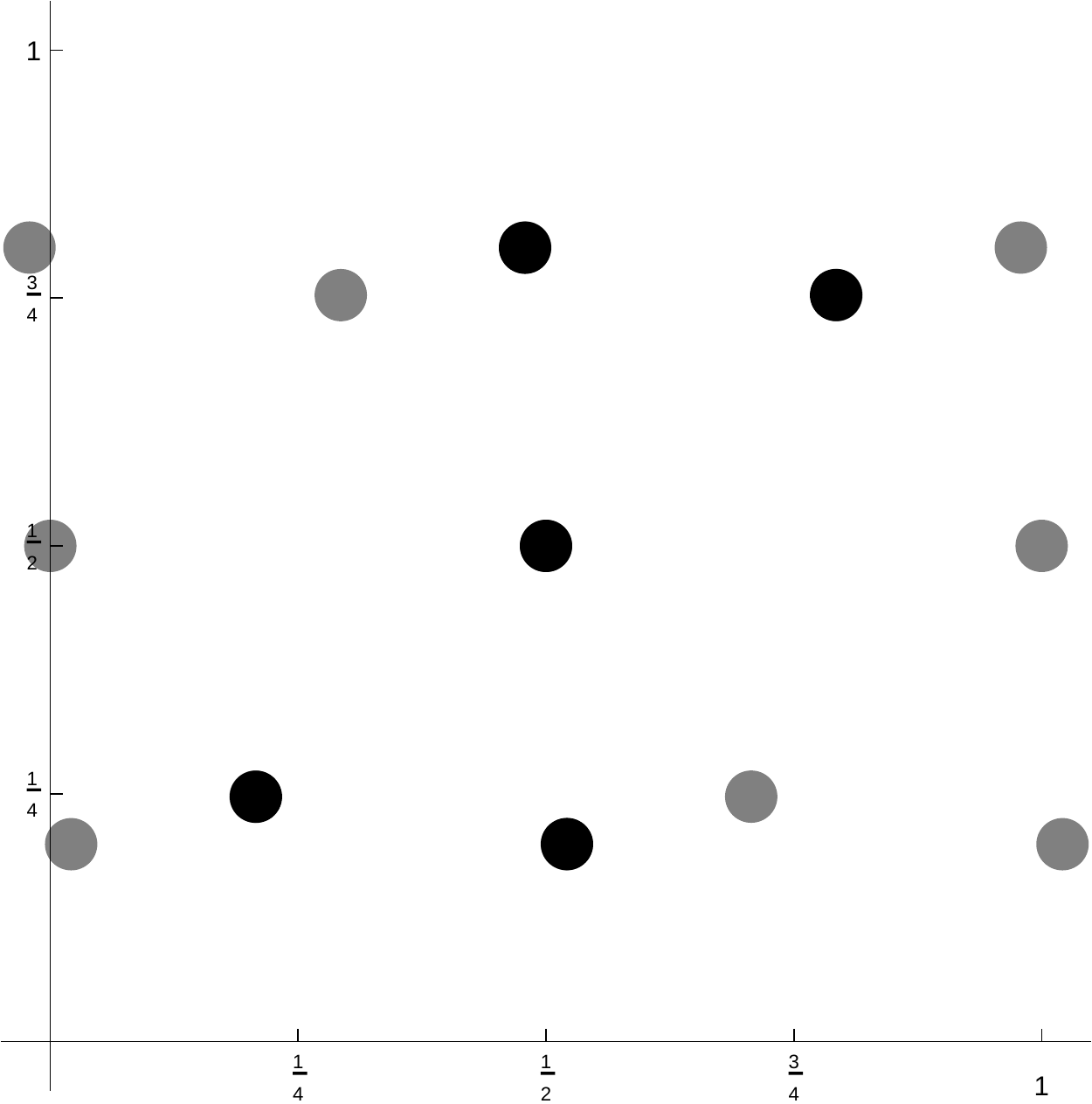}}%
}%
\begin{center}
{Extremum 10}
\end{center}

&

{%
\setlength{\fboxsep}{8pt}%
\setlength{\fboxrule}{0pt}%
\fbox{\includegraphics[width=3.5cm]{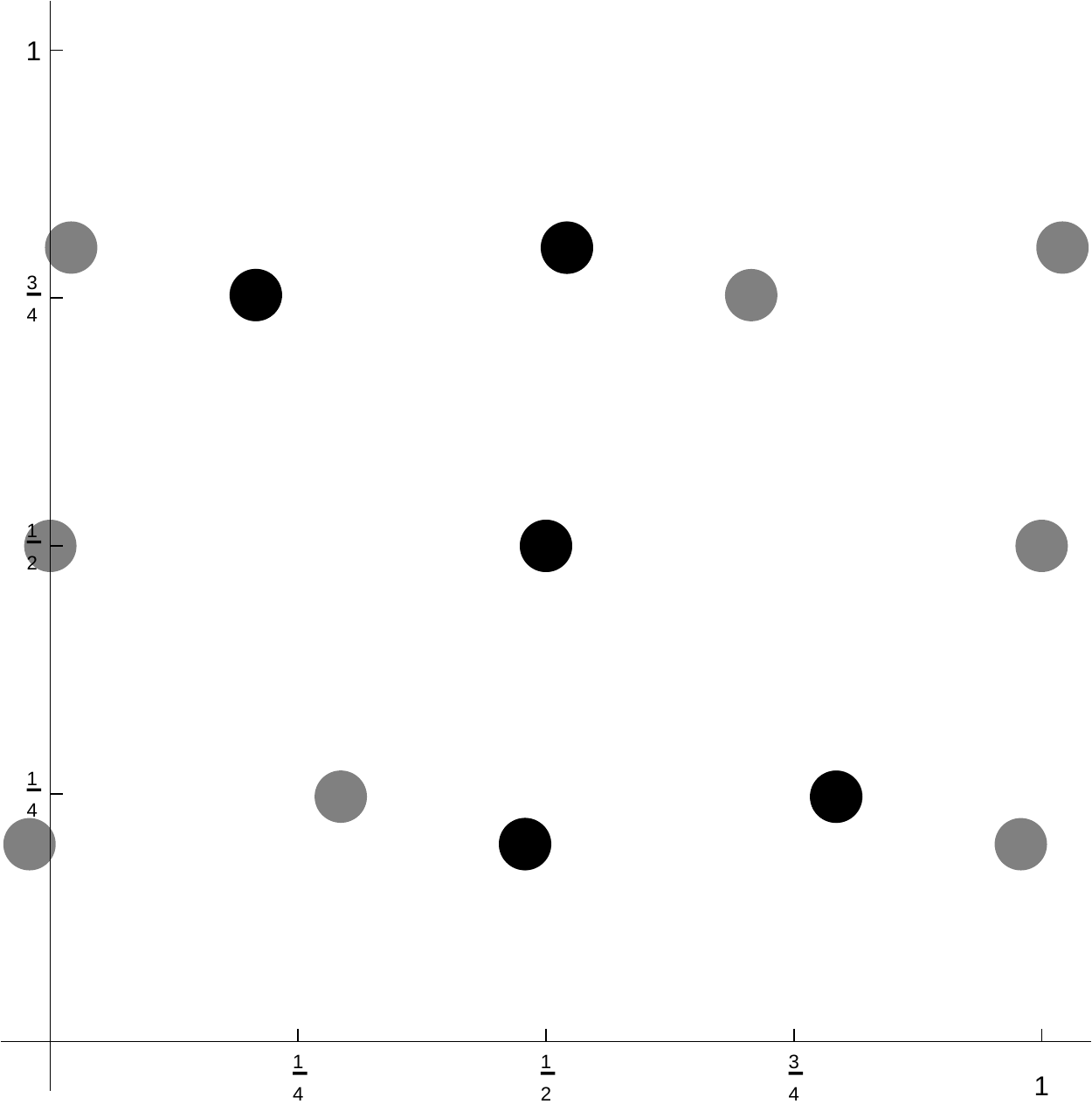}}%
}%
\begin{center}
{Extremum 11}
\end{center}

&

{%
\setlength{\fboxsep}{8pt}%
\setlength{\fboxrule}{0pt}%
\fbox{\includegraphics[width=3.5cm]{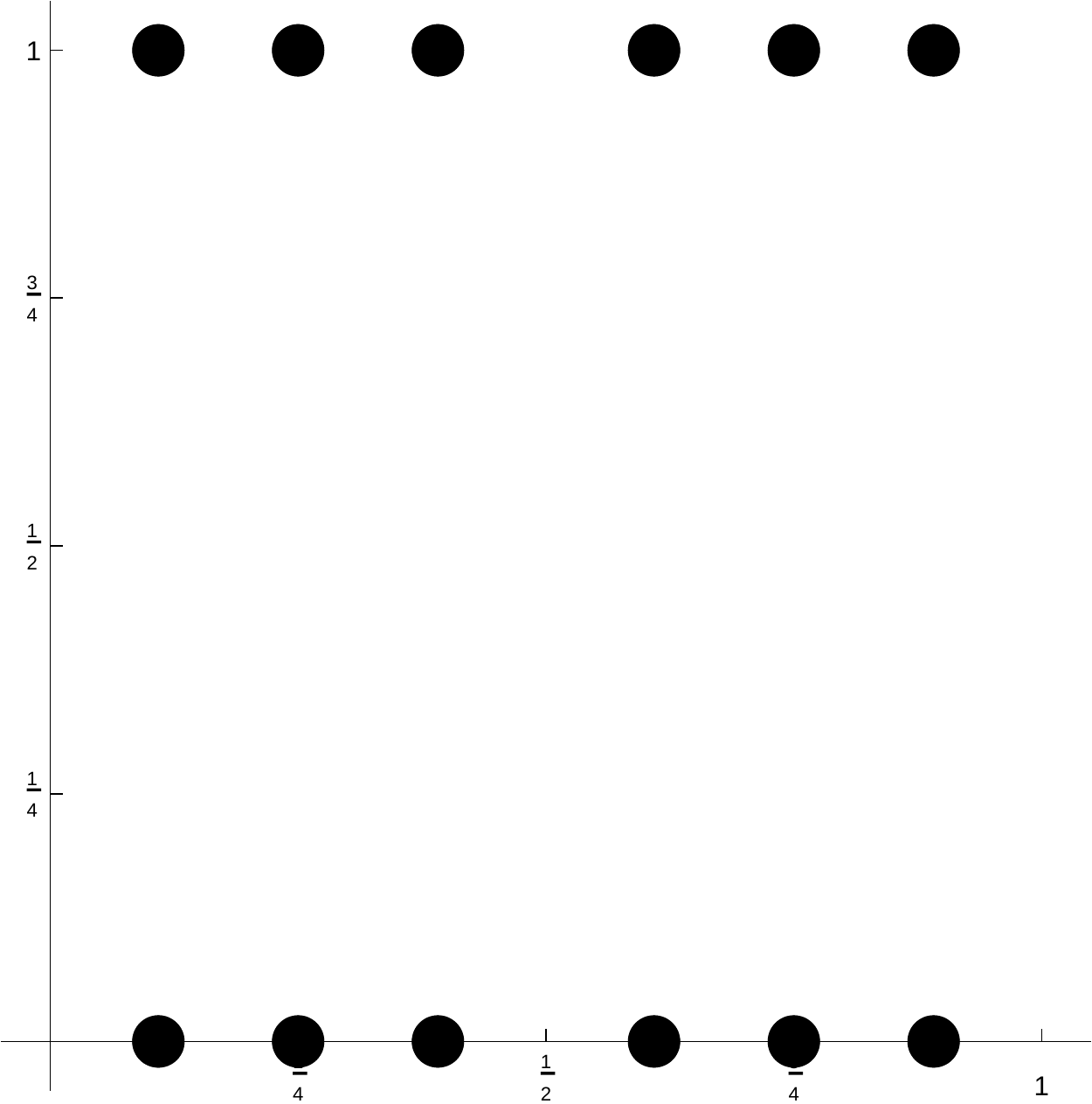}}%
}%
\begin{center}
{Extremum 12}
\end{center}

\end{tabular}
\end{minipage}

\begin{minipage}{\linewidth}
\begin{center}
Extrema at $\tau=i$ for $so(7)$
\end{center}
\vspace{1em}
\begin{tabular}{p{5cm}p{5cm}p{5cm}}

{%
\setlength{\fboxsep}{8pt}%
\setlength{\fboxrule}{0pt}%
\fbox{\includegraphics[width=3.5cm]{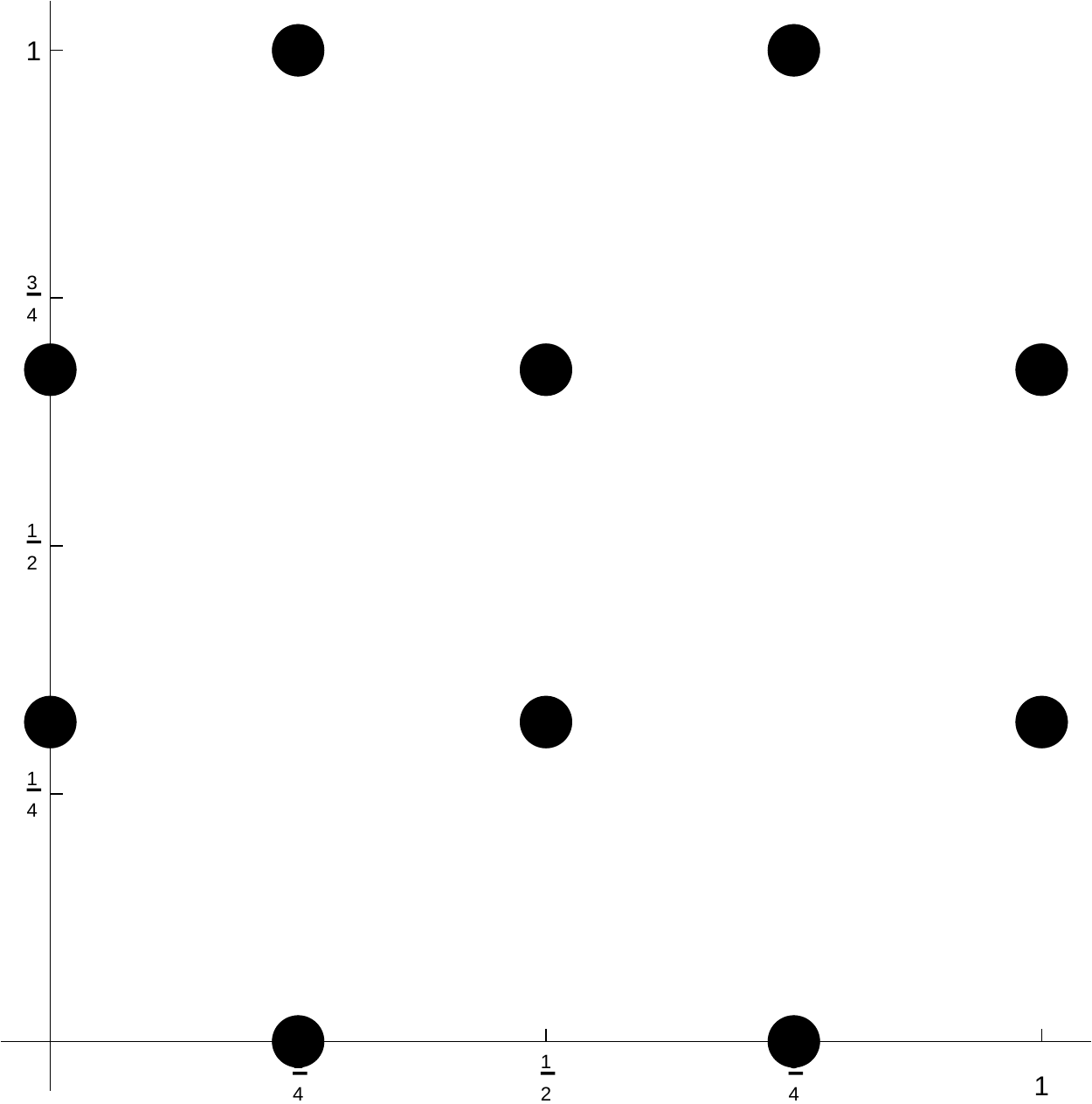}}%
}%
\begin{center}
{Extremum 13}
\end{center}

&

{%
\setlength{\fboxsep}{8pt}%
\setlength{\fboxrule}{0pt}%
\fbox{\includegraphics[width=3.5cm]{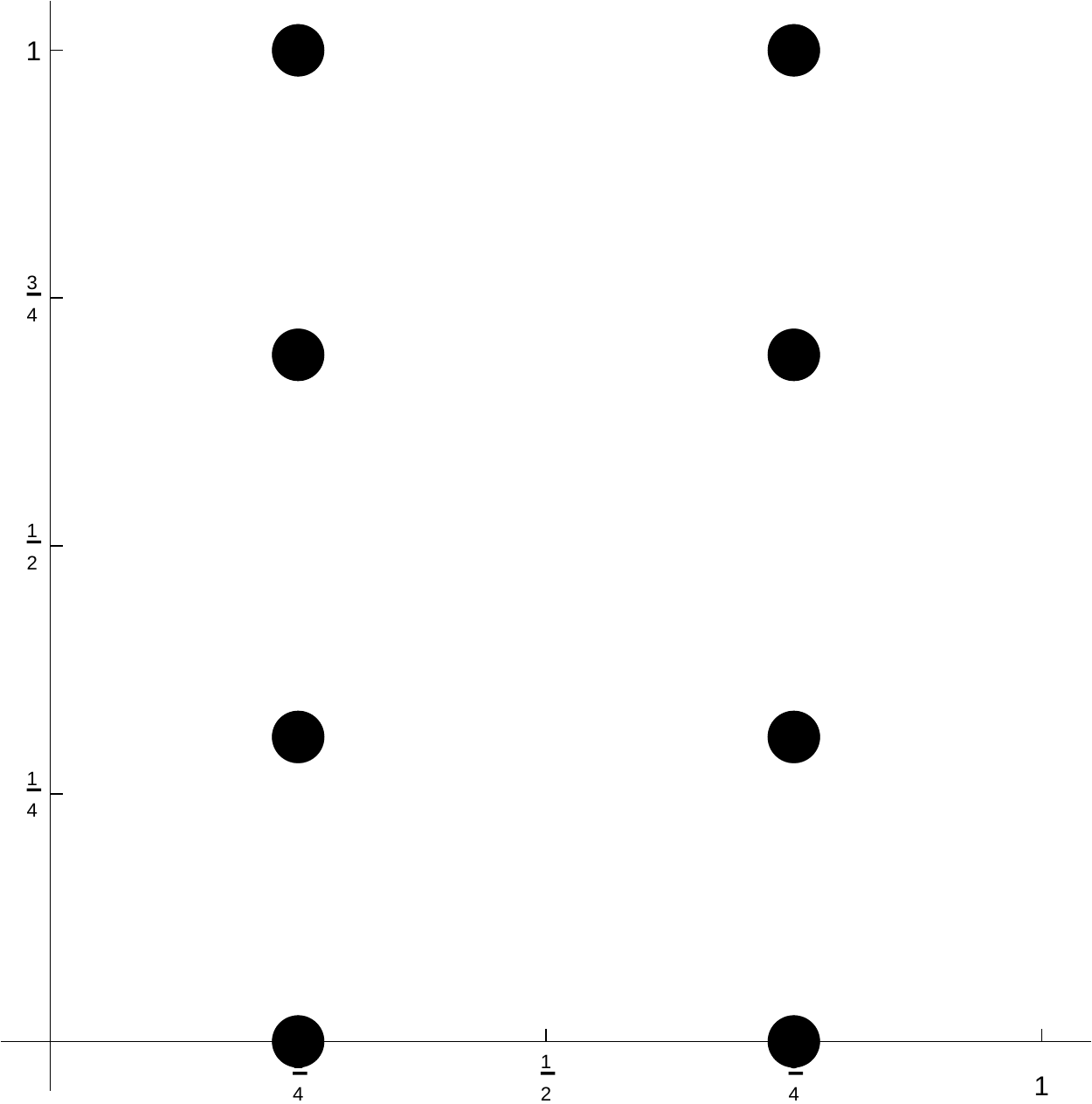}}%
}%
\begin{center}
{Extremum 14}
\end{center}

&

{%
\setlength{\fboxsep}{8pt}%
\setlength{\fboxrule}{0pt}%
\fbox{\includegraphics[width=3.5cm]{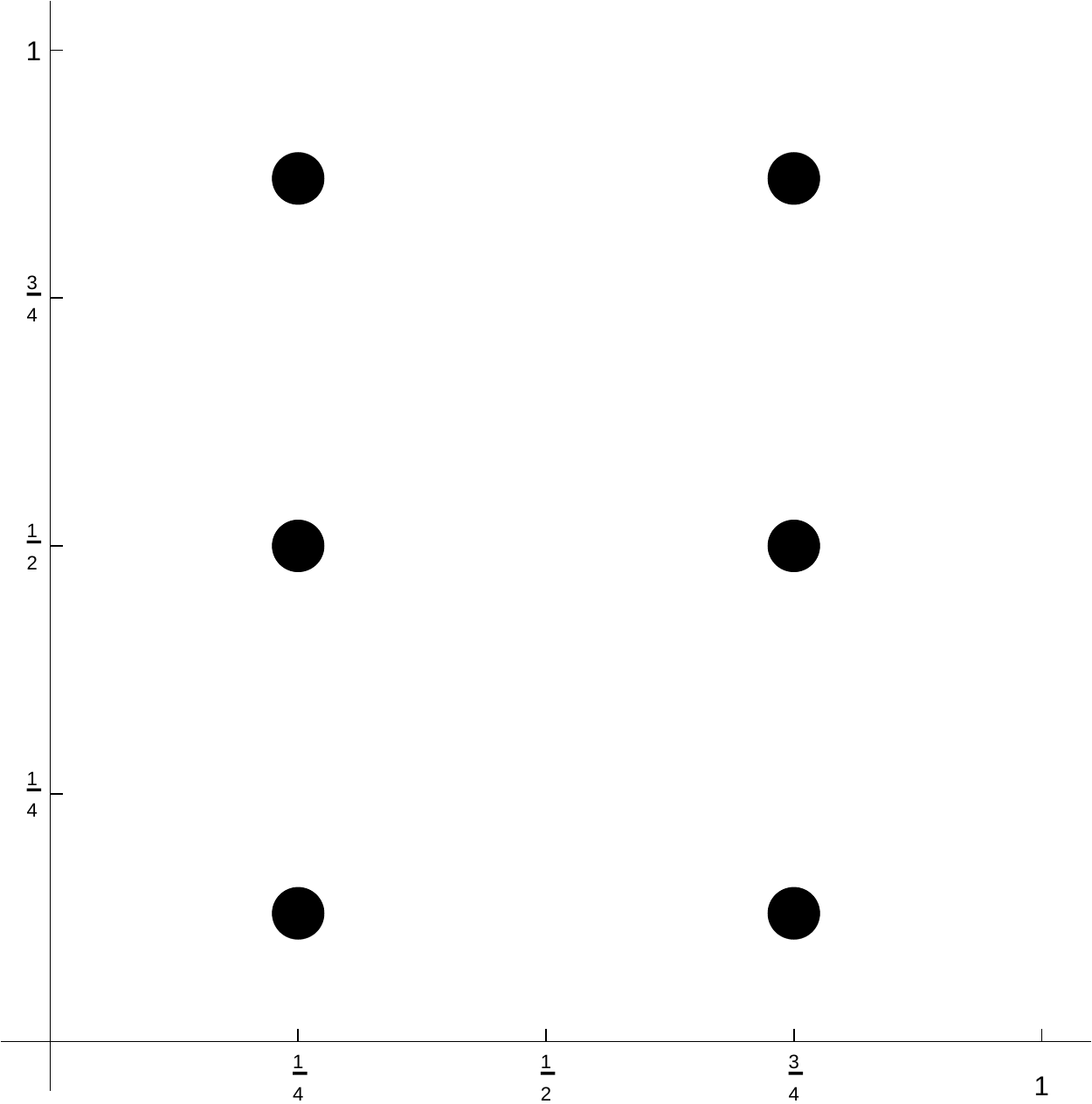}}%
}%
\begin{center}
{Extremum 15}
\end{center}
 
\\

{%
\setlength{\fboxsep}{8pt}%
\setlength{\fboxrule}{0pt}%
\fbox{\includegraphics[width=3.5cm]{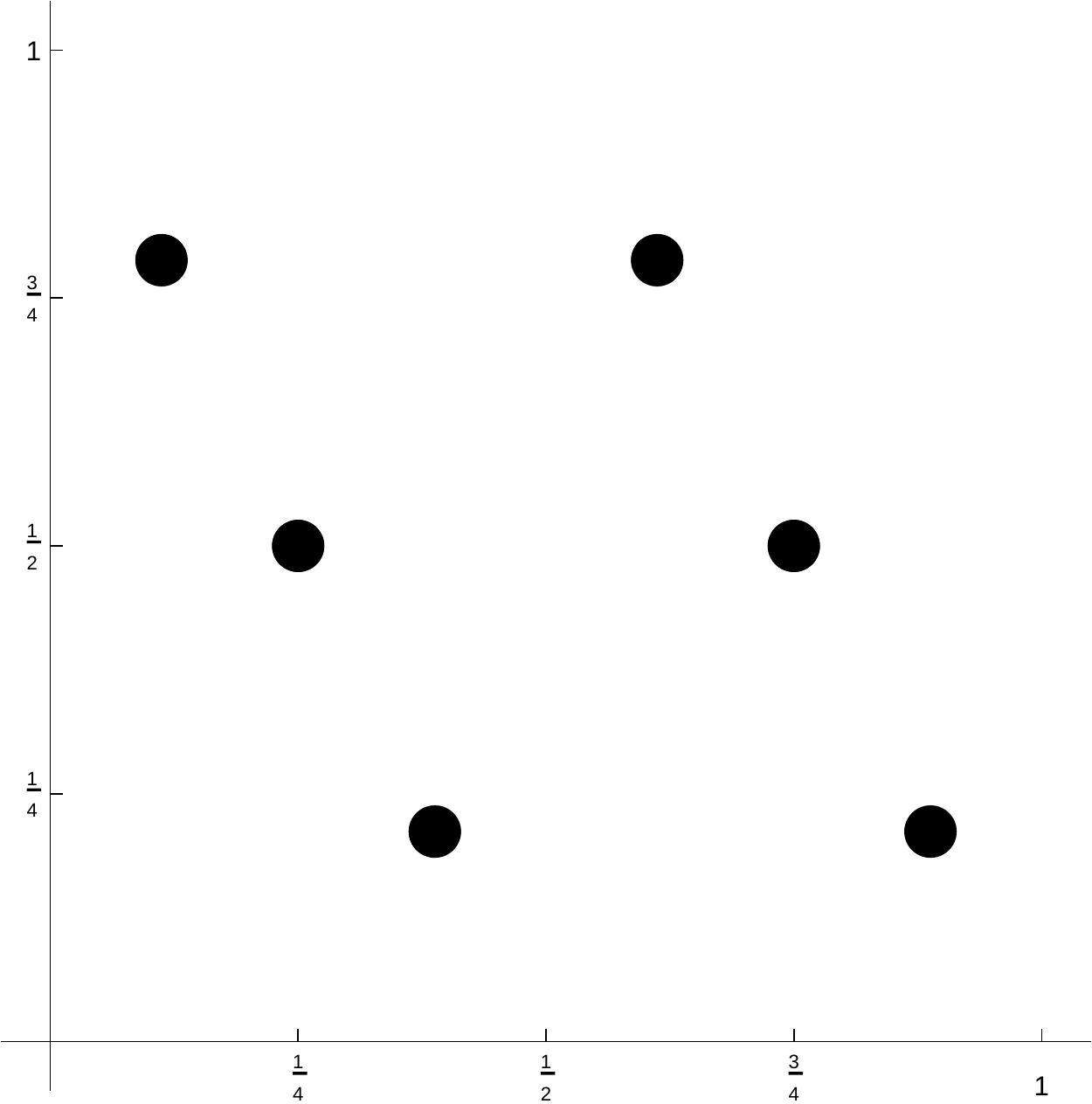}}%
}%
\begin{center}
{Extremum 16}
\end{center}

&

{%
\setlength{\fboxsep}{8pt}%
\setlength{\fboxrule}{0pt}%
\fbox{\includegraphics[width=3.5cm]{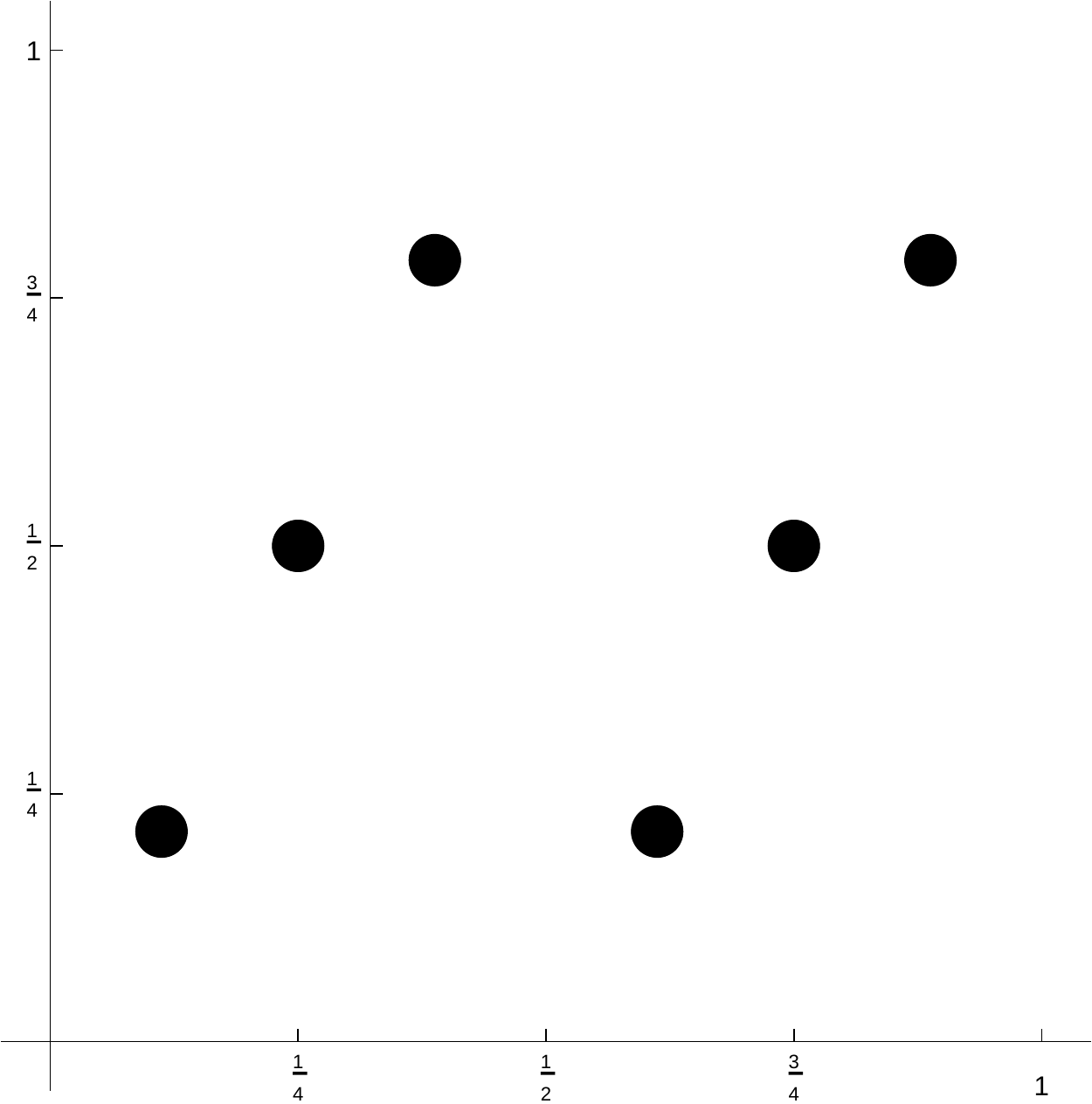}}%
}%
\begin{center}
{Extremum 17}
\end{center}

&
 
\end{tabular}
\end{minipage}

\end{document}